\documentclass[prd,epsf,aps,twocolumn,letterpaper,preprintnumbers]{revtex4}
\usepackage{amssymb}
\usepackage{amscd}
\usepackage{amsfonts}
\usepackage{epsf}
\usepackage{comment}
\usepackage{graphicx}
\begin{document}
\preprint{UTTG-02-02}
\preprint{gr-qc/0202074}

\title{ Basic Principles of 4D Dilatonic Gravity and
        Some of Their Consequences for Cosmology, Astrophysics and
        Cosmological Constant Problem}
\author{P.~P.~Fiziev% 
\footnote{
        On leave from Department of Theoretical Physics, 
	University of Sofia, 5 James Bourchier Blvd., 
	1164 Sofia, Bulgaria.
	}%
\footnote{
	Electronic address: {\tt fiziev@phys.uni-sofia.bg}\\
	\hbox{}\hskip 2truecm and \,\,\,{\tt fiziev@zippy.ph.utexas.edu}.
}
}
\address{
Theory Group, Department of Physics, University of Texas at Austin,
Austin, Texas 78712-1081, U.S.A.
}
\begin{abstract}
We present a class of simple scalar-tensor models of gravity
with one scalar field (dilaton $\Phi$) and {\em only one}
unknown function (cosmological potential $U(\Phi)$).
These models might be considered as a stringy inspired ones
with  broken SUSY. They have the following basic properties:
1) Positive dilaton mass, $m_{{}_\Phi}$,
and positive cosmological constant, $\Lambda$,
define two extremely different scales.
The models under consideration are consistent
with the known experimental
facts if $m_{{}_\Phi}>10^{-3}\,{\rm  eV}$ and
$\Lambda =\Lambda^{obs}\sim 10^{-56}\,{\rm cm}^{-2}$.
2) Einstein weak equivalence principle is strictly satisfied
and extended to scalar-tensor theories of gravity by using a novel
form of principle of ``constancy of fundamental constants''.
3) The dilaton plays simultaneously roles of an inflation field
and a quintessence field and yields
a sequential {\em hyper}-inflation with a graceful exit to
asymptotic de Sitter space-time, which is an attractor,
and is approached as $\exp(-\sqrt{3\Lambda^{obs}}\,ct/2)$.
The time duration of the inflation is
$\Delta t_{infl}\sim m_{{}_\Phi}^{-1}$.
4) Ultra-high frequency ($\omega_{{}_\Phi}\sim m_{{}_\Phi}$)
dilatonic oscillations take place in the asymptotic regime.
5) No fine tuning.  (The Robertson-Walker solutions of
{\em general type} have the above properties.)
6) A novel adjustment mechanism for the cosmological constant problem
seems to be possible: the huge value of the cosmological constant
in the stringy frame is rescaled to its observed value
by dilaton after transition to the {\em phenomenological frame}.

\noindent{PACS number(s): 04.50.+h, 04.40.Nr, 04.62.+v}
\end{abstract}
%%%%%%%%%%%%%%%%%%%%%%%%%%%%%%%%%%%%%%%%%%%%%%%%%%%%%%%%%%%%%%%%%%%
\maketitle
%
%\draft
\sloppy
%\scrollmode
%%%%%%%%%%%%%%%%%%%
\newcommand{\lfrac}[2]{{#1}/{#2}}
\newcommand{\sfrac}[2]{{\small \hbox{${\frac {#1} {#2}}$}}}
\newcommand{\ben}{\begin{eqnarray}}
\newcommand{\een}{\end{eqnarray}}
\newcommand{\la}{\label}
\newcommand{\BBox}{{\square}}
\newcommand{\Si}{{\rm Si}\,}
%
%%%%%%%%%%%%%%%%%%%%%%%%%%%%%%%%%%%%%%%%%%%%%%%%%%%%%%%%%%%%%%%%%%%%%%%%%
\section{Introduction}
The recent astrophysical observations of the type
Ia supernovae \cite{Ia}, CMB \cite{CMB},
gravitational lensing  and galaxies clusters' dynamics
(see the review articles \cite{CosmTri}
and the references therein)
gave us {\em strong and independent}
indications of existence of a new kind of dark energy
in the Universe needed to explain the accelerated
expansion and other observed phenomena.
Although we are still not completely confident
in these new observational results,
it is worth trying to combine
them with the old cosmological problems.
Most likely, the conclusion one would reach
is that a further generalization
of the well established fundamental laws of physics
and, in particular, of laws of gravity,
is needed \cite{Snowmass}.

At present, general relativity (GR) is the most  successful theory
of gravity at scales of laboratory, Earth-surface, Solar-System
and star-systems.
It gives quite good description of gravitational phenomena in
the galaxies and at the scale of the whole Universe \cite{WW}.
Nevertheless, without some essential changes
of its structure and basic notions,
or without introducing some unusual matter and/or energy,
GR seems to be unable to explain:

$\bullet$ the rotation of galaxies \cite{C},

$\bullet$ the motion of galaxies in galactic clusters \cite{C},

$\bullet$ physics of ultra-early Universe  \cite{WW,C},

$\bullet$ the inflation \cite{Snowmass, C, inflation},

$\bullet$ the initial singularity problem,

$\bullet$ the famous vacuum energy problem \cite{Weinberg}, and

$\bullet$ the present-day accelerated expansion of the
          Universe \cite{Ia}-\cite{Snowmass}.

The most promising modern theories of gravity,
like super gravity (SUGRA)
and (super)string theories ((S)ST) \cite{Strings},
having a deep {\em theoretical basis},
incorporate naturally GR. Unfortunately,
they are not developed enough to allow
{\em a real experimental test},
and introduce a large number of new
fields without any {\em direct experimental evidence}
and phenomenological support.

Therefore, it seems meaningful to look for some {\em minimal
extension} of GR which is compatible with the known gravitational
experiments, promises to overcome at least some of the above
problems, and may be considered as a {\em phenomenologically supported}
and necessary part of some more general modern theory.

In the present article we consider such a minimal model,
which we call a {\em four-dimensional-dilatonic-gravity} (4D-DG).
Up to now, this model has not attracted much attention.
The investigation of 4D-DG was started by O'Hanlon
as early as in 1974 \cite{OHanlon}
in connection with Fujii's theory of the massive dilaton
\cite{Fujii}, but without any relation to the cosmological
constant problem or other problems in cosmology and
astrophysics. A similar model
appears in the $D=5$ Kaluza-Klein theories \cite{Fujii97}.
The relation of this model with cosmology and the cosmological
constant problem was studied in \cite{F00},
where it was named {\em a minimal dilatonic gravity} (MDG).
Possible consequences of 4D-DG for boson star structure
were studied in \cite{PF2}.
Some basic properties of 4D-DG were considered briefly in
\cite{E-F_P} in the context of general scalar tensor
theories. There, the exceptional status of the 4D-DG
among other scalar-tensor theories was stressed and
a theory of cosmological perturbations for 4D-DG was sketched.

A wider understanding of
dilatonic gravity as a metric theory of gravity in different
dimensions with one non-matter scalar field
can be found in the recent review article \cite{Odintsov}.
There, one can also find many examples
of such models and a description of corresponding quantum
effects. In contrast, we use the term 4D-DG only for
our specific model.

In this article, we give a detailed consideration of the basic
principles of the 4D-DG model, its experimental grounds,
and some of its possible applications to astrophysics,
cosmology and the cosmological constant problem.
We believe that further developments of this model will yield
a more profound understanding both of theory of gravity
and of modern theories for unifying fundamental physical
interactions. The 4D-DG model seems to give an interesting
alternative for further development of these theories on
real physical grounds.

In Section II, we consider briefly the modern foundations
of scalar-tensor theories of gravity. In particular,
we outline their connection with the universal sector of
string theories and introduce our basic notations.

Section III is devoted to the role of Weyl's conformal
transformations outside the tree-level approximation
of string theory. We discuss in detail the choice of frame
and consider three distinguished frames: Einstein frame,
cosmological constant frame and twiddle frame. Then, after
a short review of basic properties of phenomenological
frame, we discuss the problem of the choice of one of these
distinguished frames as a phenomenological frame.
Using some novel form of principle of ``constancy of
fundamental constants,'' we choose the twiddle frame for
phenomenological frame, thus arriving  at our 4D-DG
model in four-dimensional space-time.

In Section IV, we describe in detail our model.

There, we introduce a new system of cosmological units
based on the observable value of cosmological constant
$\Lambda^{obs}$ and dimensionless Planck number
$P=\sqrt{\Lambda^{obs}} L_{{}_{Pl}}\approx 10^{-61}$,
where $L_{{}_{Pl}}$ is Planck length.

Then we consider the basic properties of vacuum states
in 4D-DG and the properties of admissible cosmological
potentials. We show that the mass of dilaton in
4D-DG must have nonzero value.

In Section V, the weak field approximation for static system
of point particles in 4D-DG is considered. We discuss
the equilibrium between Newtonian gravity and weak
anti-gravity, the constrains on the mass of dilaton
from Cavendish-type experiments, the basic Solar
System gravitational effects (Nordtvedt effect, time delay
of electromagnetic pulses, perihelion shift) and
possible consequences of big dilaton mass for star structure.

Section VI is devoted to some applications of 4D-DG
in cosmology. We consider Robertson-Walker metric in
4D-DG, different forms of novel basic equations for
evolution of Universe, energetic relations and some
mathematical notions, needed for analysis of this
evolution.

Then we derive the general properties of solutions in
4D-DG Robertson-Walker Universe and show the existence of
asymptotic de Sitter regime with ultra-high frequency
oscillations for all solutions, and the existence of
initial inflation with dilaton field as inflation field.
We obtain novel 4D-DG formulae for the number of e-folds
and time duration of the inflation,
the latter turns out to be related with the mass of dilaton
via some new sort of quantum-like uncertainty relation.

In sharp contrast to standard inflation models and
known quintessence models, the mass of the scalar field
in 4D-DG (i.e., the mass of dilaton)
is supposed to be very large, most probably in the TeV domain.

In addition, we give a solution of the inverse
cosmological problem in 4D-DG. This solution differs
significantly from the ones in other cosmological models.

The history of science teaches us that in the cases
when a solution of some problem is not found for a long
time, it is useful to reformulate the problem
and look for some new approach to it.
The essence of the cosmological constant problem
is to find a physical explanation of the extremely small
value of Planck number.
This number connects the observed small value of the cosmological
constant and the huge value of this quantity predicted
by quantum field theory . On the other hand, it turns out that the same
Planck number is related to the ratio of the classical action
in the Universe and the Planck constant $\hbar$.
In Section VII, we give very crude estimates for the amount
of classical action accumulated during the evolution of
the Universe after inflation in the matter sector and in 4D-DG
gravi-dilaton sector. Then we describe qualitatively
a novel idea for solution of the cosmological constant
problem. It turns out that one can have a huge cosmological
constant in basic stringy frame, due to the quantum
vacuum fluctuation, but after transition to phenomenological
frame this value is rescaled by the vacuum value of the
dilaton field to the observed small positive cosmological
constant through Weyl conformal transformation.

In the concluding Section VIII, we discuss some open
problems of 4D-DG.

Mathematical proofs of some important statements
are given in Appendices A and B.

\section{The Scalar-Tensor Theories of Gravity
and Their Modern Foundations}

Most likely, the  minimal extension of GR
must include at least one new scalar-field-degree
of freedom.
Indeed, such a scalar field is an unavoidable part of all
promising attempts to generalize GR,
starting with the first versions of Nordstr\"om and 
Kaluza-Klein-type theories, scalar-tensor theories of gravity,
SUGRA, (S)ST (in all existing versions), M-theory, etc.
In these modern theories, there is a {\em universal sector},
which we call in short a {\em gravi-dilaton sector}.
Using the well known Landau-Lifschitz conventions, we write
its action in some basic frame (BF)
in the following most general form:
\ben
{\cal A}_{g\phi}=
\!-{c \over{2 \kappa}}\!\int\!\! d^D x\sqrt{|{\bf g}|}
\bigl(F(\phi) R\! -\!4Z(\phi) (\nabla\phi)^2\!+
\! 2\Lambda(\phi) \bigr).
\la{A_GD}
\een
The contribution of the scalar field $\phi$ to the action
of the theory can be described in different
(sometimes physically equivalent)
ways, by choosing different functions
$F(\phi), Z(\phi)$ and $\Lambda(\phi)$
(which are not fixed a priory).
If the basic frame is to be considered as a physical frame,
the coefficients $F(\phi)$ and $Z(\phi)$ have to obey the
general requirements $F(\phi)>0$ and $Z(\phi) \geq 0$.
These conditions ensure non-negativity of the
kinetic energy of graviton and dilaton.
(The negative values of the function $F(\phi)$
correspond to anti-gravity,
and a zero value yields infinite
effective gravitational constant.)

In addition to the gravi-dilaton sector,
we assume that there exists some matter sector
with spinor fields $\psi$, gauge fields $A$, $\ldots$,
relativistic fluids, etc.,
and action:
\ben
{\cal A}_{matt}=
{1\over c}\!\int\!\! d^D x
\sqrt{|{\bf g}|}\,{\cal L}(\psi,\nabla\psi;A,\nabla A;
...;g_{\mu\nu},\phi).
\la{Amat}
\een
Then the variation of the total action,
${\cal A}_{tot}={\cal A}_{g\phi}+{\cal A}_{matt}$,
with respect to the metric $g_{\mu\nu}$ and the dilaton $\phi$
(after excluding the scalar curvature $R$ from the variational
equation for scalar field in the case $F_{,_\phi}\neq 0$)
yields the following field equations
\footnote{In the case $F={\rm const}=1$, we have the usual GR
system of field equations for the metrics $g_{\mu\nu}$
and scalar field $\phi$.}:
\ben
F G_{\mu\nu}= {\kappa\over {c^2}}
T_{\mu\nu}+\hskip 5,65truecm \nonumber \\
4 Z\!\left(\!\phi_{\!,_\mu}\phi_{,_\nu}\!-
\!{1\over 2}\!\left(\nabla\phi\right)^2\!g_{\mu\nu}\!\right)
\!\!+\!\left({\nabla_\mu\nabla_\nu\!-
\!g_{\mu\nu}\BBox}\right)F\!+\!
\Lambda(\phi)g_{\mu\nu},\nonumber \\
J\,\BBox \phi + {1\over 2}J_{,_\phi}
\left(\nabla\phi\right)^2 +V_{\!,_\phi}=
{1\over{D\!-\!1}}\,{\kappa\over {c^2}}\,
F_{\!,_\phi}\Theta. \hskip 1.95truecm
\la{GFE}
\een
Hereafter, the comma denotes partial derivative
with respect to the corresponding variable and
\ben
J(\phi)=F^2_{\!,_\phi}+ 4{{D\!-\!2}\over{D\!-\!1}}F Z,
\nonumber \hskip 4truecm\\
V(\phi)={{D\!-\!2}\over{D\!-\!1}}F(\phi)\Lambda(\phi)-
2 \int \Lambda(\phi)F_{\!,_\phi}(\phi)d\phi,
\nonumber \hskip .95truecm\\
\Theta=T +
(D\!-\!2) \left(\ln{}_{{}_S\!}
F\right){}_{\!\!,_\phi}{\cal L}_{\!,_\phi}.
\hskip 3.7truecm
\la{JVT}
\een
The tensor $T_{\nu\mu} = {2\over {\sqrt{|{\bf g}|}}}
{{\delta{\cal L}}\over{\delta {g^{\mu\nu}}}}$
is the standard energy-momentum tensor of matter,
and $T$ is its trace.

In addition, we have two relations:
\ben
{{D\!-\!2}\over 2}\!\left(4Z\left(\nabla\phi\right)^2\!-
\!FR\right)
\!+\!(D\!-\!1)\BBox F\!-\!D\Lambda
\!=\!{\kappa\over{c^2}}T,
\la{R_rel1}
\een
\ben
F_{\!,_\phi}R + 8Z\BBox\phi+
4Z_{\!,_\phi}\left(\nabla\phi\right)^2 +
2\Lambda_{\!,_\phi}=
2 {\kappa\over {c^2}}{\cal L}_{\!,_\phi}.
\la{R_rel2}
\een
The first one is obtained from the trace of
the generalized Einstein equation in (\ref{GFE}).
One can derive the second one from the system
(\ref{GFE}), but actually it is a direct
result of the variation of the total
action with respect to the dilaton field $\phi$.
Nevertheless, in the case $F_{,_\phi}\neq 0$,
we consider the second of the relations in (\ref{GFE})
as a field equation for the dilaton $\phi$,
instead of the relation (\ref{R_rel2}).

There have been many attempts to construct a realistic theory
of gravity with action (\ref{A_GD}),
starting with Jordan-Fierz-Brans-Dicke theory of
variable gravitational constant and
its further generalizations.
The so called scalar tensor
theories of gravity \cite{BD}
have been considered as a most natural extension of GR
\cite{Damoor+} from phenomenological point of view.
Different models of this type have been used in the
inflationary scenario \cite{inflation} and in the more recent
quintessence models \cite{Q}. For the latest developments of the
scalar-tensor theories in connection with the
accelerated expansion of the Universe, one
can consult the recent article \cite{E-F_P}.

It is natural to look for a more fundamental
theoretical evidence in favor of the action (\ref{A_GD}).

For example, a universal gravi-dilaton sector
described by action of type (\ref{A_GD})
appears in minimal $D=4$, $N=1$ SUGRA.
(See the recent article \cite{Townsend} and the references therein.)
In this model, the scalar field $\phi$ belongs to chiral supermultiplet,
$F(\phi)=1$, $Z(\phi)= 2$ and,
to obtain a general potential in a form
$\Lambda(\phi) = (w^\prime)^2 - \beta^2 w^2 + \xi^2 h$,
one needs to include one vector multiplet,
which is coupled to the scalar field
through the real function $h(\phi)$.
The function $w(\phi)$ describes the
corresponding real superpotential,
$\beta$ is a phenomenological constant
related to the matter equation of state,
and $\xi$ is Fayet-Illiopoulos constant.
Similar potentials $\Lambda(\phi)$ appear in $N=8$ SUGRA,
as well as in the brane world picture \cite{Townsend}.

The action of type (\ref{A_GD}) is common
for all modern attempts to create a unified
theory of all fundamental interactions based on
stringy idea.

Indeed, consider the {\em universal sector} of
the low energy limit (LEL) of (S)ST in stringy frame (SF),
which is the basic frame in this case.
The gravi-dilaton Lagrangian is:
$${}_{{}_S\!}L_{{}_{LEL}}^{(0)} \sim \sqrt{|{}_{{}_S\!}{\bf g}|}e^{-2\phi}
\left({}_{{}_S\!}R + 4\,
{}_{{}_S\!}(\nabla \phi)^2 +2 {}_{{}_S\!}C_\Lambda^{(0)}\right).$$
We use the upper index
${(0)}$ to label the tree-level-approximation quantities,
\ben
{}_{{}_S\!}F^{(0)}(\phi)=-{}_{{}_S\!}Z^{(0)}(\phi) = e^{-2\phi},\,\,
\Lambda^{(0)}(\phi)= {}_{{}_S\!}C_\Lambda^{(0)} e^{-2\phi} ,
\la{Stree}
\een
and the ``cosmological constant'' is
${}_{{}_S\!}C_\Lambda^{(0)}={\frac {D-26} {3\alpha^\prime}}$
for bosonic strings, and
${}_{{}_S\!}C_\Lambda^{(0)}\sim (D-10)$
for superstrings,
and $\alpha^\prime$ is Regge slope parameter.
Including the contribution of all loops,
one arrives at the following general form
of LEL stringy gravi-dilaton Lagrangian:
\ben
{}_{{}_S\!}L_{{}_{LEL}} \sim \hskip 3truecm\nonumber \\
\sqrt{|{}_{{}_S\!}{\bf g}|}
e^{-2\phi}\left({}_{{}_S\!}C_g(\phi){}_{{}_S\!}R +
4\,{}_{{}_S\!}C_\phi(\phi)(\nabla \phi)^2 +
2\,{}_{{}_S\!}C_\Lambda(\phi)\right),
\la{SL}
\een
where
\ben
{}_{{}_S\!}C_{...}(\phi)\!=
\!\sum_{n=0}^{\infty}{}_{{}_S\!}C_{...}^{(n)}\!\exp(2n\phi)
\la{C}
\een
are {\em unknown} functions with
${}_{{}_S\!}C_g^{(0)}=1$, ${}_{{}_S\!}C_\phi^{(0)}=1$.
(See the references \cite{Pol_Dam} where functions
$B_{...}(\phi)=e^{-2\phi}C_{...}(\phi)$ were introduced.
Here dots {...} stand for $g$, or $\phi$.)

The SF cosmological potential
${}_{{}_S\!}\Lambda(\phi)$
{\em must be zero in the case of  exact
supersymmetry}, but in the real world such
nonzero term may
originate from SUSY breaking due
to super-Higgs effect,
gaugino condensation, or may appear in some
more complicated, still unknown, way.
Its form is {\em not known exactly}, too.
At present, the only clear thing is that
{\em we are not living
in the exactly supersymmetric world}
and one must somehow break down the SUSY.
We consider this {\em phenomenological fact}
as a sufficient evidence in favor of the assumption that
in a physical theory which describes
{\em the real world},
${}_{{}_S\!}\Lambda(\phi)\neq 0$
both for critical and for non-critical
fundamental strings.
Thus, we use the nonzero cosmological potential $\Lambda(\phi)$
to describe pnenomenologically the SUSY breaking.

Hence, to fix the LEL gravi-dilaton Lagrangian in SF,
we have to know the {\em three} dressing functions of the dilaton:
${}_{{}_S\!}C_g(\phi)$, ${}_{{}_S\!}C_\phi(\phi)$,
and ${}_{{}_S\!}C_\Lambda(\phi)$.

This way, we arrive at a scalar-tensor theory
of gravity of most general type (\ref{A_GD})
with some specific stringy-determined functions:
\ben
{}_{{}_S\!}F(\phi)={}_{{}_S\!}C_g(\phi)e^{-2\phi},\,\,\,
{}_{{}_S\!}Z(\phi)=-{}_{{}_S\!}C_\phi(\phi)e^{-2\phi},\nonumber \\
{}_{{}_S\!}\Lambda(\phi)={}_{{}_S\!}C_\Lambda(\phi)e^{-2\phi}.
\hskip 1.5truecm
\een
In this paper, we consider general scalar theories
of gravity in this stringy context.
Although we use stringy terminology,
our considerations are valid for
all scalar-tensor theories.
We choose this language for describing our model
simply because the (S)ST,
their brane extensions, and M-theory
at present are the most popular candidates for
``theory of everything''.

In addition to the gravi-dilaton sector,
in these modern theories, there are many other fields:
axion field, gauge fields, different spinor
fields, etc., which we do not consider here in detail.
For spinor fields $\psi$ and for gauge fields $A_\mu$,
one has to add to the total Lagrangian of the theory terms
that in flat space-time ${\cal M}^{D}$ have the form
\ben
{}_{{}_S\!}C_\psi(\phi)e^{-2\phi}\,\,
\bar\psi\gamma^\mu\partial_\mu\psi=
{}_{{}_S\!}B_\psi(\phi)\,\,\bar\psi\gamma^\mu\partial_\mu\psi,
\nonumber \\
{}_{{}_S\!}C_m(\phi)e^{-2\phi}\,\,m \bar\psi\psi=
{}_{{}_S\!}B_m(\phi)\,\,m \bar\psi\psi, \nonumber \\
{}_{{}_S\!}C_e(\phi)e^{-2\phi}\,\,e\,
A_\mu \bar\psi\gamma^\mu\psi=
{}_{{}_S\!}B_e(\phi)\,\,e\,
A_\mu \bar\psi\gamma^\mu\psi, \nonumber \\
{}_{{}_S\!}C_{{}_F}(\phi)e^{-2\phi}\,\,F_{\mu\nu}F^{\mu\nu}=
{}_{{}_S\!}B_{{}_F}(\phi)\,\,F_{\mu\nu}F^{\mu\nu}
\la{psi}
\een
with unknown coefficients ${}_{{}_S\!}C_{...}(\phi)$
of type (\ref{C}). The connection of these terms
with the {\em real matter} is not clear at present.
Therefore, we describe the real matter phenomenologically,
i.e., at the same standard manner as in GR,
using the available experimental information.

%%%%%%%%%%%%%%%%%%%%%%%%%%%%%%%%%%%%%%%%%%%%%%%%%%%%%%%%%%%%

\section{The Transition to New Frames Using
Weyl Conformal Transformations}

After Weyl conformal transformation:
\ben 
g_{\mu\nu} \rightarrow e^{ -2\sigma(\phi)} g_{\mu\nu}
\la{Weyl}
\een
to some {\em new conformal frame},
ignoring a surface term which is proportional to
$2(D-1)\sqrt{|{\bf g}|}\,{}_{{}_S\!}Fe^{(D\!-\!2)
\sigma}g^{\mu\nu}\sigma_{,_\nu}$,
we obtain the stringy
LEL Lagrangian (in $D$ dimensions) in the form
\ben
L_{{}_{LEL}}\sim \sqrt{|{\bf g}|}
\left(F(\phi)R-4 Z(\phi)(\nabla \phi)^2+
2\Lambda(\phi)\right)
\la{L}
\een
where
\ben
F(\phi)={}_{{}_S\!}F(\phi)e^{(D-2)\sigma(\phi)},\,
Z(\phi)= {}_{{}_S\!}\tilde Z(\phi)e^{(D-2)\sigma(\phi)}, \nonumber\\
\Lambda(\phi)={}_{{}_S\!}\Lambda(\phi)e^{D\sigma(\phi)},\hskip 4.6truecm
\la{S_FZL}
\een
\ben
{}_{{}_S\!}\tilde Z(\phi)=
{}_{{}_S\!}Z(\phi)+ \Delta{}_{{}_S\!}Z(\phi),
\hskip 3.4truecm \nonumber\\
\Delta\!{}_{{}_S\!}Z(\phi)\!= -{{D\!-\!1}\over 2}\,{}_{{}_S\!}F\!
\left(\left(\ln\!{}_{{}_S\!}F\right){}_{\!\!,_\phi}\,
\sigma_{\!,_\phi}\!+\!
{{(D\!-\!2)}\over 2}\,\sigma_{\!,_\phi}^2\!\right).
\la{ZZ}
\een
Combining relations (\ref{S_FZL}) and (\ref{ZZ}),
we obtain the transformation law
\ben
{Z\over F}\!=\!{{{}_{{}_S\!}Z}\over {{}_{{}_S\!}F}}
\!-\!{{D\!-\!1}\over 2}
\left(\ln\!{}_{{}_S\!}F\right){}_{\!\!,_\phi}\sigma_{\!,_\phi}\!-\!
{{(D\!-\!1)(D\!-\!2)}\over 4}\left(\sigma_{\!,_\phi}\right)^2.
\la{Z/F}
\een

The transition functions $\sigma$ have the following
{\em pseu\-do-group} property:

If ${}_{{}_1\!}\sigma_{\!{}_0}$ and  ${}_{{}_2\!}\sigma_{\!{}_0}$
describe transitions from some initial frame
0F to some new frames 1F and 2F according Eq.~(\ref{Weyl}),
(i.e., ${}_{{}_I}g_{\mu\nu} = 
e^{ -2{}_{{}_I}\!\sigma\!{}_{{}_J}(\phi)}{}_{{}_J}g_{\mu\nu}$ 
for $I,J=0,1,2;$) 
then the transition from 1F to 2F is given by
\ben
{}_{{}_2\!}\sigma_{\!{}_1} =
{}_{{}_2\!}\sigma_{\!{}_0} - {}_{{}_1\!}\sigma_{\!{}_0} \, .
\la{sigma21}
\een
The relations (\ref{S_FZL})--(\ref{Z/F}) give
a specific induced representation of Weyl
transformations (\ref{Weyl}) which acts
on coefficients $F$, $Z$, and $\Lambda$
in the Lagrangian (\ref{L}) and has
the corresponding pseudo-group property.

As seen from the above relations,
for known stringy-dressed coefficients
${}_{{}_S\!}F(\phi)$, ${}_{{}_S\!}Z(\phi)$
and given ${}_{{}_S\!}\tilde Z(\phi)$,
one obtains in general (i.e., for $D>2$)
{\em two} transition functions:
\ben
\sigma^\pm(\phi)={{1}\over {D\!-\!2}}\left(-\ln {}_{{}_S\!}F(\phi)
\pm {2\over{\sqrt{D\!-\!1}}}
\tilde S(\phi)
\right) \, ,
\la{sigma}
\een
where
$\tilde S(\phi)=\int d\phi \sqrt{\Delta({}_{{}_S\!}F(\phi),
{}_{{}_S\!}Z(\phi)-{}_{{}_S\!}\tilde Z(\phi))}$,
and the following important combination of functions
have been introduced:
\ben
\Delta(F,Z)={{D\!-\!1}\over {4 F^2} }J={{D\!-\!1}\over {4} }
\left(\ln F\right)_{\!,_\phi}^2
+(D\!-\!2){Z \over F}
\la{Delta}
\een
with normalization
\ben
\Delta\left({}_{{}_S\!}F^{(0)}(\phi),
{}_{{}_S\!}Z^{(0)}(\phi)\right)
\equiv 1
\la{NDelta}
\een
and  basic property
\ben
\Delta(F,Z)\geq 0 \,,
\la{Delta0}
\een
when $ 4(D\!-\!2) {Z\over F}\geq
-(D\!-\!1) \left(\ln F\right)_{\!,_\phi}^2$.

\subsection{The Choice of Frame}

Now the following question arises: what frame to choose --
stringy, Einstein, or some other frame ?

This is still an open problem and in the literature
one can find basically different statements
(see the first article in \cite{Faraoni}
for a large amount of references and their detailed analysis).
In the present article, we try to answer this question
by analyzing the situation from different points of view and
making a series of simple steps in the direction which seems
to us to be the right one from {\em phenomenological}
point of view.

\subsubsection{Are All Frames Equivalent ?}

Some authors consider the change of frame as
a {\em formal mathematical procedure}
which is physically irrelevant.
According to this point of view,
all frames are physically equivalent,
at least up to possible singularities
in the corresponding transition functions.

The whole wisdom in this statement is related to
the rather trivial observation, that
if {\em we are given} some physical theory,
we have the freedom to change {\em locally}
variables in any convenient way.
Then we can transform to the new frame any physical law,
sometimes ignoring the fact that in the new frame this
law may have a strange and unusual form
from physical point of view.
This means that one may consider every {\em given}
physical theory in different {\em local} coordinates
in the corresponding (field) phase space.

This physically trivial statement
neglects one of the most important features of
the physical problems,
even when they are well formulated.
Namely, for each problem there exists,
as a rule, a unique ``coordinate system''
which is {\em proper for the solution} of the problem.
It is well known that the most important technical
issue for solving any physical problem is to find
this ``proper coordinate system''.

In the language of mathematics, this means that
we have to find the ``unique'' global uniformization variables
for the problem under consideration.
For real problems, this might be a nontrivial
and very complicated mathematical issue.

The naive change of frame may alter the global
properties of the physical system because
of the following reasons:

1) Weyl transformations (\ref{Weyl}) do not form a group,
but a pseudo-group and, in general, they do change the global
structure of space-time and of the physical theory.
Typically, only a part of the space-time manifold
${}_{{}_1}{\cal M}^{(D)}$ of the initial frame 1F is
smoothly mapped onto some part of the space-time manifold
${}_{{}_2}{\cal M}^{(D)}$ of the frame 2F.
In addition, (as seen, e.g., from formula (\ref{sigma})),
the mapping may be not one-to-one.

2) Under Weyl transformation (\ref{Weyl}), the Lagrangian
acquires a surface term proportional to
$$
2(D-1)\oint d\Sigma^\mu \sigma_{,_\mu}\sqrt{|{\bf g}|}\,
{}_{{}_S\!}F\,e^{(D\!-\!2)\sigma}
$$
that we ignore. In space-times with boundary, this may lead to
a physically non-equivalent theory.

There exists one more argument for using different frames.
In the case of the theories we consider in the present article,
the very physical problem is still {\em not completely fixed}.
It seems quite possible that, looking at it in different frames,
one can find some new physical grounds which can help
to restrict in a proper way the {\em a priory existing}
possibilities and to justify the unknown theoretical ingredients.

\subsubsection{Is the Basic  Frame Enough for Doing Physics ?}

If one firmly believes in beautiful theoretical
constructions like (super)strings, branes,
or in some other physical theory,
one may intend to prescribe a {\em direct} physical meaning
to the variables in which these theories
look beautiful and simple.
Therefore, one may consider the basic frame
(SF -- for string theories) as a physical one, i.e.,
as a frame in which we see directly
the properties of the real world.
It seems obvious that one need not accept
such additional hypotheses, i.e.,  the basic variables
of the fundamental theory may have only {\em indirect}
relation with the real world.
Then defining the basic principles
for choosing physical variables becomes
an important theoretical issue.
These principles must be based on some
{\em phenomenological} facts.

In the case of string theory, the wrong sign of the kinetic
term in the SF-LEL-Lagrangian is enough to consider
the basic stringy frame as a non-physical one.
Otherwise, the theory would not have a stable ground state.

\subsection{Three Distinguished Frames for Scalar-Tensor Theories}

Looking at the basic formulae (\ref{S_FZL}) and (\ref{ZZ}), it is not hard
to understand that three simple choices of frame are possible:
since under Weyl conformal transformation
the functions ${}_{{}_S\!}F(\phi)$ and ${}_{{}_S\!}\Lambda(\phi)$
have {\em a linear and homogeneous}
transformation law, one can choose the function $\sigma(\phi)$
in such a way that:

i) $F(\phi)={\rm const}>0$; or

ii) $\Lambda(\phi)={\rm const}$
($>0$ when ${}_{{}_S\!}\Lambda(\phi)>0$).

The third possibility for a simple choice of conformal gauge
is to use the {\em non-homogeneous} linear
transformation law (\ref{ZZ}) for the function ${}_{{}_S\!}Z(\phi)$
and to impose the conformal gauge fixing condition:

iii) $Z(\phi)={\rm const}=0$.

It is remarkable that in each of these three cases
one can reduce the number of the unknown
functions in the gravi-dilaton sector to one
(by using a proper re-definition of the dilaton field).
That is why, before discussing the choice of some frame
as the physical one, we describe briefly
their properties.

\subsubsection{Einstein Frame}

The most popular and well known frame is Einstein frame (EF),
defined according to the first choice (i) when $D\!>\!2$.
(If $D=2$ and ${}_{{}_S\!}F(\phi)\neq {\rm const}$, EF does not exist.)
The transition from SF to EF is described by transition function
${}_{{}_E\!}\sigma{}_{\!{}_S}=
-{{1}\over{D\!-\!2}}\ln {}_{{}_S\!}F$ and coefficients
\ben
{}_{{}_E\!}F(\phi)\equiv 1,\,\,
{}_{{}_E\!}Z(\phi)=
{1\over {D\!-\!2}}
\Delta\left({}_{{}_S\!}F(\phi),{}_{{}_S\!}Z(\phi)\right),\,\,
\nonumber \\
{}_{{}_E\!}\Lambda(\phi)= {}_{{}_S\!}F(\phi)^{-{{D}\over{D-2}}}
{}_{{}_S\!}\Lambda(\phi). \hskip 1.3truecm
%\la{EF1}
\een

For tree-level string approximation, one easily obtains
the familiar LEL coefficients
$${}_{{}_E\!}Z^{(0)}(\phi)= {{1}\over{D-2}}={\rm const}>0,\,
{}_{{}_E\!}\Lambda^{(0)}(\phi)=
{}_{{}_S\!}C^{(0)}e^{{4\phi}\over{D-2}}.$$
For the general case of a dressed LEL Lagrangian, one has to re-define
the dilaton field, introducing {\em dimensionless} EF-dilaton $\varphi$
according to the formula \footnote{We prefer not to include
the dimensional factor ${c\over\kappa}$
in the definition of EF-dilaton.
More often one prefers to use a definition
like $\varphi(\phi)=\pm\sqrt{{4\over{D\!-\!2}}{c\over\kappa}} S(\phi)$.
Then the dimensional EF-dilaton is measured in Planck-mass units.}:
\ben
\varphi(\phi)=\varphi^\pm(\phi)=\pm\sqrt{{4\over{D\!-\!2}}} S(\phi) \ ,
\la{Ephi}
\een
where
\ben
S(\phi)=\int d\phi \sqrt{\Delta
\left({}_{{}_S\!}F(\phi),{}_{{}_S\!}Z(\phi)\right)}
\la{S}
\een
is real if the condition (\ref{Delta0}) is fulfilled
for the dressed coefficients ${}_{{}_S\!}F(\phi)$
and ${}_{{}_S\!}Z(\phi)$.

The existence of {\em two} solutions
$\varphi^{+}(\phi)=-\varphi^{-}(\phi)$
reflects the well known S-duality of string theory.
In EF this duality corresponds to the invariance of the
metric ${}_{{}_E\!}g_{\mu\nu}$ under
a simple change of the sign of the dilaton field $\varphi$.

This way, we reach the final form (\ref{A_GD})
of the EF-LEL stringy Lagrangian
with coefficients
\ben
{}_{{}_E\!}F(\varphi)= 1,\,\,
{}_{{}_E\!}Z(\varphi)= {1\over 4},\,\,
{}_{{}_E\!}\Lambda(\varphi)=
\Lambda^{obs}{}_{{}_E\!}U(\varphi) \ ,
\la{EF}
\een
where $\Lambda^{obs}$ is a constant
which we choose to  equals the positive
{\em observed} cosmological constant,
and the dimensionless EF cosmological potential is
\ben
{}_{{}_E\!}U(\varphi)=
{{{}_{{}_S\!}\Lambda(\phi(\varphi))}\over {\Lambda^{obs}}}
\bigl({}_{{}_S\!}F(\phi(\varphi))\bigr)^{-{D\over{D\!-\!2}}},
\la{EU}
\een
$\phi(\varphi)$ being the inverse function to the function (\ref{Ephi}).

Now we have a standard EF-representation
not only for the tree-level LEL,
but for the entire {\em dressed} LEL stringy Lagrangian (\ref{SL}).
Its EF representation reads
\ben
{}_{{}_E\!}L_{{}_{LEL}} \sim \sqrt{|{}_{{}_E\!}{\bf g}|}
\left({}_{{}_E\!}R - {}_{{}_E\!}(\nabla \varphi)^2 +
2\Lambda^{obs}{}_{{}_E\!}U(\varphi)\right).
\la{EL}
\een

In this frame

i) Dilaton degree of freedom is separated
from Hilbert-Einstein term ($\sim R$) in the Lagrangian,
and, to some extend,
but not exactly, EF-fields' coordinates play the role
of normal coordinates for the gravi-dilaton sector.

ii) The EF-dilaton $\varphi$ looks like a normal {\em matter}
scalar field with the right sign of its kinetic energy
in the action (\ref{A_GD}) if the inequality (\ref{Delta0})
is fulfilled.

In the case of a negative function
$\Delta({}_{{}_S\!}F(\phi),{}_{{}_S\!}Z(\phi))<0$, one is not able
to introduce {\em positive} kinetic energy for a {\rm real} EF-dilaton
and the string theory in EF will not have a consistent
physical interpretation.

This observation raises the question,
is really EF the proper physical
frame for (S)ST outside the tree-level approximation.
To answer this question, one needs to know the total stringy dressed
coefficients ${}_{{}_S\!}F(\phi)$ and ${}_{{}_S\!}Z(\phi)$,
or at least one needs to have an independent
proof of validity of condition (\ref{Delta0}).
This is still an open problem, and further on we accept
the hypothesis that the condition (\ref{Delta0})
is valid in the scalar-tensor theories under consideration.

iii) As a normal matter field, the EF-dilaton $\varphi$
is {\em minimally} coupled to gravity
(i.e., to EF-metric tensor ${}_{{}_E\!}g_{\mu\nu}$),
and respects Einstein WEP.

Hence, {\em the EF-dilaton $\varphi$ may enter
the matter Lagrangian of other matter fields
in a rather arbitrary way without violation of WEP}.
The only consequence one can derive from WEP in this case
is the {\em metric character of gravity},
described only by the EF metric ${}_{{}_E\!}g_{\mu\nu}$.
One can consider {\em a priory}
arbitrary interactions of the EF-dilaton $\varphi$
with other matter.
For example, the theory does not exclude
{\em a priory} interactions, described in EF by formulae
analogous to Eq.~(\ref{psi}) with proper coefficients
${}_{{}_E\!}C_{...}(\varphi)$. Because of the interpretation
of the EF-dilaton $\varphi$ as an ordinary matter field,
in this case one would be forced to explain the deviations
of particle motion from geodesic lines (with respect
to the metric ${}_{{}_E\!}g_{\mu\nu}$) by introducing some
specific ``dilatonic charge''
(often called ''an interaction parameter'')
which determines the interaction
of dilaton with other matter fields.

iv) The cosmological potential ${}_{{}_E\!}U(\varphi)$
remains {\em the only} unknown function
in the EF-gravi-dilaton sector,
but the dependence of the matter Lagrangian
${}_{{}_E\!}{\cal L}_{...}(...,\varphi)$
on the EF-dilaton $\varphi$ is {\em a new physical problem}
which one must solve to fix the theory.
Here dots stay for other matter fields.

v) In the presence of additional matter of any other
(i.e., different from dilaton $\varphi$) kind with action
$${}_{{}_E\!}{\cal A}{...}=
{1 \over c}\!\int\!\! d^D x\sqrt{|{}_{{}_E\!}{\bf g}|}
{}_{{}_E\!}{\cal L}_{...}(...,\varphi)\ ,$$
the usual GR field equations,
\ben
G_{\alpha\beta}={\kappa\over{c^2}}{}_{{}_E\!}T_{\alpha\beta},
\nonumber \\
{}_{{}_E\!}{\BBox}\,\varphi+
\Lambda^{obs}\!{}_{{}_E\!}U_{\!,_{\varphi}}
={\frac {\kappa} {c^2}} ({}_{{}_E\!}{\cal L}_{...})_{,_{\varphi}}
\hskip 0truecm
\la{EFFEq}
\een
yield the usual energy-momentum conservation law,
$\nabla_\alpha\,{}_{{}_E}\!T^\alpha_\beta=0$,
for the total ener\-gy-mo\-men\-tum of the matter
\ben
{}_{{}_E\!}T_{\alpha\beta} =
{2\over {\sqrt{|{}_{{}_E\!}{\bf g}|}}}
{{\delta {}_{{}_E\!}{\cal L}_{...}}\over{\delta {}_{{}_E\!}{g^{\alpha\beta}}}}+
\hskip 4.1truecm \nonumber \\
{{c^2}\over \kappa}\left(\varphi_{\!,_\alpha}\varphi_{\!,_\beta}\!-\!
{1\over 2}\left(\nabla\varphi\right)^2\!{}_{{}_E\!}g_{\alpha\beta}\!+\!
\Lambda^{obs}\!{}_{{}_E\!}U(\varphi){}_{{}_E\!}g_{\alpha\beta}\right)
\la{EFEMT}
\een
and an additional relation -- the EF version of Eq.~(\ref{R_rel1}):
\ben
R-\left(\nabla\varphi\right)^2 +
{{2D}\over{D\!-\!2}}\Lambda^{obs}\!{}_{{}_E\!}U(\varphi)
+{\kappa\over{c^2}}{2\over{D\!-\!2}}{}_{{}_E\!}T_{...}=0,
\la{ER_rel2}
\een
where ${}_{{}_E\!}T_{...}$ is the trace of energy-momentum tensor
of the additional matter in EF.

vi) As seen from Eq.~(\ref{EFFEq}),
as a matter field in a {\em fixed} metric ${}_{{}_E\!}g^{\mu\nu}$,
the  EF-dilaton $\varphi$
has its own nontrivial dynamics
determined by corresponding Klein-Gordon equation
with cosmological potential ${}_{{}_E\!}U(\varphi)$
in a (curved) space-time ${\cal M}^{(1,3)}\{{}_{{}_E\!}g^{\mu\nu}\}$.
Therefore, the EF-dilaton may be a  variable field in homogeneous
space-times with constant curvature
(in particular, in a flat space-time).

vii) Because of the conservation of the {\em total} energy-momentum,
without taking into account the EF-dilaton $\varphi$,
we have to expect a violation of the conservation of
energy-momentum of other matter
if $({}_{{}_E\!}{\cal L}_{...})_{,_{\varphi}}\neq 0$,
i.e., when the matter is a source for EF-dilaton $\varphi$
according to Eq.~(\ref{EFFEq}).
Hence, the dilaton $\varphi$ is a source of other matter:
\ben
\nabla_\mu\,{}_{{}_E}\!T_{...}{}^\mu_\nu=
-({}_{{}_E\!}{\cal L}_{...})_{,_{\varphi}}\varphi_\nu \,.
\la{EEMC}
\een

\subsubsection{Brans-Dicke-Cosmological-Constant Frame}

We call this new frame a $\Lambda$-frame ($\Lambda$F)
and define it by using the second distinguished possibility
for choice of frame (see subsection B).
Now we impose the conformal gauge condition,
${}_{{}_\Lambda\!}\Lambda(\phi)={\rm const}$,
and by choosing this constant equal to
the observable value $\Lambda^{obs}$, we obtain
${}_{{}_\Lambda\!}\sigma
= -{1\over D}\ln\left({{{}_{{}_S\!}\Lambda(\phi)}
\over{\Lambda^{obs}}}\right)$ and
\ben
{}_{{}_\Lambda\!}F(\phi)=
{}_{{}_S\!}F(\phi)
\left({{{}_{{}_S\!}\Lambda(\phi)}/{\Lambda^{obs}}}
\right)^{-{{D\!-\!2}\over D}},\,\,
{}_{{}_\Lambda\!}\Lambda(\phi)=
\Lambda^{obs},\nonumber \\
{{{}_{{}_\Lambda\!}Z}\over{{}_{{}_\Lambda\!}F}}=
{{{}_{{}_S\!}Z}\over{{}_{{}_S\!}F}}+
{{D\!-\!1}\over{2D}}
{{{}_{{}_S\!}\Lambda_{,_\phi}}\over{{}_{{}_S\!}\Lambda}}
\left({{{}_{{}_S\!}F_{,_\phi}}\over{{}_{{}_S\!}F}}-
{{D\!-\!2}\over{2D}}
{{{}_{{}_S\!}\Lambda_{,_\phi}}\over{{}_{{}_S\!}\Lambda}}
\right) \,.
\la{LF_}
\een
After a re-definition of the $\Lambda$F-dilaton
according to the formula
\ben
\chi={}_{{}_\Lambda\!}F(\phi) \,,
\la{chi}
\een
we obtain the final form of the $\Lambda$F-LEL stringy
Lagrangian coefficients:
\ben
{}_{{}_\Lambda\!}F(\chi)=\chi,\,\,\,
{}_{{}_\Lambda\!}Z(\chi)=\omega(\chi)/\chi,\,\,\,
{}_{{}_\Lambda\!}\Lambda(\chi)=
\Lambda^{obs} \,,
\la{LF}
\een
where the Brans-Dicke coefficient is
\ben
\omega(\chi)= 4\left(\phi^\prime\right)^2
{{{}_{{}_S\!}Z}\over{{}_{{}_S\!}F}}+\nonumber \hskip 4.5truecm\\
2{{D\!-\!1}\over{D}}\left(\ln{}_{{}_S\!}\Lambda \right)^\prime
\left(\ln{}_{{}_S\!}F \right)^\prime -
{(D\!-\!1)(D\!-\!2)\over{D^2}}
\left(\left( \ln{}_{{}_S\!}\Lambda \right)^\prime\right)^2,
\la{BDomega}
\een
$\phi(\chi)$ is the inverse to the function (\ref{chi}),
and prime denotes differentiation with respect to $\ln\chi$.

For the tree-level LEL approximation, one obtains
\ben
\phi\!=\!-{D\over 4}\ln\chi\!+\!{\rm const},\,
{}_{{}_S\!}F^{(0)}\!=\!-{}_{{}_S\!}Z^{(0)}\!\sim\!\chi^{D\over 2},\,
{}_{{}_S\!}\Lambda^{(0)}\!\sim\!\chi^{D\over 2},\nonumber \\
\omega^{(0)}={{D\!-\!2}\over 4}.\hskip 3truecm
\la{omega0}
\een

Now we see that in $\Lambda$F the gravi-dilaton sector looks
precisely like Brans-Dicke theory
with {\em nonzero} cosmological constant, i.e., we have
\ben
{}_{{}_\Lambda\!}L_{{}_{LEL}} \sim \sqrt{|{}_{{}_\Lambda\!}{\bf g}|}
\left(\chi\,{}_{{}_\Lambda\!}R - \chi^{-1}\omega(\chi)
{}_{{}_\Lambda\!}(\nabla \chi)^2 +2\Lambda^{obs}\right).
\la{BD_L}
\een
Hence, we can apply all well-studied properties of Brans-Dicke
theory \cite{WW,BD}.
to the part of (S)ST under consideration.
We shall stress some well known properties of this theory
which we need later:

i) In contrast to the EF-dilaton $\varphi$,
the interactions of the $\Lambda$F-dilaton $\chi$
with the matter are completely
fixed by  $\Lambda$F Einstein WEP
in the simplest possible way:
to satisfy WEP in $\Lambda$F, the dilaton $\chi$
must not enter the $\Lambda$F-matter Lagrangian.
Its influence on the matter is only indirect --
it is due to the interaction (\ref{BD_L}) with the metric
${}_{{}_\Lambda\!}g_{\mu\nu}$
(which, in turn, must enter $\Lambda$F-matter Lagrangian minimally,
i.e., as in GR).

Hence, in the entire $\Lambda$F-theory we have
only {\em one} unknown function related to dilaton,
namely the Brans-Dicke function $\omega(\chi)$.

ii) One obtains the field equations for $\Lambda$F theory
by replacing in Eq.~(\ref{GFE}) the variable $\phi$ with
$\chi$ and using (\ref{LF})
and  the relations $ {}_{{}_\Lambda\!}V_{,_\chi}\equiv 0$,
$ {}_{{}_\Lambda\!}J=1+4{{D\!-\!2}\over{D\!-\!1}}\omega(\chi) $,
and $ {}_{{}_\Lambda\!}\Theta = {}_{{}_\Lambda\!}T$.

The additional relation (\ref{R_rel2}) now reads
\ben
R+8\,\chi^{-1}\omega\,\BBox\chi +
4\left(\chi^{-1}\omega\right)_{\!,_\chi}(\nabla\chi)^2=0 \,.
\la{LR_rel2}
\een

iii) The $\Lambda$F-dilaton $\chi$ is {\em not}
a matter field, but rather a part of the description of gravity.
We have arrived at a purely dynamical metric theory of gravity
with {\em one} scalar gravitational field \cite{Will}.
It plays the role of a variable effective gravitational
constant: $G_{eff}= G_N/\chi$.

This seems to be much more in the spirit of string theory
where graviton and dilaton appear in the same physical sector.

iv) If considered as a specific scalar field
in a fixed metric ${}_{{}_\Lambda\!}g_{\mu\nu}$
(according to standard Brans-Dicke dynamics),
$\Lambda$F-dilaton $\chi$ may still have
space-time variations.
For example, in homogeneous space-times
and even in flat space-time
${\cal M}^{D}\{{}_{{}_\Lambda\!}g_{\mu\nu}\}$,
one can have a variable field $\chi$.

In addition, in our stringy-inspired approach to
Brans-Dicke theory with a cosmological constant, we
obtain one more novel general property:

v) As seen from formula (\ref{BDomega}),
the observable cosmological constant $\Lambda^{obs}$
does not enter explicitly Brans-Dicke function
$\omega(\chi)$.
The same holds for any common constant scale factor in the SF
cosmological potential ${}_{{}_S\!}\Lambda(\phi)$.
This is due to the dependence of $\omega(\chi)$ on the
derivatives of $\ln {}_{{}_S\!}\Lambda$ and other functions
with respect to $\ln\chi$.
The two factors are absorbed in the $\Lambda$F
metric ${}_{{}_\Lambda\!}g_{\mu\nu}$ and in the dilaton $\chi$
as described by formulae (\ref{Weyl}), (\ref{LF_}), and (\ref{chi}).
As a result, in the $\Lambda$F-LEL Lagrangian (\ref{BD_L}),
the only remaining ``free'' parameter is $\Lambda^{obs}$.

As a consequence, when $D=4$, we discover a new {\em symmetry}:
the Lagrangian (\ref{BD_L}) is form-invariant under
rescaling of the SF cosmological potential ${}_{{}_S\!}\Lambda$ if
$\omega(\chi)$ does not depend on the  $\Lambda$F-dilaton $\chi$.

Indeed, let us consider a rescaling of
the cosmological potential ${}_{{}_S\!}\Lambda$
with a constant factor $\lambda$:
\ben
{}_{{}_S\!}\Lambda \longrightarrow \lambda \,{}_{{}_S\!}\Lambda.
\la{L_resc}
\een
Then, according to formulae (\ref{Weyl}), (\ref{LF_})--(\ref{BDomega}),
instead of Lagrangian  (\ref{BD_L}) we obtain
the rescaled one:
\ben
{}_{{}_\Lambda\!}L_{{}_{LEL}} \,\sim\,\, \lambda^{-{{2(D\!-\!2)}\over D}}
\sqrt{|{}_{{}_\Lambda\!}{\bf g}|}\times\nonumber \hskip 3.7truecm\\
\left(\chi\,{}_{{}_\Lambda\!}R -
\chi^{-1}\omega\left(\lambda^{-{{(D\!-\!2)}\over D}}\chi\right)
{}_{{}_\Lambda\!}(\nabla \chi)^2 +
2\lambda^{{{(D\!-\!4)}\over D}}\Lambda^{obs}
\right).\,
\la{BD_L_resc}
\een
Hence, in the important case $D=4$,
the observable value $\Lambda^{obs}$ remains
invariant under rescaling of the SF cosmological potential
${}_{{}_S\!}\Lambda$. If we include the common factor
$\lambda^{-{{2(D\!-\!2)}\over D}}$
in the Einstein constant $\kappa$ of the  corresponding
$\Lambda$F-action of theory, and in addition
$\omega(\chi)={\rm const}$
(as in the original Brans-Dicke theory),
we obtain a theory which is
invariant under the transformations (\ref{L_resc}).

\subsubsection{Twiddle Frame}

At the end, let us try the third  distinguished possibility
for choice of frame (see subsection B),
i.e., let us impose the conformal gauge condition $Z(\phi)=0$.
Such a frame has been used very successfully in the so called
2D-dilatonic gravity models, both for classical and quantum
problems \cite{twiddle}.
There, it was called a {\em twiddle frame} (TF).
We shall use this name, although the case $D=2$
is a singular one \footnote{For $D\!=\!2$
and ${}_{{}_S}F(\phi)\!\neq\! {\rm const}$,
Einstein frame does not exist,
the stringy S-duality is lost,
and, instead of the quadratic relation (\ref{Z/F}),  one obtains
a linear one, but we still can introduce a twiddle frame,
by using the relation
${}_{{}_T\!}\sigma(\phi)=
2\int\left( {{{}_{{}_S\!}Z}/{{}_{{}_S\!}F_{,\phi}}}\right)d\phi$,
and changing properly all formulae \cite{twiddle}.},
and we do not consider this case in present article.

Now from Eq.~(\ref{Z/F}), one obtains
\ben
{}_{{}_T\!}\sigma{}_{\!\!{}_S}(\phi)=
{}_{{}_T\!}\sigma{}_{\!\!{}_S}^\pm(\phi)=
{1\over{D\!-\!2}}
\left(-\ln{}_{{}_S\!}F \pm\sqrt{{4\over{D\!-\!1}}} S(\phi)\right)
\nonumber \\
= {}_{{}_E\!}\sigma{}_{\!\!{}_S}(\phi)
+{{\varphi(\phi)}\over{\sqrt{(D\!-\!1)(D\!-\!2)}}} \,, \hskip 2.3truecm
\la{Tsigma}
\een
where the formula (\ref{Ephi}) is used.
Due to the stringy S-duality, we have two solutions,
${}_{{}_T\!}\sigma{}_{\!\!{}_S}^\pm(\phi)$,
i.e., two different TF
for given coefficients
${}_{{}_S\!}F$ and ${}_{{}_S\!}Z$.
But when we express the TF-relations in terms of the EF-dilaton
$\varphi$, the S-duality becomes implicit, and the twofold
correspondence between SF and TF is hidden.
This way, we can lose some global properties of (S)ST
if we use {\em local} TF field variables,
or we can expect some specific TF-singularities
which are not present in SF.
In some situations, this indeed yields
catastrophic-type singularities
in the TF cosmological potential \cite{F00}.
Nevertheless, we use the EF-dilaton description
of the transition to TF because it looks simpler.

This way, we obtain
\ben
{}_{{}_T\!}F(\varphi) =
e^{{\sqrt{{{D\!-\!2}\over {D\!-\!1}}}\varphi}},\,\,\,
{}_{{}_T\!}\Lambda(\varphi)=
\Lambda^{obs} {}_{{}_E\!}U(\varphi)\,e^{{{D}\over{\sqrt{(D\!-\!1)(D\!-\!2)}}}\varphi}
\la{T_FL}
\een
and, by introducing the TF-dilaton $\Phi$ according to definition
\ben
\Phi={}_{{}_T\!}F(\varphi)=
e^{{\sqrt{{{D\!-\!2}\over {D\!-\!1}}}\varphi}} \,,
\la{Phi}
\een
we obtain the TF-LEL stringy dressed Lagrangian in the form
\ben
{}_{{}_T\!}L_{{}_{LEL}} \sim \sqrt{|{}_{{}_T\!}{\bf g}|}
\left(\Phi\,{}_{{}_T\!}R +2\Lambda^{obs}{}_{{}_T\!}U(\Phi)\right)
\la{TL}
\een
with ${}_{{}_T\!}U(\Phi)\!=\!
{}_{{}_E\!}U\!\left(\!\ln\!\left(\!\sqrt{{D\!-\!1}\over{D\!-\!2}}
\Phi\!\right)\!\right)\!
\Phi^{{D}\over{D\!-\!2}}$.

For example, for the tree-approximation we obtain
\ben
{}_{{}_T\!}\sigma_{\!{}_S}^{(0)}=
{{2\phi}\over{D\!-\!2}}\left(1\pm\sqrt{{1}\over{D\!-\!1}}\right)
\hskip 2.truecm\nonumber\\
\varphi^{(0)}=\pm\sqrt{{4\over{D\!-\!2}}}\phi,
\hskip 2.5truecm\nonumber \\
\Phi^{(0)}=e^{{\pm \sqrt{{4\over{D\!-\!1}}}\phi}}=
e^{{\sqrt{{{D\!-\!2}\over {D\!-\!1}}}\varphi^{(0)}}},
\hskip 1.7truecm \nonumber \\
{}_{{}_E\!}U^{(0)}=U^{(0)}e^{{{4}\over{D\!-\!2}}\phi}=
U^{(0)}e^{\pm \sqrt{4 \over{D\!-\!2}}\varphi^{(0)}} =
U^{(0)}\Phi^{\pm2{{\sqrt{D\!-\!1}} \over{D\!-\!2}}},
\nonumber \\
{}_{{}_T\!}U^{(0)}(\Phi)=U^{(0)}\Phi^{n_\pm(D)}\,, \hskip 2.truecm
\la{TFU0}
\een
where $n_\pm(D)\!=\!{ {D\pm2\sqrt{D\!-\!1}} \over{D\!-\!2}}$
are the solutions of the equation
$n^2-2{{D}\over{D\!-\!2}}n+1=0$ and
$U^{(0)}\!=\!{{{}_{{}_S\!}C_\Lambda^{(0)}}/{\Lambda^{obs}}}$.
We see that

i) TF-theory is a special kind of Brans-Dicke theory
with $\omega(\Phi)\equiv 0$,
i.e., {\em without} standard kinetic
term for TF-dilaton $\Phi$ in the Lagrangian (\ref{TL}).

ii) In order to satisfy TF-Einstein WEP,
the dilaton $\Phi$ must not enter the matter Lagrangian.
Its influence on the matter is only indirect --
through the interaction (\ref{TL}) with the metric
${}_{{}_T\!}g_{\mu\nu}$
(which, in turn,
must enter TF-matter Lagrangian minimally).

Thus, we have arrived at a specific dynamical
metric theory with one gravitational scalar $\Phi$
that determines the effective
gravitational constant $G_{eff}=G_N/\Phi$.
To avoid the semantic inconvenience,
when we speak about
the ``(non)constancy of gravitational constant'',
we shall call the quantity $G_{eff}$
``a gravitational factor''.

In the entire theory, we have only one unknown
function of the TF-dilaton -- the cosmological
potential ${}_{{}_T\!}U(\Phi)$.

iii) The field equations (\ref{GFE}) become simpler
because now ${}_{{}_T\!}J\equiv 1$
and ${}_{{}_T\!}\Theta \equiv T$.

iv) This version of the theory has
the following unique property.
Only in TF, the basic relation (\ref{R_rel2})
becomes an {\em algebraic} one:
\ben
R + 2\Lambda^{obs}{}_{{}_T\!}U_{,_\Phi}(\Phi)=0.
\la{R_relA}
\een
In all other frames, the corresponding relations
are {\em non-local}
because of the presence of derivatives of the dilaton field
(see formulae (\ref{R_rel2}), (\ref{ER_rel2})
and (\ref{LR_rel2})).
This property is extremely important for us,
and may be used as a definition of TF. Only in
this frame, the first variation of the action
(\ref{A_GD}) with respect to dilaton $\Phi$ gives
the algebraic relation (\ref{R_relA})
instead of dynamical field equation.

The property (\ref{R_relA}) justifies our
specific choice of the second equation of
(\ref{GFE}) as a classical field equation
for the dilaton, instead of Eq.~(\ref{R_rel2}).
Doing this, we neglect the fact that
Eq.~(\ref{R_rel2}) (for BF)
and Eq.~(\ref{R_relA}) (for TF)
give precisely the condition
for vanishing first variation of the corresponding
action with respect to the dilaton.

In addition, the relation (\ref{R_relA}) yields
the following basic properties of the TF-dilaton $\Phi$:

v) The TF-dilaton $\Phi$ does not have its own dynamics
independent of TF-metric ${}_{{}_T\!}g_{\mu\nu}$.
In particular, in a space-time with a constant scalar
curvature ${}_{{}_T\!}R=const$, we have a constant
TF-dilaton $\Phi=const$.  Hence, in homogeneous
space-times and in Einstein space-times
with ${}_{{}_T\!}R=0$, we have
a constant effective gravitational factor $G_{eff}$
independently of the field dynamics, described by
Eq.~(\ref{GFE}).

vi) Moreover, when ${}_{{}_T\!}U_{,_{\Phi\Phi}}\neq 0$,
the dilaton $\Phi$, as a physical degree of freedom,
may be included in the metric,
thus becoming a {\em scalar} part of
geometrical description of gravity.
This is possible because in this case
the action (\ref{A_GD}) with Lagrangian (\ref{TL})
may be considered as a Helmholtz
action for some nonlinear theory of gravity (NLG).
(See, for example, \cite{NLG} and
the large amount of references on NLG therein.)
The Lagrangian of nonlinear gravity which corresponds
to (\ref{TL}) is
\ben
L_{{}_{NLG}}\sim- 2 f(R)=\Phi(R)R+2\Lambda^{obs}U(\Phi(R))\,.
\la{L_NLG}
\een
The function $\Phi(R)$ can be determined from Eq.~(\ref{R_relA})
only if ${}_{{}_T\!}U_{,_{\Phi\Phi}}\neq 0$,
by the implicit function theorem.

The inverse correspondence -- from NLG
to TF-theory, may be described in a simple way as well.
For any non-constant function
$f(R)$ with $f_{,_{RR}}(R)\!\neq\!0$,
one has to solve the algebraic equation
$\Phi+2f_{,_R}(R)=0$ with respect to
$R$ and to obtain the function $R(\Phi)$.
Then ${}_{{}_T\!}\Lambda(\Phi)=
-{1\over 2}\int R(\Phi) d\Phi=
-{1\over 2}\Phi R(\Phi)-f(R(\Phi))$.

Mathematically, the correspondence between the two
descriptions, (\ref{TL}) and (\ref{L_NLG}), of the
model may be represented in a more symmetric
way by the relation
\ben
f(R)+{}_{{}_T\!}\Lambda(\Phi)+{1\over 2}R\Phi=0\,.
\la{R_Phi}
\een
These two descriptions are equivalent if and only if
\ben
f_{,_{RR}}(R)\times
{}_{{}_T\!}\Lambda_{,_{\Phi\Phi}}(\Phi)\neq 0\,.
\la{Nzero}
\een

vii) For metrics ${}_{{}_T\!}g_{\mu\nu}$  and dilaton fields
$\Phi$ that obey the relation (\ref{R_relA}), one obtains
the following simple form of the TF-gravi-dilaton action:
\ben
{\cal A}_{g\Phi}={{c\Lambda^{obs}}\over \kappa}
\int d^Dx \sqrt{|{}_{{}_T\!}{\bf g}|}
\left( \Phi\,{}_{{}_T\!}U_{\!,\Phi} -{}_{{}_T\!}U\right).
\la{TFA_GD}
\een
This useful form of the action will have important
consequences for the quantum version of the theory.
Most probably, it is the ground for
the simple {\em exact quantization} of
D2-dilatonic-gravity models \cite{twiddle}
for arbitrary potentials ${}_{{}_T\!}U(\Phi)$.
The study of its consequences in dimensions $D>2$
is a new interesting issue and may create important
results both in classical and in quantum problems.

\subsection{The Phenomenological Frame}

\subsubsection{Basic Physical Properties
of the Phenomenological Frame}

Now it is easy to recognize that the choice of frame
is a {\em physical} problem and the predictions of the theory
make sense only after the physical frame is fixed.
Below, we justify our understanding of this important
issue and apply it to the problem at hand.

We define the {\em phenomenological} frame (PhF)
as a frame in which all {\em real} physical
measurements and observations are performed,
i.e., as a frame in which the metric $g_{\mu\nu}$
is measured by
laboratory roads and clocks,
made of {\em real (fermionic)} matter,
where the accelerator physics is developed,
the space missions take place, etc.
For example, exactly in this frame we observe
the well known expansion of the Universe
with the known values of Hubble constant and
cosmological constant.

The PhF has several basic and well established physical
properties which are important for us:

1) In PhF, the space-time looks like a four dimensional (4D)
smooth manifold with Lorentzian type of signature of the metric.
Locally, the special relativistic kinematics takes place
with high precision \cite{Will}.

We are not able to present quantitative
estimates for the level of our confidence
in the four-dimensional nature of the real space-time.
The higher dimensions of space-time,
predicted by different theoretical models,
simply do not show up in any experiments until now.

Of course, it is not impossible that,
like the people in Plato' philosophical doctrine,
being confined in our four-dimensional cavern,
we are able to observe only some faint true light
which comes from the outer multi-dimensional
world and only some four-dimensional shadows
of the existing true objects are accessible even
for our most precise experimental equipment.
Therefore, it may be useful to develop different
types of multi-dimensional theories and
to look for new predictions that allow
confrontation with the real physics.
There exist a large number of such models,
each of which yielding different predictions
depending on the procedure chosen to make
the extra-dimensions (almost) invisible.

There is one more unappealing general feature
of such type of theories.
Namely, a lot of new field degrees of freedom and, hence,
an infinite number of new dynamical parameters,
are introduced in the theory,
without serious phenomenological motivation.

It is not excluded, too, that the space-time
${\cal M}^{(D)}$ is {\em not} a smooth manifold.
It may have a fractal structure at extremely
small distances. In this case, the space-time
dimension $D$ may even be non-integer.

But one thing is clear:
all admissible corrections (if any)
of our description of the real space-time
due to higher dimensions, or due to other
possible unusual features, must be small enough
to prevent their experimental observation at
the level of our present-days abilities.
Hence, {\em to the best of our real knowledge},
the phenomenologically reasonable approximation
for the space-time dimension $D$ is simply $D=4$.

2) Einstein weak equivalence principle.

The most important for us and experimentally
well checked is Einstein WEP in PhF.
At present, we know that it is valid up
to $10^{-13}$ relative error. The best available data
are obtained from
$\left({{\Delta a}\over{a}}\right)_{Moon-Earth}=
(-3.2\pm 4.6)\times 10^{-13}$
\cite{Will}, \cite{DamourExp}.
Up to now, we have no experimental
indications of any kind of violation of WEP.

3) Constancy of the interaction constants
in the matter Lagrangian.

The basic non-gravitational properties of matter
are described by the Standard Model (SM) of particle physics
which has to be taken into account
when one tries to construct a consistent theory
of gravity and to reach understanding
of physics at all stages of development of the Universe.
In SM we have, as an input, several fundamental
constants of different interactions, as well as
different masses of fundamental particles.

At present, we have most tight restrictions for
the time evolution of the fine structure constant $\alpha$.
According to the recent careful analysis
of the Oklo Natural Reactor data \cite{Chiba},
during the last 1.8 billion years we have
a limitation for the time variations
of the fine structure constant,
described by relative rate of change
${{\dot \alpha} / {\alpha}} =
(-0.2 \pm 0.8)\times 10^{-18} \,{\rm yr}^{-1}$.
Other precise measurements give an upper limit for
$|{{\dot \alpha} / {\alpha}}|$ between
$10^{-12}\,{\rm yr}^{-1}$ and $10^{-17} \,{\rm yr}^{-1}$
\cite{Will,Chiba}.
Then for the time of existence of the Oklo Reactor
$\left({{\Delta \alpha}\over{\alpha}}\right)_{Oklo}=
(-1\pm 4)\times 10^{-10}$.
If one assumes that the same rate limitations
for the time variations of fine structure constant has held
during the whole time history of the Universe,
then for the cosmological time scale, $\sim 13$ billion years,
$\left({{\Delta \alpha}\over{\alpha}}\right) < 10^{-8}$.

Weak constraints on the ratio $\frac {\Delta \alpha} \alpha$
from BBN and CMB coming from the latest observational data
can be found in \cite{Avelino}.

Besides, there are some doubts about possible
variations of the fine structure constant in the
course of cosmological evolution in the form
$\left({{\Delta \left(\alpha^2 g_p\right)}\over{\alpha g_p}}
\right)= (-0.20\pm 0.44)\times 10^{-5}$ for $z=.2467$,
and $(-0.16\pm 0.54)\times 10^{-5}$ for $z=.6847$
\cite{MWFD}, where $g_p$ is the proton g-factor.
These observations need further independent verification --
see \cite{IRPV}, where independent observational
indications about a possible cosmological time variation
of proton-to-electron mass ratio $\mu=m_p/m_e$ at a level
${{\Delta \mu}\over \mu}=(5.7 \pm 3.8)\times 10^{-5}$
were reported.

Then the unification of gauge couplings of SM
would imply that time variations of the fine structure
constant are accompanied by significant time
variations of other QCD constants and masses
\cite{QCDconstants}.

An independent derivation of the behavior of QCD effects in 
unified theories with varying couplings was given in \cite{Dent}. 
The authors of this article pointed out that the electroweak and fermion 
mass sectors could be strongly sensitive to a varying unified coupling, 
depending on the mechanisms of electroweak symmetry-breaking and 
fermion mass generation. 
In some cases the effects due to a changing Higgs vacuum expectation value, 
dynamically determined by the unified coupling,
are even larger than the QCD effects, and would significantly 
affect predictions for the variation of $\mu=m_p/m_e$, 
due to a large variation of $m_e$.

Even taking into account these preliminary results,
it seems that we can safely accept
as an experimentally established property
of the PhF that in this frame we have indeed
a space-time constancy of the fundamental interaction
constants and of the masses of the physical
particles at least at level
$\left({{\Delta \alpha}\over{\alpha}}\right) \leq 10^{-5}$
and ${{\Delta m}\over m} \leq 10^{-5}$.

4) The Cosmological Principle.

According to the basic Cosmological Principle (CP)
\cite{WW}, our Universe is 3D-spatially homogeneous
and isotropic at large enough scales,
i.e., after averaging of the large structures
at scales of several hundred $Mpc$.
This is a kinematic principle of very general nature,
and for metric theories of gravity it implies
constancy of the 3D-space scalar curvature
at such large scales, together with the 3D-space constancy of
mass-energy density and of the {\em gravitational constant},
because all cosmic quantities must be invariant
with respect to the corresponding isometries of
constant-cosmic-time surfaces in the space-time \cite{WW}.

At present, we know from direct CMB measurements that this
principle is valid within an accuracy of $10^{-4}$.
The observed spatial temperature variations of CMB are
of order of ${{\delta T}\over{T}}\sim 10^{-5}$.

Usually the ``constancy of gravitational constant''
at large space-scales is not discussed
because in GR the effective gravitational factor
in Hilbert-Einstein term is pre-supposed to be constant
at all scales. But in theories of Brans-Dicke
type, the basic cosmological principle yields
the constancy of effective gravitational factor
as a special kind of scalar field.

5) Hilbert-Einstein action for gravity.

As we know, GR is based on two basic assumptions:

i) Einstein WEP, which determines the metric
interaction of matter with gravity, and

ii) Hilbert-Einstein Lagrangian
$L_G\sim {1\over\kappa}R$ which determines the dynamics and
other physical properties of gravity.
At present, the second assumption of GR
is checked experimentally
with precision {\em only} $10^{-3}$ or at most $10^{-4}$
\cite{Will}, \cite{DamourExp}:

a) For the weak field approximation, the best restriction
was recently achieved for PPN parameter
$\gamma$ given by
${{\gamma + 1}\over 2} = 0.99992\pm 0.00014$ \cite{Will}.

b) The data for the binary pulsar PSR 1913 +16  offer
the best test of Hilbert-Einstein Lagrangian
in strong-field regime,
where the nonlinear character of the theory and the effects
of gravitational radiation are essential \cite{DamourExp}.
The lowest order GR approximation for the orbital-period
evolution rate,
${\dot P_b}/P_b= -2 \dot G/G+3 \dot l/l-2\dot m/m$,
includes time variations of the gravitational constant $G$,
the angular momentum $l$ and the reduced mass $m$ of the two-body
system, i.e., it controls the total Hilbert-Einstein term
(possible variations of speed of light have been neglected).
The best present data give
${\dot P_b}/P_b=1.0023\pm0.0041[obs]\pm0.0021[gal]$,
and confirm both the $\sim R$ form of the Hilbert-Einstein
Lagrangian and the constancy of the gravitational factor at
level of $10^{-3}$ relative error. If one ascribes the
entire experimental uncertainty of this quantity to
time variations of the gravitational constant, one obtains
$\dot G/G = (1.0\pm 2.3) \times 10^{-11} \,{\rm yr}^{-1}$
\cite{Chiba}.
Other precise experiments and observations
give about one order of magnitude more tight
restrictions \cite{Will,Chiba}.
The best estimates available at the moment are
$|\dot G/G| \leq 1.6\times 10^{-12} \,{\rm yr}^{-1}$
from Helioseismology, and
$\dot G/G = (-0.6\pm 4.2)\times 10^{-12} \,{\rm yr}^{-1}$
(at 95\% confidence level) from measurements of
neutron star masses \cite{Thorsett}.
Both of these estimates are model-dependent and may be
weakened.

Therefore, our confidence in the exact form
of Hilbert-Einstein Lagrangian must be about nine-ten
orders of magnitude smaller than in the WEP.
It seems quite possible to find experimentally
some deviations from the simplest Hilbert-Einstein
Lagrangian due to different corrections
(quantum corrections, stringy corrections,
variable gravitational constant, etc.),
although this Lagrangian has been widely
recognized as a corner stone of GR as a theory
of gravity in low energy limit.

6) The Hubble accelerating expansion of the Universe.

This is the last of the basic features of PhF which we wish
to stress as very important for our further discussion.
At present, we know the value of Hubble constant
$h_0=0.72\pm 0.08$ only at around 10\%
level of accuracy \cite{CosmTri}.
Then the value of the cosmological parameter
$\Omega_\Lambda = 0.7\pm 0.1$ gives for
the observable cosmological constant a value of
\ben
\Lambda^{obs}=3\Omega_\Lambda H_0^2 c^{-2}=
(1.27\pm 0.46)\times 10^{-56} \,{\rm cm}^{-2} \,,
\la{Lambda}
\een
which is known within around 36\% of accuracy.

One more confirmation of the expansion of the
Universe at the level of error $<10^{-3}$ gives
(within the interpretation related to the Big Bang scenario)
the observed CMB temperature $T_{{}_{CMB}}=2.725\pm0.001\,{\rm K}$,
which is the best known cosmological parameter
\cite{CosmTri}.

Although certain doubts about the absolute validity of
the above six properties ever exist,
these properties are at present among the basic and
most well established physical facts.
Therefore, in our opinion, one has to try to preserve
them as much as possible in any new theory of gravity.

Of course, it is not impossible that some day in the future
we find reliably deviations from these features
of the real physics in phenomenological frame.
But at the moment, we have nothing  better
to use as a foundation of our theoretical constructions.
Moreover, we believe that future investigations may
result only in small corrections to these six
properties in the framework of the present-day
experimental limits.
Therefore, we accept them
as {\em phenomenologically established} first principles,
and believe that this is the {\em most realistic} approach
to the problem.

\subsubsection{The General Strategy for Choice of a Frame}

It is clear that without some essential
changes in the above six principles,
it will be impossible to solve the problems
listed in the introduction.
Hence, one is forced to decide which of these
principles have to be changed, and which is the most
appropriate direction for new theoretical developments.
There exist two physically different possibilities:

I. One may introduce new (i.e., outside the SM)
kind(s) of matter with exotic properties.

II. One may try to change properly the very theory of gravity.

Naturally, some combination of these
two possibilities may turn out to be necessary,
but one has to investigate first more
simple theories which use only one of them.
Hence, we do not consider in the present
article the combination of I and~II.

The next problem will be to justify the new
model and to look for experimental
evidences which support our choice.

For scalar-tensor theories of gravity
with only one additional scalar field,
the choice between the
possibilities I and II is reduced to the
interpretation of the role of the scalar dilaton
field. This interpretation actually depends
on our decision which of the theoretically
possible frames we consider as
a phenomenological frame. We shall try to chose
for PhF one of the previously discussed
distinguished frames:

\vskip 0.2truecm
\paragraph{Einstein Frame as a Phenomenological Frame.
\,\,\,\,\,\,\,\,\,\,\hskip 2truecm}
\vskip 0.2truecm

Suppose, one insists on preserving the {\em exact form}
of Hilbert-Einstein action for describing gravity
at least in the low energy limit.
Then one is forced to consider the EF-dilaton
as a new matter field that can be used for
explanation of the new observed phenomena.
This is the most widely used approach,
and in its framework one has
a standard form of GR with one new {\em matter} field
$\varphi$ which may be massless if
${}_{{}_S\!}C_\Lambda(\phi)\equiv 0$.

The models of this type were studied
in great detail during the last two decades.
There exist a huge number of attempts,
which choose different cosmological potentials
${}_{{}_E\!}U(\varphi)$
and corresponding interactions of such a scalar field
\footnote{Often, the authors of the corresponding articles
do not try to relate the scalar field $\varphi$
with the stringy dilaton, but in our treatment of the subject,
this is a matter of conventions and terminology.}
with other fields, to use it as:

1) Inflation field (see \cite{inflation} and references therein);

2) Quintessence field (see \cite{Q} and references therein);

3) Universal field which simultaneously serves both
for inflation and quintessence field
(see \cite{DV} and references therein).

The main problem which still remains unsolved
is to reach a realistic theory of that kind --
all existing models seem to suffer from essential
difficulties.

The common difficulty for most models is that,
one has to prescribe an extremely small mass
(typically $m_{{}_\varphi}$ between $10^{-33}\,{\rm eV}$
and $10^{-30} \,{\rm eV}$)
to the ``cosmon'' scalar field $\varphi$,
in order to avoid obvious contradictions
with astrophysical observations.
Such a small mass is too far from any real
experimental and theoretical domain of masses in SM.
As an example of our present-day abilities,
we remind the reader that the best experimental
restriction for the mass of photon
(which we believe to be massless) is $<10^{-27} \,{\rm eV}$.

Hence, the physical interpretation of the extremely
light scalar field $\varphi$ will need some new
kind of physics which, being complete unknown,
is too far from the SM.
Moreover, its laboratory experimental investigation
seems to be impossible in foreseen future.

Another difficulty appears if one allows a big mass
$m_{{}_\varphi}$ of order $10^{-5}$--$10^{-2} \,m_{Pl}$,
i.e., $10^{14}$--$10^{17}\,{\rm GeV}$. Then, cosmological
defects of different type will be present in the theory
and will have to be somehow avoided, etc.

There exists one more theoretical inconvenience
in this approach. If one wishes to prevent the
theory from
the fast decay of the EF-dilaton
into other matter particles or radiation, and
from the violation of energy-momentum
conservation of other matter fields,
one must accept as
an {\em independent} assumption
that the EF-dilaton $\varphi$ is sterile \cite{DV}.
This means that it does not interact directly
with SM-matter and
$({}_{{}_E\!}{\cal L}_{...})_{,_{\varphi}}\equiv 0$.

Instead, one can introduce some special interaction
of the EF-dilaton $\varphi$ with other matter fields,
and in addition, a ``charge'' which corresponds to this interaction.
This approach is used, e.g., if one wishes
to have some ``preheating mechanism'' in the early
Universe \cite{C, inflation}.
In any case, one is forced to fix {\em independently}
the existing ambiguity in
the dependence of the matter Lagrangian on the EF-dilaton.

The situation changes essentially if one chooses
$\Lambda$F, or TF for the role of PhF,
because then the corresponding dilaton field would {\em not}
be allowed to be included in the matter Lagrangian by WEP.
But in these two cases, we have to change
the Hilbert-Einstein description of gravity.

\vskip 0.2truecm
\paragraph{$\Lambda$-Frame as a Phenomenological Frame.
\,\,\,\,\,\,\,\,\,\,\hskip 2truecm}
\vskip 0.2truecm

If we choose for PhF the $\Lambda$F, it is easy to see that:

1) At the level of stringy-tree-approximation,
our theory would contradict the observations
if the number of space-time dimensions is $D<14,000$,
since in this case the Brans-Dicke parameter $\omega^{(0)}$
would yield experimentally inadmissible values of the
PPN parameter $\gamma$ (see Eq.~(\ref{omega0})).
It seems to be unrealistic to increase the number of
space-time dimensions to such large values.

The presence of cosmological constant term (\ref{Lambda})
in the Lagrangian (\ref{BD_L})
cannot improve the situation
because the influence of pure cosmological constant
on gravitational phenomena at scales of the Solar System
and star systems is too small \cite{BDLambda}.

2) Suppose that we are able to find some special function
$\omega(\chi)$ which yields an {\em attractor} behavior
for the dilaton: $\chi \longrightarrow \chi_\infty$ as
$t \longrightarrow \infty$ for general solutions of the field
equations and without fine tuning.
Then we can comply with the experimental value of $\gamma$
in the framework of $\Lambda$F-theory
if $\omega(\chi_\infty)\gg 3,500$.

But even if we succeed in constricting such a model,
some theoretical shortcomings will still remain:
it seems to be strange to allow {\em a priory}
variations of the effective gravitational constant
which are independent of space-time geometry.
For example,
in Brans-Dicke theories of general type, one can have
a variable gravitational factor $G_{eff}= G_N/\chi$
in homogeneous space-times with $R={\rm const}$,
in Einstein space-times with $R=0$, and even in
Minkowski space-time.
This obviously contradicts the spirit of Einstein's
idea for {\em purely geometrical} description of gravity
based on WEP.
In Brans-Dicke theories of general type, the gravitational
factor is an important part of the description of gravity,
and is an additional physical degree of freedom.
This additional degree of freedom is independent from
other gravitational degrees of freedom that are
described geometrically by metric.

As a result,
in $\hbox{PhF} \equiv \Lambda \hbox{F}$-theory,
as well as in $\hbox{PhF} \equiv \hbox{EF}$-theory,
we are forced to apply the CP both for metric and
for dilaton {\em independently}.

\vskip 0.2truecm
\paragraph{Twiddle Frame as a Phenomenological Frame.
\,\,\,\,\,\,\,\,\,\,\hskip 2truecm}
\vskip 0.2truecm

The only way to avoid the above shortcomings of
both $\hbox{PhF} \equiv \Lambda \hbox{F}$
and $\hbox{PhF} \equiv \hbox{EF}$ choices
seems to be the third one, $\hbox{PhF} \equiv \hbox{TF}$.
Then, because of the properties i)-vi) of TF-dilaton
(see Section III.B.3), we will have a theory in which:

1) New exotic matter with unknown properties does not exist.

2) WEP allows only one unknown function
${}_{{}_T\!}U(\Phi)$ in the entire theory.
It also guarantees the metric character of interaction
of gravity and matter, including the independence
of matter Lagrangian $L_{matt}$ on the dilaton $\Phi$
and the absence of "dilatonic charge",
or other exotic properties of standard matter.

3) As a result of 2), and because
of the existence of {\rm only one} additional field
-- the dilaton $\Phi$ -- in the minimalistic model
at hand, we obtain a constancy of SM
interaction constants and masses.
The specific values of the coefficients
\ben
  {}_{{}_T\!}B_{...}(\Phi)\equiv {\rm const}
\la{TFC}
\een
in the TF-terms (\ref{psi}) depend on the conventions
for normalization of the matter fields.

This is a modern realization of Dirac's pioneering idea
to preserve the constancy of the fine structure
constant and the masses of fundamental particles,
but to allow variability of the gravitation constant
\cite{Dirac,WW}.

If the present-day observational doubts
(Section III.C.1) in the existence
of small variations of the SM fundamental constants
and masses are reliably confirmed,
the above property may be considered as a good
{\em first approximation} to the real physics
in $\hbox{PhF} \equiv \hbox{TF}$.
To explain the small variations, one may need to introduce
some additional fields,
like the totally anti-symmetric stringy field
$H_{\mu\nu\rho}$.
Considering a {\em minimal} extension of GR, we completely
ignore this field in the present article,
although it belongs to the same stringy LEL
sector and appears together with the dilaton and the metric.
Another possibility is to use other scalar moduli fields
in (S)ST, etc.

4) We have an automatic fulfillment of the CP
for the dilaton $\Phi$ when CP is valid for the metric
as a result of condition (\ref{R_relA})
when ${}_{{}_T\!}U(\Phi)_{,_{\Phi\Phi}}\neq 0$.
The last inequality yields {\em a nonzero mass}
$m_{{}_\Phi} \neq 0$ for dilaton $\Phi$,
as we shall see later. In this sense, one may conclude
that CP implies nonzero mass of dilaton in TF-theory.

5) The nonzero mass of the TF-dilaton $\Phi$ provides
us with the possibility of including the degree of freedom of $\Phi$
into the geometrical description of gravity using nonlinear
metric representation of TF-theory given by the
Lagrangian (\ref{L_NLG}).
We wish to stress that the condition $m_{{}_\Phi} \neq 0$
is critical for the very existence of such possibility.
We relate the dilaton to the scalar curvature $R$ of
the Riemannian space-time ${\cal M}^{(D)}$,
and the only problem we have to solve
is to find the exact form of the algebraic relation
(\ref{R_relA}), i.e., the form of the TF-cosmological potential.
This geometrical interpretation is more economical
and more physical than the one suggested
in \cite{SS} (in which the dilaton was related to a possible
non-metricity of space-time metric).

6) The dynamics of the dilaton $\Phi$ and its propagation
are deeply related to the metric and the space-time curvature.
This follows from the absence of standard kinetic term for the dilaton
field in the Lagrangian (\ref{TL}). Just because of this circumstance,
in flat space-time $\Phi$ has no dynamics.
Moreover, the second order dynamical equation
for the dilaton $\Phi$ in the system (\ref{GFE})
is obtained by two integrations by parts of
the corresponding terms in the variation of the action
not with respect to the dilaton,
but with respect to the metric.
This specific feature of TF-theory makes natural the identification
of the dilaton $\Phi$ as a scalar part of gravity,
as opposed to its interpretation as a new sort of scalar matter field.

7) The price one has to pay for these new possibilities
is the specific modification of Hilbert-Einstein Lagrangian
for gravity by introduction of a variable gravitational factor,
as described by Eq.~(\ref{TL}).

As we saw in Section III.C.1, such a modification seems
to be acceptable from experimental point of view.
It is in the spirit of the early article by Fiertz \cite{BD}
who was the first to point out that the extremely high precision of WEP
suggests that the coupling of gravity with matter must have
an {\em exact metric} form, but there is still a room to change
the Hilbert-Einstein Lagrangian of GR.
(See also the recent article \cite{DamourExp}.)
Nowadays, we have much more experimental
evidence in favor of Fiertz proposal.

Therefore, in the present article we accept as a basic hypothesis
that $\hbox{PhF}\equiv \hbox{TF}$, and decide to apply Einstein
WEP just in TF. If successful, this hypothesis can help
further developments of string theory as a physical description
of the real world.

This way, we have arrived at a modification of GR
which is maximally close to the original Einstein idea
to describe gravity purely geometrically
by using WEP and Riemannian space-time geometry with a metric
dynamically determined by the {\em usual} matter.
It is remarkable that the above {\em specific extension} of WEP
to scalar-tensor theories of gravity
definitely requires a violation of SUSY
via a cosmological potential with {\em nonzero mass of the dilaton}.

8. Another interesting fact we wish to emphasize
is the recovering of the geometrical meaning of
the EF-dilaton $\varphi$.
As seen from formulae (\ref{sigma21})
and (\ref{Tsigma}), the transition function from EF to TF is
\ben
{}_{{}_T\!}\sigma{}_{\!\!{}_E}=
{{\varphi(\phi)}\over{\sqrt{(D\!-\!1)(D\!-\!2)}}}\,. \hskip 1.3truecm
\la{ETsigma}
\een
Hence, up to a normalization which can be chosen
in any convenient way,
the EF-dilaton $\varphi$ coincides with the transition
function ${}_{{}_T\!}\sigma{}_{\!\!{}_E}$. It is
possible to chose the normalization in such a way
that the EF-dilaton will resemble a matter field in
the corresponding gravi-dilaton Lagrangian (\ref{EL}),
but a purely geometrical interpretation of
the field $\varphi$ seems to be the most plausible
\footnote{
From this point of view, it is natural to
use in EF the scalar field
$\sigma(\phi)=\sqrt{(D\!-\!1)(D\!-\!2)}\varphi(\phi)$.
Then
${}_{{}_E\!}L_{{}_{LEL}}\!\sim\!\sqrt{|{}_{{}_E\!}{\bf g}|}\!
\left(\!{}_{{}_E\!}R\!-\!(D\!-\!1)(D\!-\!2){}_{{}_E\!}(\nabla \sigma)^2
\!+\!
2\Lambda^{obs}{}_{{}_E\!}U(\sigma)\!\right)$,
where ${}_{{}_E\!}U(\sigma)\!=\!{}_{{}_T\!}U\!
\left(e^{(D\!-\!2)\sigma}\right)\!
e^{D\sigma}$.}. It is consistent with geometrical 
interpretation of the field
$\Phi=\exp((D-2)\,{}_{{}_T\!}\sigma{}_{\!\!{}_E})$. 

Our extended WEP does not forbid the use
of EF for purely technical purposes.
For example, from naive point of view,
in order to use EF, one needs only to transform
the TF cosmological potential ${}_{{}_T\!}U(\Phi)$
into the EF one,
$${}_{{}_E\!}U(\varphi)\!=\!{}_{{}_T\!}U\!
\left(\!\exp\!\left(\sqrt{{D\!-\!2}\over{D\!-\!1}}
\varphi\right)\!\right)\!
\exp\!\left(\!{{-D\varphi}\over
{\sqrt{(D\!-\!1)(D\!-\!2)}}}\!\right),$$
and to perform a simple calculation of the coefficients
${}_{{}_E\!}B_{...}(\varphi)$ for matter
terms (\ref{psi}) by using the relations
(\ref{TFC}) and the conformal dimensions
of the corresponding matter fields.
However, there exist more subtle problems in such a transition.
One of them is preservation of
the global properties of the theory.
Another one has been stressed in \cite{E-F_P}: in EF frame, the
helicity-0 and the helicity-2 degrees of freedom
(i.e., $\varphi$ and ${}_{{}_E\!}g_{\mu\nu}$)
are separated to some extent.
Therefore, the EF-Cauchy problem is well posed.
The correct translation of this useful mathematical
property of EF in the language of TF variables is still
an open problem.

In Sections IV--VII, we present some consequences
of our basic hypotheses.

\vskip .2truecm
\paragraph{Experimental Fixing of Phenomenological Frame.
\,\,\,\,\,\,\,\,\,\,\,\,\,\,\,\hskip 2truecm}
\vskip 0.2truecm

Here we wish to make one more remark on the strategy
for the frame choice.
Recently, it was suggested to find the real functions
$F$, $Z$, and $\Lambda$, in PhF by using astrophysical data
\cite{Starobinsky} instead of looking for theoretical
arguments in favor of some specific choice of these
functions. According to our interpretation,
this means an experimental fixing of PhF
for scalar-tensor theories of gravity.

Unfortunately, the realization of this
interesting idea is strongly model-dependent.
First it was applied to the problem of determining
the EF-cosmological potential as the only unknown
function in the gravi-dilaton sector of EF.
In this case, it is sufficient
to use the observational  data for the luminosity
distance $D(z)$ as a function of the red-shift $z$.
In a series of subsequent articles, the field of
theoretical investigation was enlarged to include
the general scalar-tensor theories with arbitrary
functions $F$, $Z$, and $\Lambda$, in PhF.
For this purpose, it was suggested to complete
the information needed for determination of these
three unknown functions, with proper CMB data
because the knowledge of the luminosity function $D(z)$
was not enough to solve the enlarged reconstruction
problem.
The most general considerations can be found in \cite{E-F_P}.
The technical procedure described there
is not directly applicable for our specific TF-theory.
It differs essentially from other scalar-tensor
theories of gravity, as it was pointed out in \cite{E-F_P}
and displayed in details in the present article.

There exist two main difficulties in this method
for fixing PhF:

1) At present, we have no good enough observational
data that would allow us to fix the functions $F$, $Z$,
and $\Lambda$ reliably even in a small interval
of values of their argument.

2) In principle, the total data for all values
of $z\in (-1,\infty)$ needed to solve
completely the reconstruction problem within this
approach will be never available.

Nevertheless, an essential information for further
development of the theory may be reached by using the above idea,
and we give the general scheme for its 4D-DG-realization
below in Section VI.C.

\section{The 4D-Dilatonic Gravity (4D-DG)}

\subsection{The Basic Equations.}

So far, our consideration was independent
of the specific value $D$ of the dimension
of the real space-time ${\cal M}^{(D)}$.
According to our present-days knowledge, $D=4$
(see Section III.C.1).
Ignoring possible higher dimensions
and adopting pseudo-Riemannian metric $g_{\mu\nu}$
with signature $\{+---\}$, i.e., accepting the assumption
${\cal M}^{(D)}={\cal M}^{(1,3)}$,
we have the following final form of
the gravi-dilaton action:
\ben
{\cal A}_{g,\Phi}=
-{\frac c {2\kappa}}\int d^4 x\sqrt{|{\bf g}|}
 \bigl(\Phi R + 2 \Lambda^{obs}\, U(\Phi) \bigr),
\la{4A}
\een
where $\kappa$ is Einstein constant.

We call this simple scalar-tensor model of gravity
{\em $4D$-dilatonic-gravity}.
To simplify our notations, further on we shall skip the
frame index ``T'' for all quantities in our
$\hbox{PhF} \equiv \hbox{TF}$-model.

The 4D-DG without cosmological term contradicts
the gravitational experiments,
because it is nothing but a Brans-Dicke
theory with $\omega=0$ which gives inadmissible value
$\gamma={{1+\omega}\over {2+\omega}}={1\over2}$
for the PPN parameter $\gamma$. This happens because
the zero cosmological term leads to zero mass of the dilaton,
which yields a long range ``fifth'' force.
A universal cure for this problem is to prescribe a
big enough mass to the dilaton.
As we shall see in Section V,
in this case the additional dilatonic force will act
only at very short distances without affecting the
standard gravitational experiments.
Hence, the presence of proper cosmological term
in the action (\ref{4A})
is critical for overcoming the experimental difficulties
with zero values of the Brans-Dicke parameter $\omega$.

In 4D-DG, we have only one unknown function --
the {\em cosmological potential} $U(\Phi)$ --
which has to be chosen to comply with
{\em all} gravitational experiments and observations
in laboratory, in star systems, in astrophysics and
cosmology, and, in addition, to solve the {\em inverse
cosmological problem} (namely, to determine $U(\Phi)$
that reproduces given time evolution of the scale
parameter $a(t)$ in Robertson-Walker (RW) model
of Universe \cite{F00}).

The field equations for the metric $g_{\alpha\beta}$
and the dilaton field $\Phi$ in usual laboratory units
are
\ben
\Phi G_{\alpha\beta}\!-\!\Lambda^{\!obs} U(\Phi)
g_{\alpha\beta} \!-\!(\nabla_\alpha
\nabla_\beta\!-\!g_{\alpha\beta}\BBox)\Phi
\!=\! {\frac {\kappa} {c^2}} T_{\alpha\beta},
\nonumber\\
\BBox\Phi\!+\!\Lambda^{obs} V_{,{}_\Phi}(\Phi)=
\! {\frac {\kappa} {3 c^2}}T.\hskip .25truecm
\la{4FEq}
\een
The relation between {\em dilatonic potential} $V(\Phi)$
and cosmological potential $U(\Phi)$ is
\ben
V(\Phi)={\frac 2 3}\Phi U(\Phi)
-2 \int U(\Phi)d\Phi + const.
\la{4V}
\een

\subsection{The Cosmological Units}

A basic component of our 4D-DG is the
positive cosmological constant $\Lambda^{obs}$.
Despite the relatively large uncertainties
in the corresponding astrophysical data,
we accept the observed value (\ref{Lambda})
of the cosmological constant
as a basic quantity which defines natural units
for all other cosmological quantities.
We call the units in which the cosmological constant
equals one {\em cosmological units} \cite{F00}.
The use of cosmological units emphasizes
the exceptional role of the cosmological constant
for the problems at hand.
We hope that these natural cosmological scales
may throw additional light on the problems.

We introduce a {\em Planck number} $P\approx 10^{-61}$
according to the relation
\ben
P^2 := \Lambda^{obs} L^2_{{}_{Pl}}\approx 10^{-122}\,,
\la{P}
\een
where $ L_{{}_{Pl}}$ is the Planck length.

Then we define a cosmological unit for length
$ A_c = 1/\sqrt{\Lambda} =
P L_{{}_{Pl}} \approx 10^{61} L_{{}_{Pl}}\approx 10^{28}\, \,{\rm cm}$,
a cosmological unit for time
$T_c= A_c/c=  P T_{{}_{Pl}}\approx 11 \, \,{\rm Gyr}$, a cosmological
unit for energy density
$\varepsilon_c= \Omega_\Lambda \varepsilon_{crit}=
\Lambda^{obs} c^2 / \kappa = P^2 \varepsilon_{{}_{Pl}}$,
and rewrite the definition (\ref{P}) in a form
$\Lambda = P^2  L_{{}_{Pl}}^{-2}$.

The cosmological unit for action,
\ben
{\cal A}_c := {{c}\over {\kappa\Lambda^{obs}}} =
P^{-2} \hbar\approx10^{122}\hbar \,,
\la{UnitA}
\een
appears naturally in the formula (\ref{TFA_GD}) for
$D=4$ when one expresses all coordinates $x^\mu$
and the scalar curvature $R$ in cosmological units.

In the next Sections,
when we will discuss the cosmological
applications of 4D-DG, we shall use dimensionless
quantities like $x^\mu/A_c$, $t/T_c$,
$\Lambda R$, $\Lambda G_{\mu\nu}$,
${\cal A}/{\cal A}_c$, $\ldots$, without changing
the notations with the only exception
$\epsilon=\varepsilon/\varepsilon_c$ \footnote{
In these cosmological units, the present value of
the critical energy density of the Universe is
$$\epsilon_{crit}=1/\Omega_\Lambda > 1.$$
Although the last quantity
is {\em not} a fundamental constant,
its important role in GR cosmology makes
natural the popular choice of $\varepsilon_{crit}$
as a unit for energy density,
i.e., the choice of normalization
$\epsilon_{crit}|{}_{{}_{GR}}=1$.
In 4D-DG, $\epsilon_{crit}$ is not of the same
importance as in GR.}.
For this purpose, it is enough to substitute
$\Lambda=1$, $\kappa=1$ and $c=1$
in the action (\ref{4A}) and in the field
equations (\ref{4FEq}).

\subsection{The Vacuum States}

Now we have to justify our model by using additional
available phenomenological information.

One can easily obtain the simplest solutions
of the 4D-DG by considering a constant dilaton field
$\Phi = \Phi^*={\rm const}$.
In this case, the field equations (\ref{4FEq}) give
\,\,\,$T\!=\!T^*\!=\!4\epsilon^*\!=\!{\rm const}$,  and
$G_{\mu\nu}^*\!=\!\left(T_{\mu\nu}^*\!+\!
U(\Phi^*)g_{\mu\nu}^*\right)/\Phi^*$.
If we assume the matter distribution
in the Universe to be isotropic, i.e., if
$T_{\mu\nu}^*\sim g_{\mu\nu}^*$,
we will have $T_{\mu\nu}^*\!=\!\epsilon^* g_{\mu\nu}^*$
and
\ben
V{\!,_\Phi}(\Phi^*)={\frac 4 3} \epsilon^*,\,\,\,
R^*= -2U{\!,_\Phi}(\Phi^*)={\rm const},\nonumber\\
G_{\mu\nu}^* = \lambda^* g_{\mu\nu}^*.
\hskip 2truecm
\la{*}
\een
Hence, in this case, ${\cal M}^{(1,3)}$
is de Sitter space-ti\-me with
dimensionless cosmological constant
$\lambda^*=\left(\epsilon^*+U(\Phi^*)\right)/\Phi^*$.
In addition, the relations $U{\!,_\Phi}(\Phi^*)=2 \lambda^*$
and $R^*=-4\lambda^*$ take place.

In contrast to O'Hanlon's model,
we require that 4D-DG reproduce
GR with {\em nonzero} $\Lambda^{obs}$,
given by Eq.~(\ref{Lambda}),
for some constant value $\bar\Phi$ of dilaton
field $\Phi$. This leads the following
normalization conditions:
\ben
\bar\Phi=1,\,\,\,U(\bar\Phi)=U(1)=1\,.
\la{Unorm}
\een
The value of $\bar\Phi$ defines
the vacuum  state of the theory with
$V{\!,_\Phi}(\bar\Phi)=0$,
as can be seen from the system (\ref{4FEq}).
The first condition in (\ref{Unorm}) reflects our convention
that the parameter $\kappa$ be the {\em standard}
Einstein gravitational constant.

Now the solution of the inverse problem
-- finding the cosmological potential for a given
dilatonic potential --
is described by formulae:
\ben
U(\Phi)=\Phi^2 + W(\Phi),\nonumber\\
W(\Phi)={3\over2}\Phi^2\int^\Phi_1 \Phi^{-3}V{\!,_\Phi}(\Phi)d\Phi.
\la{4WV}
\een
In EF these formulas yield relations ${}_{{}_E}U=1 +{}_{{}_E}W$, 
${}_{{}_E}W=\Phi^{-2}W$ which show, that the term $\Phi^2$ in $U(\Phi)$
describes a pure cosmological constant in EF and the term $W(\Phi)$
represents an additional cosmological potential with basic properties
$W(1)=W{\!,_\Phi}(1)=0$. 

In addition, we define the total energy-momentum tensor
of the matter,
$TT_{\mu\nu} = T_{\mu\nu}^{vac} +T_{\mu\nu} =
\epsilon^{vac}_{matt}\,\,g_{\mu\nu}  + T_{\mu\nu}$,
as a sum of the vacuum-energy tensor (with a density
$\epsilon^{vac}_{matt})$ and the standard part
$T_{\mu\nu}= {2\over {\sqrt{|{\bf g}|}}}
{{\delta {\cal A}_{matt}}\over{\delta {g^{\mu\nu}}}}$
corresponding to matter excitations above the
vacuum ones. As seen from Eq.~(\ref{4FEq}),
the vacuum energy of matter is already included
in the cosmological potential $U(\Phi)$.
Nevertheless, we will see that the consideration of
the total energy-momentum tensor $TT_{\mu\nu}$ is useful.

Since {\em two different} potentials
-- the cosmological potential, $U(\Phi)$,
and the dilaton potential, $V(\Phi)$, --
enter the field equations,
in 4D-DG we have {\em two different} types
of vacuum states of the Universe
-- one with $U{\!,_\Phi}(\Phi)=0$,
and another one with $V{\!,_\Phi}(\Phi)=0$.
This fact -- specific for 4D-DG --
is possible due to the absence of a standard kinetic
term for dilaton field in the action (\ref{4A}).

\subsubsection{De Sitter Vacuum}

We define de Sitter vacuum (dSV)
as a physical state in which $T_{\mu\nu}=0$, i.e.,
\ben
\epsilon^*=\bar\epsilon=0\,.
\la{dSV}
\een
Then the constant dilaton $\bar\Phi$
must be an extremal point of the dilaton potential:
\ben
V{\!,_\Phi}(\bar\Phi)=0\,.
\la{PhidS}
\een
We choose the normalization (\ref{Unorm})
for the ground state of de Sitter type.
Then, from equations (\ref{*}), we obtain the values
$\bar\lambda=1,\,\,\,U^\prime(1)=2, \,\,\,\bar R=-4$,
and the equation for the space-time metric,
\ben
\bar G_{\mu\nu}=\bar g_{\mu\nu},
\la{GdS}
\een
i.e., ${\cal M}^{(1,3)}$ is de Sitter space-time with
$\lambda^*=\bar\lambda=1$.

We interpret dSV state as a {\em physical} vacuum.
In this state of Universe, the space-time is
curved by the vacuum energies of matter and gravitation
(the sum of which is $\epsilon_c=1$).

\subsubsection{Einstein Vacuum}

We define the Einstein vacuum (EV) as a state in which
both total energy-momentum tensor
and the scalar curvature are zero:
\ben
\epsilon^*=\epsilon_0=-\epsilon^{vac}_{matt}\,,
\quad R_0=0\,.
\la{EV2}
\een
Then, because of the relation (\ref{R_relA}),
the constant dilaton $\Phi_0$ must be
an extremal point of the cosmological potential:
\ben
U{\!,_\Phi}(\Phi_0)=0\,.
\la{Phi0}
\een
From equations (\ref{*}), we obtain the values
$U(\Phi_0)=\epsilon^{vac}_{matt}$, $\lambda_0=0$,
and the Einstein equations for the metric,
\ben
(G_0)_{\mu\nu}=0\,.
\la{GE}
\een
Hence, in this case ${\cal M}^{(1,3)}$
is Einstein space-time.
We see that EV is a {\em nonphysical} vacuum
and corresponds to an empty space-time with
``turned off'' quantum vacuum fluctuations.
To reach such state, one has to prescribe
a fixed (nonphysical) negative energy density
to the standard matter, designed to compensate
the quantum vacuum fluctuations which are
included in the cosmological potential.
In spite of the nonphysical character of EV,
it is useful for fixing the 4D-DG parameters.

\subsection{The Simplest Cosmological Potentials}

\subsubsection{The Quadratic Potential}

The quadratic cosmological potential of general form,
$U(\phi)= U_0+{{U_2}\over 2}(\Phi-\Phi_{const})^2$,
contains at most three constant parameters:
$U_1$, $U_2$, and $\Phi_{const}$.
Using the above normalization conditions,
one can easily check that
it must have the form
$U(\Phi)= \Phi_0 + {\sfrac 1 {1-\Phi_0} }(\Phi-\Phi_0)^2=
\Phi^2+ {\sfrac {\Phi_0} {1-\Phi_0}}(\Phi - 1)^2$.
Then $V(\Phi)={\sfrac 2 3}\Phi_0(\Phi-1)^2/(1-\Phi_0)+{\rm const}$,
where ${\rm const}=V(1)$ is an inessential parameter.
The only parameter in the cosmological potential
which remains to be fixed is $\Phi_0=\epsilon^{vac}_{matt}\in (0,1)$.
The restriction on the range of $\Phi_0$
reflects the stability requirement
$U(\Phi)>0$ for all admissible values $\Phi>0$, $\Phi_0>0$.

To recover a new basic relation,
from the second equation of the system (\ref{4FEq})
we obtain in linear approximation
a standard wave equation in de Sitter
space-time for small deviations of the dilaton from its dSV
expectation value, i.e., for the field $\zeta=\Phi-1$:
\ben
{\BBox}\zeta + p_{{}_\Phi}^{-2}\zeta=0\,.
\la{ZEq}
\een
Here $p_{{}_\Phi}$ is the dimensionless
Compton length of the dilaton in cosmological units,
defined by equation
\ben
p^2_{{}_\Phi}=\Lambda^{obs}\, l_{{}_\Phi}^2\,,
\la{p}
\een
where $l_{{}_\Phi}={\hbar\over{c m_{{}_\Phi}}}$
is the usual Compton length of the dilaton.
The relation (\ref{p}) is analogous to Eq.~(\ref{P}) for
Planck number and introduces a new {\em dilatonic scale}
in 4D-DG \cite{F00}. The equation
\ben
\Phi_0= \epsilon^{vac}_{matt}=
\left(1+ {\frac 4 3}p^2_{{}_\Phi} \right)^{-1}
\la{pPhi0}
\een
relates the values of $\Phi_0$ and $p_{{}_\Phi}$,
which yields
\ben
\epsilon_c=
\epsilon^{vac}_{matt}\left({1+
{\frac 4 3}p^2_{{}_\Phi}}\right)\equiv 1 \,.
\la{EV}
\een
In our model, we have only a matter sector
and a gravi-dilaton sector. Hence, the total
cosmological energy density $\epsilon_c$
can include contributions only of these two sectors.
This forces us to interpret the term
$\epsilon^{vac}_{grav}:=
{\sfrac 4 3}\, p^2_{{}_\Phi}\,\epsilon^{vac}_{matt}$
as a vacuum energy of the gravi-dilaton sector.
Then Eq.~(\ref{EV}) reads
$\epsilon_c =  \epsilon^{vac}_{matt}+\epsilon^{vac}_{grav}=1$.
This equation, together with conditions
$\epsilon^{vac}_{matt}>0$, $\epsilon^{vac}_{grav} > 0$
imply a convenient description of the separation
of the cosmological energy density $\epsilon_c$
into two parts by using a new angle variable
$\gamma_{{}_\Phi}$:
$\epsilon^{vac}_{matt}= \cos^2\gamma_{{}_\Phi}$
and $\epsilon^{vac}_{grav}= \sin^2\gamma_{{}_\Phi}$.
Now we obtain
$ p_{{}_\Phi}={\sfrac {\sqrt{3}} 2}\tan\gamma_{{}_\Phi}$,
and under the normalization $V(1)=0$,
the two quadratic potentials acquire the form
\ben
U(\Phi)=\Phi^2+(\Phi-1)^2 \cot^{2}\gamma_{{}_\Phi}, \nonumber \\
V(\Phi)={\frac 2 3} (\Phi-1)^2 \cot^{2}\gamma_{{}_\Phi}.
\la{UV2}
\een

It is clear that the above consideration is
approximately valid in a vicinity of any proper
minimum of the cosmological potential $U(\Phi)$
of general (non-quadratic) form.
Eq.~(\ref{ZEq}), which is exact for
quadratic potentials, gives the linear approximation
for the field $\zeta$ in a vicinity of dSV of any dilaton
potential $V(\Phi)$.

However, if we consider the potentials (\ref{UV2})
globally, i.e., if we accept these formulae to be
valid for all values of the field $\Phi$, we will
encounter a physical difficulty.
Namely, the quadratic potentials (\ref{UV2}) allow
unwanted negative values of $\Phi$ which correspond
to negative energy of gravitons
and to anti-gravity instead of gravity,
or zero value of $\Phi$ which leads to an infinite
gravitational factor. Such values contradict
the fifth basic principle (Section III.C.1).
We are not able to exclude
non-positive values of $\Phi$
in the very early Universe or in astrophysical
objects with extremely large mass densities.
However, to exclude this possibility
for standard physical situations, we need to assume that
only positive values of $\Phi$ are admissible.

One has to emphasize that the zero value of the dilaton
field $\Phi$ in principle may cause some
mathematical problems in solving Cauchy
problem for basic equations (\ref{4FEq}) \cite{E-F_P}.
This problem certainly needs a careful investigation.

\subsubsection{The Dilatonic Potentials
$\sim\left(\frac 1{\nu_+}\Phi^{\nu_{+}}
+ \frac 1{\nu_-} \Phi^{-\nu_{-}}\right)$}

The simplest way to avoid non-positive values of $\Phi$
is to chose a proper form of the dilatonic potential
$V(\Phi)$ in the second equation in (\ref{4FEq})
which forbids {\em dynamically} the zero value of
$\Phi$ and transitions to negative values of this field.
This way, if we start with positive values of the dilaton,
we will have positive $\Phi$ in the entire space-time.

The simplest pair of one parametric potentials
of this type is
\ben
V_{1,1}(\Phi)={1\over 2}p^{-2}_{{}_\Phi}
\left(\Phi+{1\over \Phi}-2\right),\nonumber \\
U_{1,1}(\Phi)= \Phi^2+{3\over {16}}p^{-2}_{{}_\Phi}
\left(\Phi-{1\over\Phi}\right)^2\,.
\la{Vsimplest}
\een

An immediate three-parameter generalization
is given by the formulae
\ben
V_{\nu_{+},\nu_{-}}(\Phi)=
{{p^{-2}_{{}_\Phi}}\over (\nu_{+}+\nu_{-})}
\left({{\Phi^{\nu_{+}}-1}\over\nu_{+}}+
{{\Phi^{-\nu_{-}}-1}\over\nu_{-}}\right),\nonumber\\
U_{\nu_{+},\nu_{-}}(\Phi)=\Phi^2+\hskip 4.8truecm\nonumber\\
{{3\,p^{-2}_{{}_\Phi}/2}\over {\nu_{+}\!+\!\nu_{-}}}
\left({{\Phi^{\nu_{+}\!-\!1}}\over{\nu_{+}\!-\!3}}\!+\!
{{\Phi^{-\nu_{-}\!-\!1}}\over{\nu_{-}\!+\!3}}\!-\!
{{(\nu_{+}\!+\!\nu_{-})\,\Phi^2} \over
{(\nu_{+}\!-\!3)(\nu_{-}\!+\!3)}}
\right).
\hskip .2truecm
\la{Vnu}
\een
The two additional parameters $\nu_{+}>0,\,\,\nu_{-}>0$ ($\nu_{+}\neq 3$)
determine different asymptotics of the potentials (\ref{Vnu})
at the points $\Phi=0$ and $\Phi=\infty$.
The corresponding EF additional cosmological potential 
in ${}_{{}_E}\!U_{\nu_{+},\nu_{-}}=1+{}_{{}_E}\!W_{\nu_{+},\nu_{-}}$ is
$$
{}_{{}_E}\!W_{\nu_{+},\nu_{-}}\!=\!
{{3\,p^{-2}_{{}_\Phi}/2}\over {\nu_{+}\!+\!\nu_{-}}}
\left({{\Phi^{\nu_{+}\!-\!3}}\over{\nu_{+}\!-\!3}}\!+\!
{{\Phi^{-(\nu_{-}\!+\!3)}}\over{\nu_{-}\!+\!3}}\!-\!
{{(\nu_{+}\!+\!\nu_{-})} \over
{(\nu_{+}\!-\!3)(\nu_{-}\!+\!3)}}
\right).
$$

\subsubsection{The Potentials of General Form}

For potential pairs $V(\Phi),\,\, U(\Phi)$ of more
general form with the same asymptotics at
$\Phi=0$ and $\Phi=\infty$ as the potential pairs
$V_{\nu_{+}\nu_{-}}(\Phi),\,\, U_{\nu_{+}\nu_{-}}(\Phi)$
given by Eq.~(\ref{Vnu})
but  with more complicated behavior,
for finite values of the dilaton field $\Phi$,
we have the following common properties
for dSV state:
\ben
V(1)=0,\,\,\,V{\!,_\Phi}(1)=0,\,\,\,
V{\!,_{\Phi\Phi}}(1)=p^{-2}_{{}_\Phi};
\hskip 1.truecm\nonumber\\
U(1)=1,\,\,\,U{\!,_\Phi}(1)=2,\,\,\,
U{\!,_{\Phi\Phi}}(1)=
2\left(1+{3\over4}p^{-2}_{{}_\Phi}\right) \,,
\la{dSVcommon}
\een
and for EV states:
\ben
U{\!,_\Phi}(\Phi_0)=0, \hskip 3truecm \nonumber\\
\epsilon^{vac}_{matt}=
U(\Phi_0)=-{3\over 4}V{\!,_\Phi}(\Phi_0),\,\,\,
\epsilon^{vac}_{gr}=1-U(\Phi_0)\,.\hskip 0.truecm
\la{EVcommon}
\een
These conditions yield a representation of
more general potentials in the form
\ben
V(\Phi)=V_{\nu_{+},\nu_{-}}(\Phi)\,\cos^2\iota +
\Delta V(\Phi)\,\sin^2\iota\,,\nonumber\\
U(\Phi)=U_{\nu_{+},\nu_{-}}(\Phi)\,\cos^2\iota +
\Delta U(\Phi)\,\sin^2\iota\,,
\hskip .1truecm
\la{VUplus}
\een
with an arbitrary constant mixing angle $\iota$,
additional dilaton potential $\Delta V(\Phi)$, and
additional cosmological potential $\Delta U(\Phi)$.

In general, the potentials (\ref{VUplus}) may have
an oscillatory behavior with more then one extremum.
The simplest example is given by the pair (\ref{Vsimplest})
and
\ben
\Delta V(\Phi)=
{{p^{-2}_{{}_\Phi}}\over {2\pi^2}}\sin^2(\pi(\Phi-1)),
\hskip 3.1truecm \nonumber\\
\Delta U(\Phi)=-{{3}\over 4} \pi p^{-2}_{{}_\Phi}
\Bigl(
\bigl(2\pi(\Si(2\pi\Phi)-\Si(2\pi))-1\bigr)\Phi^2+
\nonumber\\
\Phi\cos(2\pi\Phi)
+{1\over{2\pi}}\sin(2\pi\Phi)
\Bigr),
\hskip .6truecm
\la{adU}
\een
where $\Si(...)$ is the integral sine function.
It is interesting that for values
$\iota=\pi/2\pm\delta$ with a small positive
$\delta \leq \delta_1\approx 0.3$,
a second minimum  of the cosmological potential
(i.e., a second EV) appears. Below some value
$\delta \leq \delta_2\approx 0.14<\delta_1$,
it becomes negative: $U_{min}^{(2)}<0$.
This corresponds to a negative cosmological constant term
in the action (\ref{4A}) and to a negative vacuum energy
of matter. The dilatonic potential $V(\Phi)$ has many
extremal points for all $\iota \neq 0$,
so there are many de Sitter vacua.
The normalization (\ref{Unorm}) is valid for the absolute
minimum of the potential $V(\Phi)$ which determines
the ground state of the theory.

Using more sophisticated additional potentials
$\Delta V(\Phi)$ in Eq.~(\ref{VUplus}),
one may expect more complicated
structure of the sets of dSV and EV states.
Obviously, the requirement for existence of a simple
physical vacuum state in the 4D-DG model may
restrict the admissible cosmological potentials.

It turns out that in our DG we have an important
restriction on the structure of the vacuum states of the
theory which follows from condition (\ref{Nzero}):
if we wish to have a DG model that is {\em globally}
equivalent to nonlinear gravity, the potentials
$V(\Phi)$ and $U(\Phi)$ must have {\em only one} extremum.
Otherwise their second derivatives with respect to
dilaton $\Phi$ will have zeros and the condition
(\ref{Nzero}) will be violated.
Then, from the stability requirements it follows that
the only extremum of these potentials must
be a minimum. This means that the dilaton field
$\Phi$ must have nonzero positive mass $m_{{}_\Phi}$.
Thus, we arrived at the following

\noindent
{\bf \em Proposition 1:}
{\em If the condition $(\ref{Nzero})$ holds globally,
then the physical vacuum in DG is unique,
and the mass $m_{{}_\Phi}$ is non-zero.
In addition, the stability requirement
implies that $m_{{}_\Phi}$ is real and positive.
}

We wish to emphasize that this conclusion
is independent of the space-time dimension $D$.

\noindent
{\bf \em Proposition 2:}
{\em As a consequence of Proposition 1 and Eq.~$(\ref{4WV})$,
the cosmological potential $U(\Phi)$ is strictly positive
in the interval $\Phi\in (0, \infty)$.}

The proof is simple: According to Proposition~1,
$V_{,_\Phi}(1)=0$ is the only minimum of $V$.
Hence, $V_{,_\Phi}(\Phi)<0$ for $\Phi\in (0,1)$,
and $V_{,_\Phi}(\Phi)>0$ for $\Phi\in (1,\infty)$.
Then Eq.~(\ref{4WV}) yields
$W(\Phi)\geq 0$ for $\Phi\in (0,\infty)$.
But $W(1)=0$ is the only zero point of $W$,
and $U(1)=1$.
As a result, $U(\Phi)>0$ for any $\Phi\in (0,\infty)$.

We need better knowledge of the field dynamics in
the 4D-DG to decide what kind of
additional requirements on the cosmological term
in the action (\ref{4A}) need to be imposed.

\section{Weak Field Approximation for a Static
System of Point Particles}

To enhance the comparison of our formulae
with the well known ones,
in this Section we use standard (instead of cosmological)
units and non-relativistic notations.

\subsection{General Considerations}

In vacuum, far from matter, 4D-DG has to allow
weak field approximation:
$\Phi=1+\zeta$,\,\,$|\zeta|\ll 1$, which we
consider in harmonic gauge. Then the field $\zeta$
obeys Eq.~(\ref{ZEq}).
This equation shows that the weak field
approximation does not depend on
the precise form of the dilaton potential,
but only on the dilaton mass $m_{{}_\Phi}$ and
(implicitly) on the cosmological constant $\Lambda^{obs}$.
Hence, within the weak field approximation,
we can obtain information only about
these two parameters of the cosmological term
in the action (\ref{4A}).

For few point particles of masses $m_a$ at rest,
which are the source of metric and dilaton fields
in Eq.~(\ref{4FEq}),
we obtain Newtonian approximation
($g_{\alpha\beta}\!=
\!\eta_{\alpha\beta}\!+\!h_{\alpha\beta}$,
$\!|h_{\alpha\beta}|\!\ll\!1$)
for the gravitational potential $\varphi({\bf r})$
and the dilaton field $\Phi({\bf r})$:
\ben
\varphi({\bf r})\!=\!- G
\sum_a\!{\frac {m_a}{|{\bf r - r}_a|}}\!
\left(\!1\!+\!\alpha(p_{{}_\Phi})
e^{-|{\bf r - r}_a|/l_\Phi} \right)
\nonumber\\
- {\frac 1 6}p_{{}_\Phi}^2 c^2
\sum_a\!{\frac{m_a} M}
\left(|{\bf r - r}_a|/l_\Phi\right)^{\!2},
\nonumber \\
\!\Phi({\bf r})\!=\!1+\!{\frac 2 3}
{\frac G{c^2(1-{\frac 4 3} p_{{}_\Phi}^2)}}
\sum_a\!{\frac {m_a}{|{\bf r - r}_a|}}
e^{- |{\bf r - r}_a|/l_\Phi},
\la{SolNewton}
\een
where
$G\!=\!{\frac {\kappa c^2}
{8\pi}}(1\!-\!{\frac 4 3}p_{{}_\Phi}^2)$ is
Newton constant, and $M\!=\!\sum_a m_a$ is the total mass.
The term
\begin{eqnarray*}
& & -{\frac {c^2} 6}p_{{}_\Phi}^2
\sum_a{\frac{m_a} M}
\left(\frac{|{\bf r-r}_a|}{l_\Phi}\right)^2\\
& & \qquad \qquad =
-{\frac {c^2} 6} \Lambda^{obs}\left|{\bf r}-
\sum_a {\frac {m_a} M}{\bf
r}_a\right|^2+{\rm const}
\end{eqnarray*}
in $\varphi({\bf r})$
is known from GR with $\Lambda^{obs} \neq 0$.
It represents a universal anti-gravitational
interaction of a test mass $m$ with any other mass $m_a$
via repulsive elastic force
\ben
{\bf F}_a=
{\frac 1 3}\Lambda^{obs} m c^2 {\frac{m_a} M}({\bf r - r}_a)\,.
\la{Fel}
\een

\subsection{The Equilibrium between Newton Gravity
and Weak Anti-Gravity}

It is instructive to evaluate the average effect of the
presence of the repulsive interaction between
matter particles described by Eq.~(\ref{Fel})
in a homogeneous and isotropic medium with mass density
$\rho_M$ and mass $M=\rho_M\times {\rm Volume}$.
The condition for equilibrium between the Newtonian
gravitational force and the new
anti-gravitational one (\ref{Fel}),
$|{\bf F}_a|=|{\bf F}_{Newton}|$, reads
\ben
\Lambda c^2 = 4 \pi G \rho_M\,.
\la{Equi}
\een

Rewritten in terms of the standard cosmological
parameters $\Omega_\Lambda = {{\Lambda c^2}\over{3 H^2}}$
and $\Omega_M={ {8 \pi G \rho_M} \over{3 H^2}}$, Eq.~(\ref{Equi})
reads $\Omega_\Lambda ={1\over 2} \Omega_M$.
According to the modern astrophysical data at the present
epoch, we have
$\Omega_\Lambda\approx 2\Omega_M>{1\over 2}\Omega_M$.
This means that at scales of several hundred ${\rm Mpc}$,
at which the CP is applicable, the repulsive force
(\ref{Fel}) dominates the Newtonian gravitational
force and confirms the conclusion that the expansion
of the Universe must be accelerating at the present epoch.

In contrast, from Eq.~(\ref{Equi}), we see that
in the case of a denser medium
(e.g., in star systems, in stars and for usual matter on the Earth),
the Newtonian Gravitational force is many orders of magnitude
larger than the anti-gravitational force~(\ref{Fel}).

\subsection{Constraints on the Mass of the Dilaton from
Cavendish-Type Experiments}

For the Solar System distances, $l\leq 1000 \,{\rm AU}$,
the whole repulsive elastic term in
$\varphi({\bf r})/{c^2}$ may be neglected since it is of 
order $\leq 10^{-24}$ \cite{BDLambda}.
Then we arrive at the known form of gravitational
potential $\varphi({\bf r})\,$ \cite{STGExp}, but with a specific
for 4D-DG coefficient:
$$\alpha(p_{{}_\Phi})=
{\frac{1/4+p_{{}_\Phi}^2}{3/4- p_{{}_\Phi}^2}}.$$
The comparison of the two existing possibilities -- 
$\alpha \geq\!{\frac 1 3}$ or $\alpha \leq -1$ -- 
with Cavendish type experiments yields the
experimental constraint $l_\Phi \leq 1.6\,{\rm mm}$
if one uses the old data from articles by
De~R\'ujula \cite{STGExp} and by Fischbach and
Talmadge \cite{FT}.
The modern data for validity of Newton law of gravitation
\cite{mum} give
$l_\Phi \leq 75 - 218 \,\mu{\rm m}$, 
hence,
\ben
p_{{}_\Phi} \leq 10^{-30}.
\la{pObs}
\een
Now we see that:

1) Formulae (\ref{SolNewton}) and (\ref{pObs}) show
that deviations from Newton law of gravity  
cannot be expected at distances greater 
then $100\,\mu{\rm m}$.

2) The 4D-DG correction of the relation between
Einstein constant $\kappa$ and Newton constant $G$, 
\ben
\kappa = {{8\pi G}\over{c^2}}\left(1-{4\over 3}
p^2_{{}_\Phi}\right)^{-1} \, ,
\la{Gkappa}
\een
is extremely small and practically inessential.

3) Finding $p_{{}_\Phi}$ is
equivalent to finding 
$m_{{}_\Phi}= (P/p_{{}_\Phi}) M_{{Pl}}$.
Thus, we obtain the constraint
\ben
E_{{}_\Phi}=m_\Phi c^2 \geq 10^{-3} \,{\rm eV}
\la{m_Phi}
\een
which does not exclude a small value of the
rest energy $E_{{}_\Phi}$ of a hypothetical
$\Phi$-particle with respect to typical rest energy 
scales for particles in SM.
But the corresponding value of the mass of dilaton $\Phi$
is strikingly different from the non-physical
small value of the mass of the scalar field in models with 
a quintessence field, or in inflation models with 
a slow rolling scalar field. Moreover, in 4D-DG, values
$m_{{}_\Phi}\sim 1\,\, {\rm GeV}$ to $m_{{}_\Phi}\sim 1\,\, {\rm TeV}$,
$m_{{}_\Phi}\sim M_{Pl}$, or even
$m_{{}_\Phi} > M_{Pl}$ (i.e., $p_{{}_\Phi}\sim P$,
or even  $p_{{}_\Phi} < P$) are not excluded at present
by the known gravitational experiments.

4) We obtain much more definite predictions
than the general relations between
$\alpha$ and the length $l_\Phi$ given in
the articles by De~R\'ujula and by Helbig
\cite{STGExp} for general scalar-tensor theories
of gravity. This is because in 4D-DG
the condition $\omega\equiv 0$ fixes the value
of Brans-Dicke parameter and we have to extract
from experiments only information about the dilaton mass.

\subsection{Basic Solar System Gravitational Effects in 4D-DG}

In the Solar System phenomena, the factor
$e^{-l/l_\Phi }$ has fantastically small values 
($<\exp(-10^{14})$ for $l$ of order of the Earth-Sun distance, 
or $< \exp(-3\times 10^{11})$ 
for $l$ of order of the Earth-Moon distance), 
so there is no hope of finding any 
differences between 4D-DG and GR in this domain.

The parameterized-post-Newtonian (PPN)
solution of equation (\ref{4FEq}) is complicated,
but because of the constraint $p_{{}_\Phi} < 10^{-30}$,
with huge precision we may neglect the second term 
in the gravitational potential
$\varphi({\bf r})$, put $\alpha\!=\!{\frac 1 3}$, 
and use  Helbig's PPN formalism which differs
essentially from the standard one \cite{Will}
for zero mass dilaton fields.

The basic gravitational effects in the Solar System are:

\subsubsection{Nordtvedt Effect}

In 4D-DG, a body with a significant gravitational self-energy 
$E_{{}_G}=\sum_{b\neq c} G{\sfrac {m_b m_c}{|{\bf r}_b - {\bf r}_c|} }$ 
will not move along
geodesics due to the additional universal {\em anti-gravitational} force, 
\ben
{\bf F}_{\!{}_N} = -{\sfrac 2 3} E_{{}_G}\nabla \Phi \, .
\la{NordtF}
\een
For usual bodies, this force is too small even at distances
$|{\bf r-r}_a|\!\leq\!l_\Phi$ because of the small factor $E_G$.
Hence, in 4D-DG there is no strict strong equivalence principle, 
although the weak equivalence principle is not violated.

The experimental data for Nordtvedt effect caused by the Sun are
formulated as a constraint $\eta=0 \pm .0013$ \cite{Will} on the
parameter $\eta$ which in 4D-DG becomes a function of the distance
$l$ to the source:
$$\eta(l)=-{\sfrac 1 2}\left(1+l/l_\Phi\right)
e^{-l/l_\Phi}\,.$$
Taking into account the value of the Astronomical
Unit (AU) $l_{{}_{\rm AU}} \approx 1.5\times 10^{11}\,{\rm m}$, we obtain
from the experimental value of $\eta$ the constraint $l_\Phi \leq
2\times 10^{10 }\,{\rm m}$.

\subsubsection{Time Delay of Electromagnetic Waves}

The standard action for electromagnetic field, 
and the Maxwell equations in 4D-DG do not depend directly
on the field $\Phi$. Therefore, the influence of
this field on electromagnetic phenomena like
the propagation of electromagnetic waves in vacuum is possible
only indirectly -- via its influence on the space-time
metric. The Solar System measurements of the time delay
of electromagnetic pulses give the value
of the post Newtonian parameter $\gamma$ \cite{Will} used above.  
In 4D-DG,  we have the relation
$\gamma^{obs}\, b( l_{{}_{\rm AU}})= 1$, which gives the
constraint $l_\Phi \leq  10^{10 }\,{\rm m}$. Here
$$b(l):=1+{\sfrac 1 3 }(1+l/l_\Phi) e^{-l/l_\Phi}.$$

\subsubsection{Perihelion Shift}

Helbig's results \cite {STGExp} applied to 4D-DG
give the following formula for the perihelion
shift of a planet orbiting around the
Sun ($M_\odot$ is the Sun's mass): 
$$
\delta \varphi_{\,per} 
= {\frac {k(l_p)}{b(l_p)}}\delta\varphi^{\,{}_{GR}}_{\,per} \, .
$$ 
Here $l_p$ is the
semi-major axis of the orbit of planet, 
and 
$$
k(l_p) \approx 1 +
{\sfrac 1 {18}}\left( 4 + {\sfrac{l_p^2}{l_\Phi^2}}{\sfrac {l_p
c^2}{GM_\odot}}\right) e^{-l_p/l_\Phi} -{\sfrac 1 {27}}
e^{-2l_p/l_\Phi}
$$ 
is obtained by neglecting the eccentricity of the orbit.  
The observed value of perihelion shift of
Mercury gives the constraint $l_\Phi \leq  10^{9 }{\rm m}$
(see the article by De~R\'ujula \cite{STGExp}).

The above weak restrictions on $l_\Phi$, derived in
4D-DG from gravitational experiments in the Solar System,
show that in presence of a dilaton field $\Phi$, 
essential deviations from GR are impossible.

\subsection{The 4D-DG and Star Structure}

The qualitative consideration of star structure,
based on equations (\ref{Equi}), (\ref{SolNewton}),
and (\ref{pObs}) shows that one cannot expect
essential changes in the static structure of usual stars
and neutron stars in 4D-DG, at least in the weak field
approximation. The strong field effects
in neutron stars are not studied yet.

The study of strong field effects due to the presence of dilaton 
$\Phi$ in boson stars is an open problem, too.
The numerical study of boson stars showed
that their structure does not depend on the
precise form of the dilaton potential, 
if the deviation of dilaton $\Phi$ from its dSV value is small.  
Under this condition the boson star structure  can 
differ from predictions of GR by few percents,
if the mass of the boson field is similar to
the mass of the dilaton:
$m_{{}_B}\simeq m{{}_\Phi}$ \cite{PF2}.
Unfortunately, boson stars are not observed
in Nature and we are not able to check this
prediction of 4D-DG.

Concluding this Section, we can say that
new phenomena that are due to the presence of a {\em static}
4D-DG dilaton $\Phi$ seem to be impossible at
the scales of the star systems. Therefore, one must
look for such phenomena at much bigger
astrophysical scales.

Another field of search for possible new phenomena that are 
due to the 4D-DG dilaton $\Phi$ are the non-static
problems.

\section{Cosmological Application of 4D-DG}

In this Section, we will show that the real domain where one
can find new phenomena predicted by 4D-DG is cosmology.
The design of a {\em realistic} model of the Universe
lies beyond the scope of the present article.
Here we would like only to outline some general features of 4D-DG
applied to cosmological problems, and to show that 
this model is able to solve some of the problems listed
in the Introduction.

We derive the equations of the inverse cosmological
problem in 4D-DG and demonstrate indications for some
unexpected new physics.

In the present Section we use only cosmological units as defined
in Section IV.B.

\subsection{Basic Equations for RW Universe in 4D-DG}

Consider RW adiabatic homogeneous isotropic Universe
with $ds^2_{RW}= dt^2 - a^2(t) dl^2_k$, 
where $dl^2_k={\frac {dl^2}{1-kl^2}+l^2(d\theta^2+
\sin^2\theta)d\varphi^2}$ (for $k=-1, 0, 1$), 
and $t$, $a$ are dimensionless, 
in the presence of matter with dimensionless energy density
$\epsilon(a)$ and dimensionless pressure $p$. 
The pressure $p$ become a function of the variable $a$
when the equation of state of the matter
is given in the form $p=p(\epsilon)$. For usual matter,
the equation of state is $p=w\epsilon$ with some
{\em constant} parameter $w$ (for dust matter $w_3=0$, 
for ultra-relativistic matter $w_4=1/3$, 
for vacuum energy of matter $w_0=-1$, etc.). Then,
introducing the parameter $n=3(1+w_n)$ and the corresponding
{\em constant} coefficients $\Phi_n$
(see Eq.~(\ref{pPhi0}) for $\Phi_0$), one obtains
\ben
\epsilon^{tot}_{matt}\!=\!\epsilon^{vac}_{matt}+\!\epsilon\!=\!
\sum_n{\frac{\Phi_n}{a^n}},\,\,\,\,\,
p^{tot}\!=\!\sum_n {\frac {n-3} 3 }{\frac{\Phi_n}{a^n}}.
\la{epsilon_p}
\een
Obviously, for normal matter $n>0$, i.e., its energy density
and pressure decrease under expansion of the space-time volume.
Moreover, for normal matter, the condition 
$n\in [3,4]$ (i.e., $w_n\in [0, 1/3]$) holds.  

\subsubsection{The Equations for Time Evolution}

The dimensionless action for RW model of Universe
\ben
{\cal A}_{a\Phi}\!=\!\!\!\int\!\!dt\,a^3\!\!
\left(\!\!-\!3\Phi\!\!\left(\!{\frac {\dot\Phi} \Phi}
{\frac {\dot a}{a}}\!+\!{\frac{\dot a^2}{a^2}}\!\right)\!
\!-\!\!\left(\!\epsilon(a)\!+\!U(\Phi)\!
-\!3k{\frac {\Phi} {a^2}}\!\right)\!\!\right)\,
\la{RWA}
\een
is obtained from the original action (\ref{4A})
per unit volume by 
substituting in it the RW metric and 
neglecting a term which is a total derivative
with respect to the time, 
and omitting the factor (\ref{UnitA}).

Introducing canonical moments
$\pi_a=-3a^2\dot\Phi -6\Phi a\dot a$
and $\pi_{{}_\Phi}=-3a^2\dot a$, we obtain
the canonical Hamiltonian of the dynamical
system with action (\ref{RWA}):
\ben
{\cal H}\!=\!{\frac 1 {3a^3}}\pi_{{}_\Phi}
(\Phi\pi_{{}_\Phi}\!-\!a\pi_a)\!+\!
a^3\left(U(\Phi)\!+\!\epsilon(a)\!-\!
3\Phi{\frac k {a^2}}\right).
\la{H}
\een
From Eq.~(\ref{R_relA}) and from the (00)-Einstein
equation in (\ref{4FEq}), we obtain the following
basic dynamical equations governing 
the time evolution of the 4D-DG-RW Universe:
\ben
{\frac {\ddot a} a}+
{\frac {\dot a^2} {a^2}}+ {\frac k {a^2}}=
{\frac 1 3}\, U_{{,_\Phi}}(\Phi),\nonumber\\
{\frac {\dot a} a} \dot\Phi+
\Phi \left({\frac {\dot a^2} {a^2}}
+{\frac k {a^2}}\right)=
{\frac1 3}\bigl(U(\Phi)+ \epsilon(a)\bigr).
\la{DERWU}
\een
The field equation for the dilaton $\Phi$
in the system (\ref{4FEq}) gives 
\ben
\ddot\Phi+3{\frac{\dot a} a}\dot\Phi+V_{,_\Phi}(\Phi)=
{\frac 1 3}\epsilon(a) - p.
\la{EqPhi}
\een

The conservation law
$\nabla_\alpha\,T^\alpha_\beta=0$,
when applied to the RW metric,
gives the well known GR relation 
${d\over{da}}\left(a^3\epsilon\right)=-3p a^2$.
It makes it possible to exclude the pressure
$p$ from Eq.~(\ref{EqPhi}). Then, just as in GR,
one can prove that Eq.~(\ref{EqPhi})
follows from the basic system of
dynamical equations (\ref{DERWU}) \cite{F00}.
Hence, any solution of the system (\ref{DERWU})
satisfies Eq.~(\ref{EqPhi}), 
and, when solving time evolution problems for the 4D-DG-RW
Universe, one has to regard Eq.~(\ref{EqPhi}) into
account only as an useful additional relationship.

\subsection{The Energetic Relations}

We would like to emphasize that the first of the equations
(\ref{DERWU}) and Eq.~(\ref{EqPhi})
are equivalent to the Euler-Lagrange
equations for the action (\ref{RWA}).
The second of the equations (\ref{DERWU}),
being a first order differential equation,
represents the corresponding energy integral
of this Euler-Lagrange system.
In canonical variables this equation reads
\ben
{\cal H} \approx 0.
\la{H0}
\een
By the symbol ``$\approx$'', 
we mean that the canonical Hamiltonian ${\cal H}$
equals zero in weak sense, i.e., on the solutions of
the field equations. This is a well known property of
all theories that are covariant under general coordinate
transformations. Accordingly, one can represent the
second of the equations (\ref{DERWU}) in a form of
a mechanical energy-conservation law
\ben
\epsilon_{total}=\epsilon_{kin}+\epsilon_{pot}={\rm const}\,(=0),
\nonumber \\
\epsilon_{kin}=3\Phi\left({\frac 1 4 }
{\frac{\dot\Phi^2}{\Phi^2}}-
\left({\frac {\dot a}{a} +{\frac 1 2}
{\frac{\dot\Phi}{\Phi}}}\right)^2\right),\nonumber \\
\epsilon_{pot}=
\epsilon(a)+U(\Phi)-3k{\frac {\Phi} {a^2}}.
\la{energy_c}
\een
An important and unusual feature of the 4D-DG-RW Universe
is that {\em the kinetic energy $\epsilon_{kin}$ is
not positive definite}. As in GR, the contribution of
the metric is related to the ${{\dot a}/ a}$-term and has
a negative sign. In GR, the RW model behaves like a mechanical
system with a definite kinetic energy (with a wrong minus
sign) because of the absence of a ${{\dot\Phi}/\Phi}$-term,
and the dynamics is much simpler than in 4D-DG. The zero
value of the total energy $\epsilon_{total}$ is physically
inessential. Its conservation describes the balance of
energy in the fields-matter system, and the positive sign
of the matter energy $\epsilon(a)>0$ defines the physically
correct signs of the other terms in relations (\ref{energy_c})
and their physical interpretation. We see that {\em the kinetic energy
of the gravi-dilaton complex may play the role of a source of energy
for matter.}

\subsubsection{Friedmann Form of Time Evolution}

One can represent the time evolution of the Universe
in 4D-DG by using the effective Friedmann equation
\ben
{\frac{\dot a^2}{a^2}}+{\frac k {a^2}}=\epsilon_{eff}(a).
\la{Friedmann}
\een
Then from Eq.~(\ref{DERWU}) one obtains the following
second order non-autonomous system of differential
equations for the effective energy density
$\epsilon_{eff}(a)$ and for the dilaton $\Phi$ as
a function of the scale parameter $a$:
\ben
a{\frac {d\epsilon_{eff}}{da}}+4\epsilon_{eff}=
2U_{,_\Phi}(\Phi), \nonumber \\
\left({\frac 1 3}\epsilon_{eff}-{\frac k {a^2}}\right)
a{\frac {d\Phi}{da}}+\Phi\epsilon_{eff}=
{\frac 1 3}\bigl(U(\Phi)+\epsilon(a)\bigr).
\la{Eeff}
\een
The effective Friedmann equation (\ref{Friedmann})
gives the time evolution in the form
\ben
\Delta t=\pm \int_{a_{in}}^a
{\frac {da}{\sqrt{{1\over 3}a^2\epsilon_{eff}(a)-k}}}\,.
\la{tInt}
\een

We have to emphasize that this form of time evolution
only resembles the Friedmann evolution in GR. Actually,
the dilaton degree of freedom is hidden in the effective
energy density $\epsilon_{eff}(a, C_1,C_2)$ 
(which depends on two integration constants
$C_{1,2}$ since it is a solution of the second order system
(\ref{Eeff})). Representing the effective energy density
in the form $\epsilon_{eff}=
{\frac 1 \Phi}\left(\epsilon +\epsilon_{{\Phi}}\right)$, 
and using (ii)-Einstein equations (\ref{4FEq}), 
one can introduce the effective dilaton energy density
$\epsilon{{}_\Phi}=U(\Phi)-3H\dot\Phi$
and effective dilaton pressure
$p{{}_\Phi}=\ddot\Phi+2H\dot\Phi-U(\Phi)$, where
$H=\dot a/ a$ is the standard Hubble
parameter. Then we obtain 
\ben
w{{}_\Phi}={\frac {p{{}_\Phi}}{\epsilon{{}_\Phi}} }=
-{\frac{U(\Phi)-\ddot\Phi-2H\dot\Phi}{U(\Phi)-3H\dot\Phi}} \, .
\la{w}
\een
In the other cosmological models with scalar field $\varphi$ 
in Einstein frame an analogous parameter $w{{}_\varphi}=
{\frac{\dot\varphi^2/2-V(\varphi)}{\dot\varphi^2/2 +V(\varphi)}}\,$,
is used. The comparison of the parameter  $w{{}_\varphi}$ with 
 $w{{}_\Phi}$ (\ref{w}) shows once again the essential 
difference between these models and 4D-DG.
Nevertheless, for static fields $\Phi={\rm const}$ and 
$\varphi={\rm const}$ we have
$w{{}_\Phi}= w{{}_\varphi}=w_0=-1$.

\subsubsection{Normal Forms of the Equations}

Considering the Hubble parameter $H$ as a function
of the scale parameter $a$, $H(a)=a^{-1}\dot a(t(a))$ 
(where $t(a)$ is the inverse function of $a(t)$),
using a new variable $x_3=\ln a$,
and denoting by prime the differentiation with respect to
$x_3$, we write down the equations (\ref{DERWU})
as a second order non-autonomous system for the functions
$\Phi(x_3)$ and $H^2(x_3)$:
\ben
{\frac 1 2}(H^2)^\prime +2 H^2 + k e^{-2x_3}=
{\frac 1 3}U_{,_\Phi}(\Phi), \hskip 1.15truecm
\nonumber \\
H^2 \Phi^\prime +\left(H^2+k e^{-2x_3}\right)\Phi=
{\frac 1 3}(U(\Phi)+\epsilon(e^{x_3}))
\la{NDE}
\een
and the relation
$\Delta t\!=\int\limits^{x_3}_{{x_3}_{in}}{\frac {dx_3}{H(x_3)}}$
describing the dependence of the scale parameter
$a$ on the cosmic time $t$.

\vskip .3truecm

\paragraph{First Normal Form of the Dynamical Equations\\}

Introducing a new regularizing variable $\tau$ by 
\ben
dt=H d\tau\,\,\,\Rightarrow\,\,\,
ds^2=H^2(\tau)d\tau^2-a^2(\tau)dl^2_k,
\la{tau}
\een
and the notations $x_1:=H^2 \geq 0$, $x_2:=\Phi > 0$,
we can rewrite the basic system (\ref{DERWU})
in standard normal form:
\ben
{\frac d {d\tau}}{\bf x}= {\bf f(x)} \,,
\la{Norm}
\een
where ${\bf x}, {\bf f}\in {\cal R}^{(3)}$ are
three-dimensional vector-columns with components
$\{x_1,x_2,x_3\}$ and $\{f_1,f_2,f_3\}$, respectively, and
\ben
f_1= 4 x_1
\left(Z_1(x_2)-x_1-ke^{-2x_3}/2\right),
\hskip 1.1truecm \nonumber \\
f_2=x_2\left( Z_2(x_2) -x_1-  k e^{-2x_3}\right)+
\epsilon(e^{x_3})/3, \nonumber \\
f_3=x_1.\hskip 5.35truecm
\la{f}
\een
Here the quantities
\ben
Z_1(\Phi)={\frac 1 6}U{\!,_\Phi}(\Phi)=
Z_2(\Phi)+{\frac 1 4}V_{\!,_\Phi}/\Phi,
\nonumber \\
Z_2(\Phi)={\frac 1 3}U(\Phi)/\Phi=
{\frac 1 3}\Phi+{\frac 1 3}W/\Phi.
\hskip .4truecm
\la{Z1Z2}
\een
are regarded as functions of $x_2=\Phi$.

Now one can derive another important relation of
a pseudo-energetic type by using Eq.~(\ref{EqPhi}).
The qualitative dynamics of its solutions is
determined by the function
\ben
\eta={1\over 2}\dot\Phi^2+V(\Phi)-{1\over 3}\Phi T,
\la{eta}
\een
where $T$ is the trace of the energy-momentum tensor.
To some extent, this function plays the role of
a (non conserved) energy-like function for the dilaton field
$\Phi$ in the 4D-DG-RW Universe, and obeys the equation
\ben
{d\over {d\tau}} \eta = N_{+} - N_{-}\,.
\la{Eq_eta}
\een
Here
\ben
N_{+}:= -{1\over 3}\Phi H^2 aT{\!,_a} \geq 0,
\nonumber \\
N_{-}:= 3 \left({d\over {d\tau}}\Phi\right)^2 \geq 0.
\la{NN}
\een
One has to use the equations (\ref{epsilon_p}) and
the definition of $T$ to derive the formula
$aT{\!,_a}=\sum_n n(n-4){\frac {\Phi_n}{a^n}}$.
For $n\in [0,4]$, i.e., by including all kinds of normal
matter, this expression yields the first inequality
in the relations (\ref{NN}).

For the case of {\em ultra relativistic matter} ($n=4$),
when $T\equiv 0$ and $N_{+}\equiv 0$, the function
$\eta$ is a Lyapunov function for the normal system of
ordinary differential equations (\ref{Norm}), and
gives a possibility of analyzing the qualitative behavior
of its solutions in the phase space. This property is of
invaluable importance for the analysis of the evolution of
the very early hot 4D-DG Universe, when all matter was
in an ultra-relativistic state.

The second limiting case 
-- in which $N_{+}\equiv 0$, and 
$\eta$ is a Lyapunov function in generalized
sense, is de Sitter Universe filled only with vacuum
energy ($n=0$).

In these two limiting cases (as well as for
$n \notin (0,4)$), the parameter $\eta(\tau)$
is a monotonically decreasing function of $\tau$.
For matter with $n\in (0,4)$, one may have a more
complicated dynamics of the function $\eta$.

\begin{figure}[htbp]
\vspace{5.5truecm}
\includegraphics{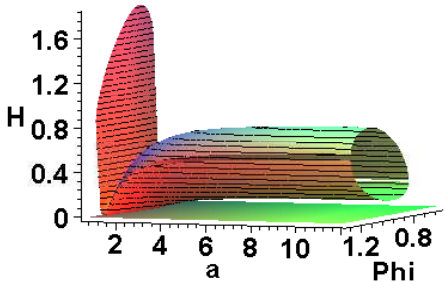}
\caption{\hskip 0.2truecm Level surface 
        of the Lyapunov function $\eta$ (\ref{eta}) 
        for the potential $V(\Phi)$ (\ref{Vsimplest}) in presence 
        of radiation and $k=+1$.
    \hskip 1truecm}
    \label{FigL1}
\end{figure}

A typical form of a level-surface of the function $\eta$
for the simple potential $V(\Phi)$ (\ref{Vsimplest})
in presence of radiation is shown in Fig.~\ref{FigL1}.
All solutions must cross this surface and go into its
interior.

One can see two main specific parts of this quite
complicated surface:

1) A high ``vertical'' part at small values of the RW scale factor
$a$. We shall see that the solutions that go into
the inner part of the $\{a,\Phi,H\}$-space through this part
of the surface describe the regime of inflation of the Universe.

2) An infinitely long  ``horizontal'' tube along the $a$-axes.
The solutions that enter the inner part of the 
$\{a,\Phi,H\}$-space through this part of surface
go to de Sitter asymptotic regime winding around the axis
of the tube and approaching it for $t \to \infty$.

The behavior of the solutions in these completely different
regimes, and the transition of given solution from inflation
to asymptotically de Sitter regime will be described in
Section VI.~C in more detail.

\vskip .5truecm

\paragraph{Second Normal Form of the Dynamical Equations\\}

We need one more normal form of the dynamical equations to study
the behavior of their solutions for $x_1,x_2\to \infty$.
The standard techniques for investigation of
the solutions in this limit \cite{Arnold} show that
the infinite point is a complex singular point,
and one has to use the so called $\sigma$-process
to split this singular point into elementary ones.
This dictates the following change
of variables:
\ben
x_1={\frac 3 {16}}p_{{}_\Phi}^{-2}z^{-4}g^{-4},\,\,\,
x_2=g^{-1},\,\,\,t={\frac 4 {\sqrt{3}}}\,p_{{}_\Phi}\Theta\,,
\la{x1x1zg}
\een
which transforms the equations (\ref{Norm})
with right hand sides (\ref{f}) into a new system:
\ben
g^\prime=g\,{\cal D}, \,\,\,
z^\prime=z{\cal Z}, \,\,\,
\Theta^\prime=z^2 g^2
\la{Norm2}
\een
where
\ben
{\cal D}(g,z,a;p_{{}_\Phi}^2,k) &=& 
1-{\frac 1 3}z^4 g^5 w(g) \nonumber \\
&-& \left(\!{\frac{4p_{{}_\Phi}}3}\!\right)^{\!2}\!\!g^3 z^4\!
\left(\!1\!+\!g^2\epsilon(a)\!-\!3g{\frac k {a^2}}\!\right)
\!     \nonumber \\
&=&{\cal D}_0(g,z)\!+\!{\cal O}_2(p_{{}_\Phi}^2)
\nonumber \\
{\cal Z}(g,z,a;p_{{}_\Phi}^2,k) &=&
{\frac 1 4}z^4g^7 v{,_g}(g)   \nonumber \\
&+& \left(\!{\frac {4p_{{}_\Phi}} 3}\!\right)^{\!2}\!g^4 z^4\!
\left(\!g\epsilon(a)\!-\!{\frac 3 2}{\frac k {a^2}}\!\right)
\!  \nonumber \\
&=&\!{\cal Z}_0(g,z)\!+\!{\cal O}_2(p_{{}_\Phi}^2) \,.
\la{D}
\een
Here we have introduced the functions
$v(g)\!=\!{\frac {16} 3}p_{{}_\Phi}^{2}\!V({\frac 1 g})$ and
$w(g)\!=\!{\frac {16} 3}p_{{}_\Phi}^{2}\!W({\frac 1 g})$
{\em which do not depend on the parameter} $p_{{}_\Phi}$ and
are simply related: $w=3/2 g^{-2}\int_1^g g^3 dv$.
The functions ${\cal D}$ and ${\cal Z}$
are defined according to formulae
\ben
{\cal D}= {\frac {d(\ln g)}{d(\ln a)}}\,,\quad
{\cal Z}= {\frac {d(\ln z)}{d(\ln a)}}\,.
\la{Ddef}
\een

The representation (\ref{D}) shows that one can
develop a simple perturbation theory for the highly
nonlinear system (\ref{Norm2}) using the extremely
small parameter $p_{{}_\Phi}^{2}$ as a
perturbation parameter.

It is interesting to emphasize that in the domain
$a\!\gg\!p_{{}_\Phi}^{2/{\bf n}}$
one can consider both the space-curvature term
${\frac k {a^2}}$ and the matter density $\epsilon(a)$
as small perturbations. 
Here ${\bf n}=\max\{2,n_{max}\}$ and $n_{max}$
is the largest degree in formula (\ref{epsilon_p}).
Thus, 4D-DG gives an immediate contribution to the solution
of the so-called {\em flatness problem} in cosmology.
Without further tuning of the model and for any
admissible cosmological potential, it becomes clear
that at the epoch with $a \sim 1$ the curvature term
can be neglected because of
the extremely small factor $p_{{}_\Phi}^{2}\alt 10^{-60}$
in (\ref{D}). At such epoch, the influence of
the matter on the dynamics of the gravi-dilaton sector
is negligible as well because of the same reason.
Only at very early stages of the evolution of the Universe
these terms may have had a significant impact on
the gravy-dilaton sector and on the space-time curvature,
according to the basic relation (\ref{R_relA}).
In a later epoch, the Universe will look spatially flat
if some space-curvature is not accumulated during
the Beginning.

From the dynamical equations (\ref{Norm}), (\ref{f})
and (\ref{Norm2}), one easily obtains
the following contour-integral representation
for the number of e-folds ${\cal N}$ and for the
elapsed time  $\Delta t_{\cal N}=
{\frac 4 {\sqrt{3}}}\,p_{{}_\Phi}\Theta_{\cal N}$:
\ben
{\cal N}(p_{{}_\Phi}^2,k)\!=\!
\int_{\tau_{in}}^{\tau_{fin}} \limits
H^2(\tau) d\tau\!=\!\!
\int_{{\cal C}_{in}^{fin}}\limits\!
{\frac {dg/g}{{\cal D}(g,z,a;p_{{}_\Phi}^2,k)}},
\la{N}
\een
and
\ben
\Theta_{\cal_N}(p_{{}_\Phi}^2,k)=
\int_{{\cal C}_{in}^{fin}}\limits\!
{\frac {z^2 g\,dg}{ {\cal D}(g,z,a;p_{{}_\Phi}^2,k)} }.
\la{TN}
\een
Here, a start from some initial (in) state of
the 3D-DG-RW Universe, followed by a motion on
a contour ${\cal C}_{in}^{fin}$ (determined
by corresponding solution of system (\ref{Norm2})), 
and an end at some final (fin) state, are assumed.

\subsection{General Properties of the Solutions
            in the 4D-DG RW Universe}

\subsubsection{Properties of the Solutions in a Vicinity of dSV}

Let us first consider the simplest case when
$k=0$ and $\epsilon=0$. The system (\ref{Norm})
in this case splits into a single equation for
$x_3$, which is solved by the {\em monotonic} function
$x_3(\tau)=x_3^0+\int_{\tau_0}^\tau x_1(\tau)d\tau$,
and the independent of $x_3$ system
\ben
{\frac d {d\tau}}{ x_1}=4 x_1\left(Z_1(x_2)-x_1\right),
\nonumber \\
{\frac d {d\tau}}{ x_2}=\,\,x_2\left( Z_2(x_2) -x_1\right).
\hskip 0.truecm
\la{kE0}
\een

Now it is clear that the curves
$\hat x_1= Z_1(\hat x_2)$ and
$\check x_1= Z_2(\check x_2)$ are the
zero-isoclinic lines for the solutions of (\ref{kE0}).
These curves describe the points of local extrema of the functions
$x_1(\tau)$ and $x_2(\tau)$, respectively,
in the domain $x_{1,2}>0$. Because of the condition
$U(\Phi)>0$  and the existence of a unique
minimum of the cosmological potential (see Propositions 1 and~2),
these lines have a unique intersection point
-- a dSV state with $\bar x_1=1/3$, $\bar x_2=1$.
This singular point represents the standard
de Sitter solution, which in usual variables reads
\ben
\bar H = 1/\sqrt{3},\,\,\,\bar \Phi=1,\,\,\,
a(t)=a_0 \exp(t/\sqrt{3}).
\la{dSS}
\een

\begin{figure}[htbp]
\vspace{5.5truecm}
\includegraphics{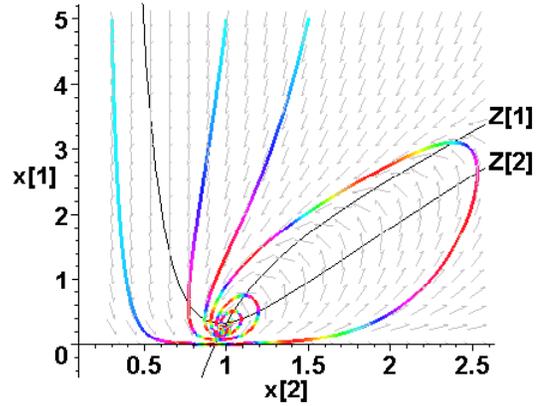}
\caption{\hskip 0.2truecm A typical phase portrait of the system (\ref{kE0}).
        The parts of solutions with the same color are covered
        by the Universe for equal $\tau$-intervals.
    \hskip 1truecm}
    \label{Fig1}
\end{figure}

One can see the typical behavior of the solutions
of the system (\ref{kE0}) in the domain $x_{1,2}>0$,
together with the curves
$Z[1] := \{\hat x_1= Z_1(\hat x_2)\}$,
and
$Z[2]:=\{\check x_1= Z_2(\check x_2)\}$,
in Fig.\ref{Fig1}, where the corresponding phase portrait
is shown for the case of the potentials (\ref{Vsimplest})
and $p_{{}_\Phi}=1/4$.

Consider the solutions of
the system (\ref{kE0}) of the form
$x_{1,2}(\tau)=\bar x_{1,2}+\delta x_{1,2}(\tau)$
that are close to dSV,
i.e., with $|\delta x_{1,2}(\tau)|\ll 1$.
Using the relations (\ref{dSVcommon}),
one obtains in linear approximation
\ben
\delta x_1(t)=
\delta x_1^0\,\exp\!\left({-{\frac {\sqrt{3}} 2}t}\right)
\cos\left(\omega_{{}_\Phi} t\right),
\hskip 1.4truecm \nonumber \\
\delta x_2(t)\!=\!{\frac 3 2}
{\frac { p_{{}_\Phi}\delta x_1^0}
{\sqrt{ p_{{}_\Phi}^2\!+\!{\frac 3 4}}}}
\exp\!\left(\!{-{\frac {\sqrt{3}} 2}t}\!\right)\!
\cos\left(\omega_{{}_\Phi}t\!-\!\psi_{{}_\Phi}\right)
\nonumber \\
\delta x_3(t)\!=\!\delta x_3^0\!+
\!\sqrt{3} p_{{}_\Phi}\delta x_1^0
\exp\!\left(\!{-{\frac {\sqrt{3}} 2}t}\!\right)\!\times
\hskip 1.2truecm \nonumber \\
\bigl(\cos\left(\omega_{{}_\Phi}t\!-\!\psi_a\right)-
\cos\left(\psi_a\right)
\bigr)\,,
\la{dxt}
\een
where
\ben
\omega_{{}_\Phi}=\sqrt{p_{{}_\Phi}^{-2}-3/4}\,,
\hskip 1.5truecm \nonumber \\
\tan\psi_{{}_\Phi}={\frac {2\sqrt{3}} 5}\,\omega_{{}_\Phi},
\,\,\,
\tan\psi_a=-{\frac {2\sqrt{3}} 3}\,\omega_{{}_\Phi},
\la{omega}
\een
$\delta x_1^0=\delta x_1(0)$ is the small
initial amplitude of $\delta x_1(t)$,
and the solution for the deviation $\delta x_3(t)$
has been added
for completeness and for later use.

The frequency $\omega_{{}_\Phi}$ is a real and
positive number if $p_{{}_\Phi} <2/\sqrt{3}$.
According to the estimate (\ref{pObs}), in 4D-DG we have
$\omega_{{}_\Phi} \geq 10^{30}$ in cosmological units,
or $\omega_{{}_\Phi} \geq  10^3\, {\rm GHz}$ in usual units,
and $\psi_{{}_\Phi}\approx \pi/2\approx -\psi_a$.

Thus, we see that dSV is a stable focus in
the phase portrait of the system (\ref{kE0}).
For $k=0$ and $\epsilon=0$, all solutions of
this system that lie in a small enough vicinity
of dSV oscillate with an ultra-high frequency
$\omega_{{}_\Phi}$ (\ref{omega}),
and approach dSV in the limit $t \to \infty$.

Now we generalize this statement for the
case of arbitrary $k=0,\pm1$ and $\epsilon\neq 0$.

\noindent
{\bf \em Proposition 3:}
{\em The de Sitter solution $(\ref{dSS})$ is an attractor
in the 4D-DG-RW Universe if for $a\to \infty$ we have
$\epsilon(a)\sim a^{-n}$ with $n >3/2$.
In this case, all solutions that are in a small
enough vicinity of the de Sitter solution tend to that solution in the
limit $t \to \infty$, oscillating with an ultra-high
frequency $(\ref{omega})$.
All solutions with an arbitrary $k=0,\pm1$, 
and an arbitrary $\epsilon$ have the asymptote 
$x_1(t)\sim {\frac 1 3}+\delta x_{1}(t)$,
$x_2(t)\sim 1 +\delta x_{2}(t)$, and
$x_3(t)\sim x_3^0+t/\sqrt{3} + \delta x_3(t)$,
with the same functions $\delta x_{1,2,3}(t)$ $(\ref{dxt})$.
In general, the constants $\delta x_{1,3}^0$
may depend on $k$ and $\epsilon$.}

This turns out to be possible even in
the presence of 3-space-curvature
and energy-density terms, because in the limit
$t \to \infty$ we have $x_3\to \infty$,
$ke^{-x_3}\to 0$ for all $k\!=\!0,\pm1$ and
$\epsilon\!\to\!0$ fast enough, according to formula
(\ref{epsilon_p}). One can find the proof of this
result in Appendix A.

We see that one important general prediction
for the 4D-DG-RW Universe is the existence of ultra-high
dilatonic oscillations with frequency $\omega_\Phi$ (\ref{omega})
in 3-spaces with any curvature and in the presence
of any kind of normal matter.

If $p_{{}_\Phi} \geq 2/\sqrt{3}$, i.e.,
if $m_{{}_\Phi}\leq 10 ^{-33}\, {\rm eV}$ 
(as in inflation models with a slow-rolling scalar field
and in quintessence models), 
the above ultra-high dilatonic oscillations do not exist
in the 4D-DG-RW Universe.
In this case, the frequency $\omega_{{}_\Phi}$ becomes imaginary
and, instead of a stable focus, we have an unstable saddle point in
the phase portrait of the system (\ref{kE0}).
Such a situation was considered first in
\cite{Starobinsky80} in a different model of nonlinear
gravity based on a quadratic with respect to scalar
curvature $R$ Lagrangian (\ref{L_NLG}),
but with some additional terms that originate from
quantum fluctuations in curved space-time \cite{QNLG}.
These additional terms vanish in the case of
RW metric but yield an essentially different theory
in other cases. Therefore, for RW Universe the model 
described in \cite{Starobinsky80} 
is equivalent to 4D-DG with the non-physical
quadratic cosmological potential (\ref{UV2}).

An immediate consequence of Proposition 3 is the existence of
ultra-high frequency oscillations of the effective
gravitational factor $G_{eff}= G_{{}_N}/\Phi$,
accompanied with an extremely slow exponential decrease
of its amplitude
$\sim \exp\left(-{\frac 1 2}\sqrt{3\Lambda^{obs}}\,ct\right)$
(in usual units). From Eq.~(\ref{Lambda}), one obtains
$\bar H^2/ H^2_0=\Omega_\Lambda$ and $\delta H^2_0 =
{\frac 1 3}{\frac {1- \Omega_\Lambda}{\Omega_\Lambda}}$.
Hence, at the present epoch with $\Omega_\Lambda \approx {\frac 2 3}$,
we have $\delta x_1^0\approx {\frac 1 6}$.
Then the second equation of the system (\ref{dxt}) gives
\ben
g(t) \approx 1 -
p_{{}_\Phi} {\frac {\sqrt{3}} 2}
\exp\!\left({-{\frac {\sqrt{3}} 2}t}\right)
\cos\left(\omega_{{}_\Phi}t-\psi_{{}_\Phi}\right)\,,
\la{Gvar}
\een
where we have introduced a dimensionless gravitational
factor, $g(t)={ {G_{eff}(t)}/{G_{{}_N}}}=1/\Phi(t)$.

Because of the extremely small amplitude
$\!p_{{}_\Phi}\!\leq\!10^{-30}$,
these variations are beyond the possibilities
of present-day experimental techniques
(see Section III.~C.).

In contrast, the oscillations of the Hubble parameter
$H$ have a relatively big amplitude
$\delta H_0=\sqrt{\delta x_0}\approx 0.4$, and the
same huge frequency $\omega_{{}_\Phi}$,
as the oscillations of gravitational factor.
It is very interesting to find possible
observational consequences of such phenomena.

High frequency oscillations of the effective gravitational
factor were considered first in the context
of Brans-Dicke field with BD parameter $\omega >1$
in \cite{Steinhardt}. These oscillations were induced
by an independent inflation field, but the analysis of the existing
astrophysical and cosmological limits on the oscillations of
$G_{eff}(t)$ is  applicable for our 4D-DG model as well.
The conclusion in  \cite{Steinhardt} is that
the oscillations in the considered frequency-amplitude range,
being proportional to $\dot g/(gH)$,
do not affect the Earth-surface laboratory measurements,
Solar System gravitational experiments,
stellar evolution, nucleosynthesis, but can produce
significant cosmological effects because the frequency
is too large and the Hubble parameter is small
(in usual units). It can be seen explicitly from Eq.~(\ref{dxt})
that this is precisely what happens in 4D-DG,
although in it the oscillations are self-induced.

As stressed in \cite{Steinhardt}, despite the fact
that the variations of the type (\ref{Gvar}) have
extremely small amplitudes, they can produce significant
cosmological effects because of the nonlinear
character of gravity. The 4D-DG version of
the corresponding formula -- analogous to the one in
the first of references \cite{Steinhardt} -- is
\ben
H={\frac 1 2}{\frac{\dot g} g }\pm
\sqrt{{\frac 1 4} \left({\frac {\dot g} g}\right)^2
+{\frac 1 3}g(U+\epsilon) -{\frac k{a^2}}}\,.
\la{H_nlin}
\een
Being a direct consequence of Eq.~(\ref{DERWU}),
this formula shows that, after averaging of
the oscillations, the term ${{\dot g}/g}$
has a non-vanishing contribution
because it enters the Hubble parameter
(\ref{H_nlin}) in a nonlinear manner.
A more detailed mathematical treatment of this new phenomenon
in 4D-DG is needed to derive reliable conclusions.
The standard averaging techniques for differential equations
with fast oscillating solutions and slowly developing modes
seem to be the most natural mathematical method for this purpose,
but the applications of these techniques
to 4D-DG lies beyond the scope of the present article.

%%%%%%%%%%%%%%%%%%%%%%%%%%%%%%%%%%%%%%%%%%%%%%%%%%%%%%%%%%%%
%
%  The text is checked up to here  -  March 26 2002, Austin
%
%%%%%%%%%%%%%%%%%%%%%%%%%%%%%%%%%%%%%%%%%%%%%%%%%%%%%%%%%%%%

\subsubsection{Inflation in 4D-DG-RW Universe}

Having in mind that: 1) the essence of inflation
is a fast and huge re-scaling of Universe, and
2) the dilaton is the scalar field responsible
for the scales in Universe, it seems natural to relate
these two fundamental physical notions instead of
inventing some specific ``inflation field''.
In this section we show that our 4D-DG model indeed
offers such a possibility.

\vskip .5truecm
\paragraph{ The Phase-Space Domain of Inflation\\}

\begin{figure}[htbp]
\vspace{6.truecm}
\includegraphics{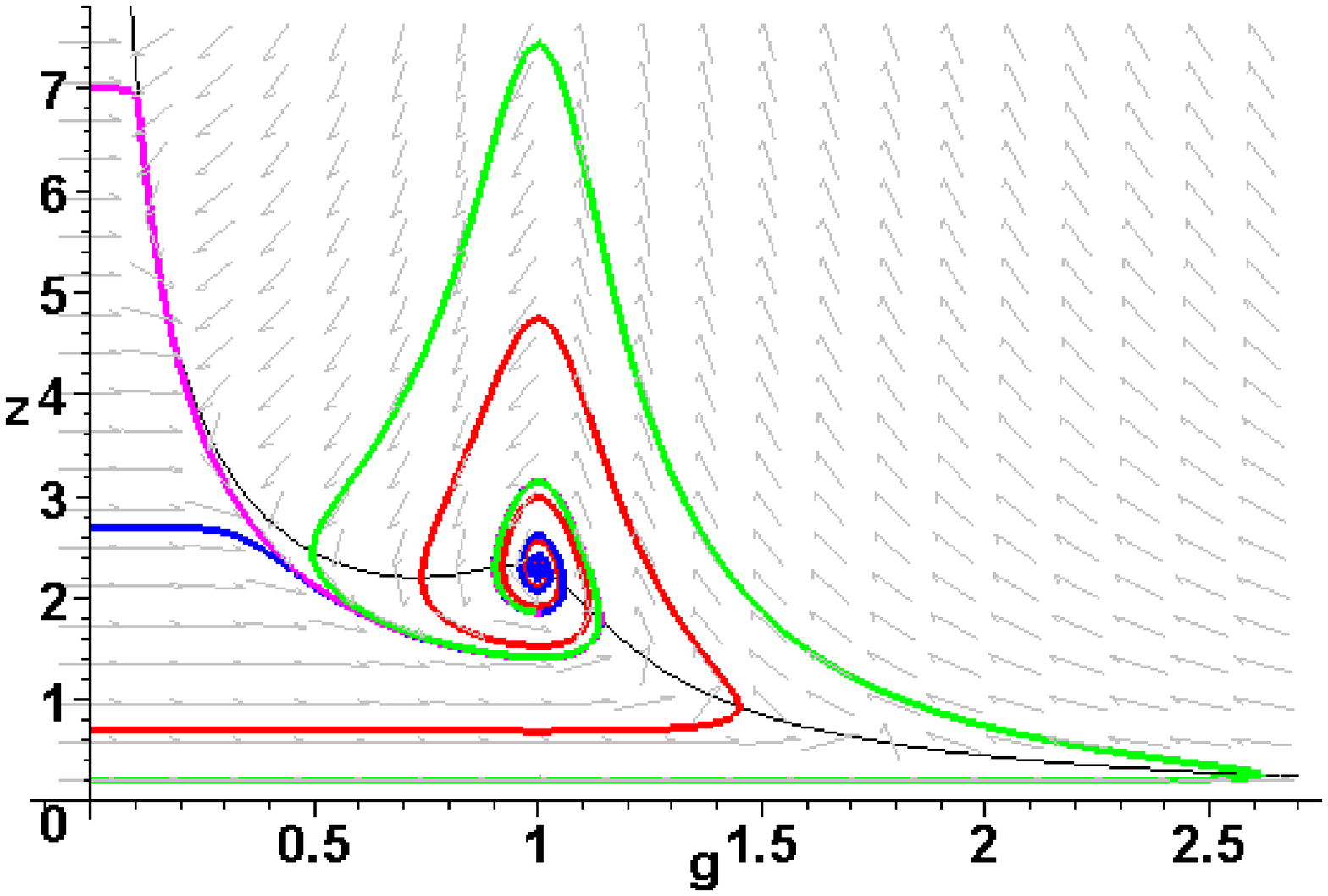}
\caption{\hskip 0.2truecm The phase portrait of Eq.~(\ref{zg}).
The black line shows the zero line Z[2] of the denominator~${\cal D}$.
    \hskip 1truecm}
    \label{Fig2}
\end{figure}

As seen from the phase portraits in Fig.~\ref{Fig1}--\ref{Fig2},
for values $H\geq H_{crit}$ and $\Phi\geq \Phi_{crit}$,
%i.e., above some critical values $\{H_{crit},\Phi_{crit}\}$,
the ultra-high-frequency-oscillations do not exist.
The evolution of the Universe in this domain of
phase space of the system (\ref{DERWU})
reduces to some kind of monotonic
expansion, according to the equation
${\frac d {d\tau}} x_3=H^2>0$.
We call this expansion an {\em inflation}.
As we shall see, it indeed has all needed
properties to be considered as an inflation
phenomenon \cite{inflation}.

The transition from inflation to high-frequency
oscillations is a nonlinear phenomenon, and
we will describe it in the present article very
approximately. Here our goal is to have some
approximate criteria for determining the end of
the inflation. It is needed for evaluation of the basic
quantities that describe the inflation.  

As seen from Eq.~(\ref{dxt}), the amplitude, $\delta \Phi_0$, 
of the oscillations of $\Phi$ is extremely small
compared with the amplitude $\delta H^2$
of the oscillations of $H^2$:\,\,
$\delta\Phi_0 \alt {\frac 3 2}p_{{}_\Phi}/
\sqrt{p_{{}_\Phi}^2+{\frac 3 4}}\,\delta H_0^2$.
An obvious crude estimate for the amplitude
$\delta H_0^2$ is $\delta H_0^2\leq 1/3\,(=\bar H^2)$.
Then, for $p_{{}_\Phi}\leq 2/\sqrt{3}$ (which is the
condition for existence of oscillations), we obtain
$\delta \Phi_0 \alt\delta H_0^2\alt 1/3$.
The last estimate is indeed very crude for
the physical model at hand, in
which $p_{{}_\Phi}\leq 10^{-30}$.
This consideration gives the constraint
$H_{crit}\alt\sqrt{2/3}$ and $\Phi_{crit}\alt 4/3$,
but, taking into account the extremely small value
of $p_{{}_\Phi}$, we will use for simplicity the
very crude estimate $H_{crit}, \Phi_{crit}\sim 1$.

Now it becomes clear that the study of the inflation
requires to consider big values of the variables $x_{1,2}$,
i.e., to use the second normal form (\ref{Norm2})
of the dynamical equations.

\vskip .5truecm

\paragraph{The Case  $k=0$, $\epsilon=0$\\}

Let us consider first the case $k=0$, $\epsilon=0$.
From Eq.~(\ref{Norm2}) one obtains the simple
first order equation
\ben
{\frac{dz}{dg}}={\frac {z^5 g^6}{4}}
{\frac{v_{,_g}(g)}{{\cal D}(g,z;p_{{}_\Phi}^2)}}
\la{zg}
\een
with 
\ben
{\cal D}(g,z;p_{{}_\Phi}^2)=1-{\frac 1 3}z^4g^5 u(g)=
\hskip 3.truecm\nonumber \\
1\!-\!{\frac 1 3}z^4 g^5 w(g)\!-\!
\left(\!{\frac{4p_{{}_\Phi}}3}\!\right)^{\!2}\!\!g^3 z^4\!=\!
{\cal D}_0(g,z)\!+\!{\cal O}_2(p_{{}_\Phi}^2),\,\,\,\,
\la{D00}
\een
where $u(g)={\frac{16}{3}}p_{{}_\Phi}^2 U(1/g)$.

Some basic properties of the solutions of Eq.~(\ref{zg}) are
derived in Appendix B.  
It turns out that, for all solutions in the case
$k=0$, $\epsilon=0$, there is a Beginning, defined as
a time instant $t=0$ at which RW scale factor vanishes, 
$a(0)=0$, as in GR. In a small vicinity of the Beginning, 
for the potentials (\ref{Vnu}), one obtains a different
behavior of the solutions, depending on the parameter
$\nu_{+}>0$ (see Appendix~B).

For $0<\nu_{+}< 6$, we have:
\ben
a(t)\sim\left({\frac t { p_{{}_\Phi}}}\right)^{1/2}\!\!
,\,\,\,g(t)={\frac 1 {z_0}}
\left({ \frac {\sqrt{3}} {4 p_{{}_\Phi}} } t\right)^{1/2}\!\!
+{\cal O}_{3/2}(t),\nonumber\\
z(t)\!=\!z_0\!-\!{\frac {4 z_0^{\nu_{+}-1}\left(\!{\frac{\sqrt{3}}
{4 p_{{}_\Phi}}}t\!\right)^{3\!-\!\nu_{+}/2}}
{3(\nu_{+}\!+\!\nu_{-})(6\!-\!\nu_{+})}}\!+\!
{\cal O}_{4\!-\!\nu_{+}/2}(t).\,\,\,
\la{nu<6}
\een
Since the behavior of the RW scale factor $a(t)$
for small $t$ is similar to its behavior in GR in
the presence of radiation, one can conclude that, if
$0<\nu_{+}< 6$,
at the Beginning the dilaton plays a role,
similar to the role of radiation.

For $6\leq\nu_{+}$, we obtain
\ben
a(t)\sim\left({\frac t { p_{{}_\Phi}}}\right)^\alpha\!\!,
\,\,\,g(t)=
\left({ \frac {\sqrt{3}} {4\alpha z_1^2 p_{{}_\Phi}} }t\right)^\alpha
\!\!+{\cal O}_{\alpha\!+\!1}(t)\!\!,\nonumber\\
z(t)={\rm const}\left(\!{\frac{t} {p_{{}_\Phi}}}\!\right)^{\gamma}\!\!+
{\cal O}_{\gamma\!+\!1}(t), \hskip 2truecm \nonumber \\
\alpha\!=\!{\frac 1 {\nu_{+}-1}}\!\in\!\left(0,{ 1/5}\right],\,\,
\gamma\!=\!{\frac{\nu_{+}-3}{2(\nu_{+}-1)}}
\!\in\!\left[3/10, 1/2\right).\,\,
\la{nu>6}
\een

For all potentials (\ref{Vnu}) in 4D-DG, we have zero
gravity at the Beginning, i.e., $g(0)=0$. This leads to
some sort of an initial power-law expansion, which
for $\nu_{+}\geq 6$ is stronger then in GR.  
The number of e-folds ${\cal N}(t) \to -\infty$
as $ t\to +0$ like $\ln t$, since ${\cal N}(t)=\ln a(t)$.
For fixed initial time instant $t_{in}>0$,
final time instant $t_{fin}>t_{in}$, and number ${\cal N}$,
one obtains for the duration of this expansion
$\Delta t_{infl}=t_{fin}(1-\alpha/{\cal N})=t_{in}({\cal N}/\alpha -1)$.
Hence, the smaller $\alpha$, the faster 
the initial (hyper)inflation.

The general conclusion is that the potentials (\ref{Vnu})
do not help to overcome the initial singularity problem
in the 4D-DG-RW Universe with $k=0$ and $\epsilon=0$.
However, we would like to emphasize that 4D-DG is not applicable
for times smaller then the Planck time
$t_{Pl}\sim 10^{-44}\,{\rm sec}$, because it is a low-energy
theory and ignores quantum corrections in (S)ST. But
one can expect 4D-DG to be valid after some initial
time instant, 
$t_{in}\sim t_{{}_\Phi}= \hbar/(m_{{}_\Phi}c^2)$.
If $m_{{}_\Phi}\ll M_{Pl}$, we will have
$t_{in}\gg t_{Pl}$ and our results for the case
under consideration may have a physical meaning,
leaving open the initial singularity  problem.

As seen from Eq.~(\ref{N}) and Eq.~(\ref{D}), one can represent 
the RW scale factor $a(t)$ in the form:
\ben
a(t)=g(t)\exp(\Delta {\cal N}(t)),
\la{a_dN}
\een
where the re-normalized number of e-folds: 
\ben
\Delta {\cal N}=
\int_{{\cal C}_{in}^{fin}}\limits\!
{\frac {dg} g}\, {\frac { 1-{\cal D}(g,z,a;p_{{}_\Phi}^2,k) }
{{\cal D}(g,z,a;p_{{}_\Phi}^2,k)}}
\la{dN}
\een
is a finite quantity  in the entire time 
interval $t\in [0,\infty)$. Obviously, 
$\Delta {\cal N}(t^{(i)})\equiv  {\cal N}(t^{(i)})$
at the special time instants $t^{(i)}$, $i=0,1,...$, 
when $g(t)$ reaches its dSV value, i.e., 
when $g(t^{(i)})=1$, for example, in the limit 
$t\to\infty$.

\begin{figure}[htbp]
\vspace{5.5truecm}
\includegraphics{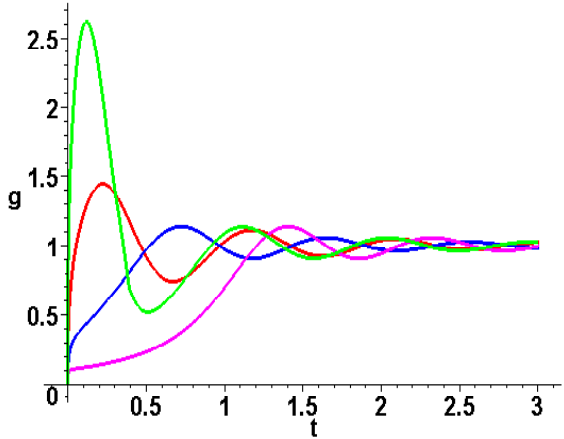}
\caption{\hskip 0.2truecm The dependence of dimensionless gravitational
            factor g on cosmic time t for solutions of Eq.~(\ref{zg}).
    \hskip 1truecm}
    \label{Fig3}
\end{figure}

The phase portrait of Eq.~(\ref{zg}) and the time-dependence
of the dimensionless gravitational factor $g(t)$
for the potentials (\ref{Vsimplest}) are shown
in Fig.~\ref{Fig2} and Fig.~\ref{Fig3}, respectively.
From Fig.~\ref{Fig3} we see that one can define
analytically the time of duration of the 
{\em initial inflation} $\Delta t_{infl}^{(0)}$ as the time 
spent by Universe from the Beginning to the first time instant 
$t^{(0)}$, when the gravitational factor $g(t^{(0)})=1$.
In addition, we see in Fig.~\ref{Fig3} that
this time interval is finite and has different
values for different solutions of Eq.~(\ref{zg}).

In Fig.~\ref{Fig1} and Fig.~\ref{Fig2} one can see
a new specific feature of 4D-DG:
the solutions may enter {\em many times}
the phase-space domain of inflation and the 
function $g(t)$ oscillates around its dSV 
value $\bar g=1$ with a variable preriod.  
Between two successive maxima of $g(t)$, 
the squared logarithmic derivative $H^2$
of $a(t)$ has its own maxima, just at the already defined
time instants $t^{(i)}$.
In the vicinity of each maximum of  $H^2$, the function a(t) 
increases very fast -- like $\exp({\rm const} \times t^\aleph)$, 
with ${\rm const}>0$, and parameter $\aleph \geq 2$.  
Therefore, we call such inflation, which is much faster
than the usual exponential de Sitter inflation,
a {\em hyper-inflation}. 
Hence, in 4D-DG, we have some sort of successive
hyper-inflations in the  Universe. For simplicity, 
we define the time-duration $\Delta t_{infl}^{(i)}=t^{(i)}-t^{(i-1)}$,
$i=0,1,...$ (by definition $\Delta t_{infl}^{(0)}=t^{(0)}$) 
of each of these periods of hyper-inflation
as the time period between two successive dSV values
of the function $g(t)$, although the 
hyper-inflation itself takes place only around 
the maxima of the function $H^2(t)$. The corresponding
number of e-folds is ${\cal N}_{infl}^{(i)}$.
It is clear that ${\cal N}_{infl}^{(i)}$
is a decreasing function of the number~$i$.
The inflation can be considered
as cosmologically significant only if
for some short total time period
$\Delta t_{infl}^{total}=
\sum_{i=0}^{i_{max}}\Delta t_{infl}^{(i)}$, 
the total number of e-folds 
${\cal N}_{infl}^{total}=
\sum_{i=0}^{i_{max}}{\cal N}_{infl}^{(i)}$
exceeds some large enough number ${\cal N}$. It is
known that one needs to have ${\cal N} \agt 60$ to be able to 
explain the special-flatness problem, the horizon problem, 
and the large-scales-smoothness problem in cosmology~\cite{C}.

\begin{figure}[htbp]
\vspace{5.5truecm}
\includegraphics{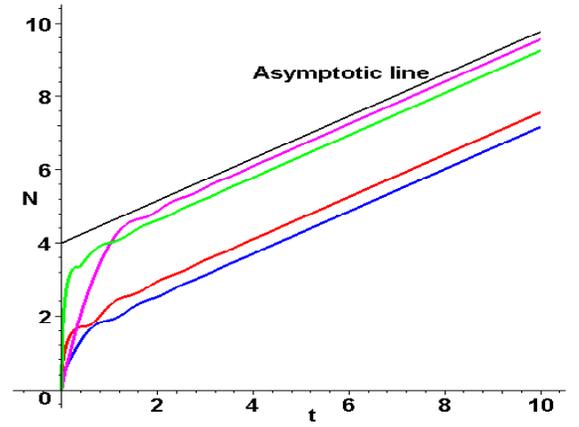}
\caption{\hskip 0.2truecm The dependence of the number of e-folds ${\cal N}$
                      on the cosmic time $t$ for solutions of Eq.~(\ref{zg}).
                      The straight black line gives an example of an 
                      asymptotic line for some solution.
    \hskip 1truecm}
    \label{Fig4}
\end{figure}

Fig.~\ref{Fig4} illustrates both the inflation and
the asymptotic behavior of the function
${\cal N}(t)$ for $t\to \infty$.
The small oscillations of this function are
``averaged'' by the crude graphical abilities
of the drawing device, and for large values of time $t$
we actually can see in Fig.~\ref{Fig4} only the
limiting de Sitter regime~(\ref{dSS}), 
when the averaged finction 
$\langle{\cal N}(t)\rangle \equiv \langle \Delta{\cal N}(t)\rangle$.
In this regime, for $t \to \infty$, we have
an obvious asymptotic of the form
\,$\langle{\cal N}(t)\rangle\sim\bar{\cal N}+\bar H t$
for the averaged with respect to dilatonic oscillation
(\ref{dxt}) function $\langle{\cal N}(t)\rangle$.
We accept the constant $\bar{\cal N}$ 
as our {\em final definition of total number
of e-folds during inflation}:
\ben
\bar{\cal N}=\lim_{t\to\infty}\left(\langle{\cal N}(t)\rangle
           -\bar H t\right)
            =\lim_{t\to\infty}\left(\langle \Delta{\cal N}(t)\rangle
           -\bar H t\right).\,\,
\la{barN}
\een
It is clear that this integral characteristic
describes  precisely the number of e-folds
{\em due to the true inflation} in 4D-DG, i.e.,
during the fast initial expansion of the Universe,
as a new physical phenomenon.
To obtain this quantity, one obviously must subtract
the asymptotic de Sitter expansion 
from the total function $\langle{\cal N}(t)\rangle$. 
According to Proposition 3, one can apply the same
definition for solutions in the general case of
arbitrary values of the space-curvature parameter $k$
and physically admissible energy densities
$\epsilon(a)$ since they have the same asymptotic.
Hence, in all cases the solutions with a cosmologically
significant inflation must have values 
$\bar{\cal N} \agt 60$.

As we see in Fig.~\ref{Fig4},
the total number of e-folds $\bar{\cal N}$
decreases, starting from the 'green''
solution (with $z_0= 0.2$) 
to the ``red'' one (with $z_0= 0.7$)
and to the ``blue'' one (with $z_0= 2.7$), 
reaches a minimum ($\approx 1$) 
for some initial value $z_0^*$, and then
increases for the ``magenta'' solution (with $z_0=7$), 
and for solutions with larger values of $z_0$.
Thus, we see that inflation is a typical behavior
for all solutions of Eq.~(\ref{zg}), 
and that most of them have large values of $\bar{\cal N}$.
(In Fig.~\ref{Fig4}, we show only solutions with
small values of $\bar{\cal N}$ that are close to its minima.)

The value of the constant $\bar{\cal N}$  depends on
the closeness of the given solution to
the zero curve $Z[2]$ of the denominator 
of the integrand in the right-hand side of~(\ref{TN}). 
The equation ${\cal D}=0$ can be explicitly
solved. Its solution reads
$z=\left({\frac 1 3}g^5u(g)\right)^{-1/4}$.
The corresponding curve is shown both in
Fig.~\ref{Fig1} and Fig.~\ref{Fig2}.

The value of the quantity ${\cal N}(t)$ increases
essentially each time when the solution
approaches this curve. 
The corresponding increment of ${\cal N}(t)$ remains finite, 
even if the solution crosses this curve,  
although in this case the denominator in the integrals (\ref{N}), 
or (\ref{dN}), 
reaches a zero value (see Appendix~B).  
Hence, an essential increase in ${\cal N}(t)$
is accumulated when the solution
becomes close and parallel to the curve $Z[2]$.
This is possible both for big values of $g$ and small
values of $z$ (the ``green'' solution), or for
small values of $g$ and big values of $z$
(the ``magenta'' solution).
All this results in a big value of
$\bar{\cal N}$. Only a small fraction of the solutions
(like the ``blue'' one and ``red'' one)
stay all the time relatively far from the line
${\cal D}=0$, and therefore acquire relatively
small number of e-folds, $1\lesssim\bar{\cal N}\lesssim 2$.

This qualitative consideration,
combined with the structure of the phase portrait 
shown in Figures \ref{Fig1} and \ref{Fig2}, 
not only explains the universal character of the 
inflation in 4D-DG, but gives us a better understanding 
of its basic characteristics.  
Hence, we do not need fine tuning of the model
to describe the inflation as a typical physical phenomenon.

Using Eqs.\ (\ref{N}), (\ref{TN}), and (\ref{D00}), we obtain
\ben
{\cal N}_{infl}^{(i)}(p_{{}_\Phi}^2)={\cal N}_{infl}^{(i)}(0)
+{\cal O}_2(p_{{}_\Phi}),
\hskip 0.2truecm \nonumber \\
t_{infl}^{(i)}(p_{{}_\Phi}^2)=
{\frac {4} {\sqrt{3}}}p_{{}_\Phi}\Theta_{infl}^{(i)}(0)
+{\cal O}_3(p_{{}_\Phi}),\nonumber
\een
where 
\ben
{\cal N}_{infl}^{(i)}(0)=\int_{{\cal C}_{in}^{fin}(i)}\limits
{\frac {dg/g}{{\cal D}(g,z;0)}},\,\,\,\hbox{and}
\nonumber \\
\Theta_{infl}^{(i)}(0)=\int_{{\cal C}_{in}^{fin}(i)}\limits
{\frac {z^2 g\,dg}{ {\cal D}(g,z;0)} }, \hskip .15truecm
\la{Gamma}
\een
are independent of the parameter $p_{{}_\Phi}$.
Taking into account the extremely small physical value
of this parameter,
one can conclude that the higher order terms in
${\cal N}_{infl}^{(i)}(p_{{}_\Phi}^2)$ and
$\Delta t_{infl}^{(i)}(p_{{}_\Phi}^2)$
are not essential. Neglecting them, we
actually ignore the contribution of the term
$\Phi/3$ in the functions $Z_{1,2}$ (\ref{Z1Z2}), or the
corresponding term $\Phi^2$, in the cosmological potentials $U(\Phi)$
(\ref{4WV}), (\ref{Vsimplest}), (\ref{Vnu}).
It is natural to ignore these terms
in the domain of inflation, because
they are essential only in a small vicinity of dSV.
In the function $W(\Phi)$, we have a term 
that dominates for $\Phi-1 \gg p_{{}_\Phi}^2$,
having a huge coefficient $\sim p_{{}_\Phi}^{-2}$.
Physically this approximation means that we are neglecting 
the small pure cosmological constant term in the cosmological
potential and preserve only the terms, which are proportional to 
the mass of dilaton.

The relation
$\Delta t_{infl}^{(i)}(p_{{}_\Phi}^2)\sim
{\frac 4 {\sqrt{3}}}p_{{}_\Phi} \Theta_{infl}^{(i)}$, 
written in physical units, reads 
\ben
E_{{}_\Phi} \Delta t_{infl}^{(i)}\sim
\hbar {\frac 4 {\sqrt{3}}}\Theta_{\cal N}^{(i)}(0).
\la{mT}
\een
It resembles some kind of a quantum
``uncertainty relation'' for the rest energy
$E_{{}_\Phi}$ of the dilaton and the time of inflation
and maybe indicates the quantum character
of the inflation as a physical phenomenon.

More important for us is the fact that Eq.~(\ref{mT})
shows the relationship between the mass of dilaton, 
$m_{{}_\Phi}$, and the time duration of inflation.
Having large enough mass of dilaton, we will have small
enough time duration of inflation.
This recovers the real meaning of the mass $m_{{}_\Phi}$
as a physical parameter in 4D-DG, and gives possibility
to determine it from astrophysical observations as
a basic cosmological parameter.

\vskip .5truecm

\paragraph{The Case $k\neq 0$ and $\epsilon\neq 0$\\ }

The nonzero space-curvature term with $k=\pm 1$, and the
presence of matter with $\epsilon(a)\neq 0$ of the form
(\ref{epsilon_p}) change drastically the behavior of
the solution for small values of $g$ and~$a$. They
yield a multitude of new possibilities. Indeed,
if one assumes the physically natural 
value $n_{max}=4$ in Eq.~(\ref{epsilon_p}),
the system of differential equations
(\ref{Norm2}) has to be rewritten in two new forms.
The first one, for the case $0<\nu_{+}<6$, is 
\ben
{\frac {dg}{d\sigma_1}}\!=\!g\!\left(\!
{\cal D}_{\nu_{+}\nu_{-}}^{(0,0)}(g,z)a^4\!-\!
\left(\!{\frac {4p_{{}_\Phi}} 3 }\!\right)^{\!2}\!\!g^4 z^4
\!\Bigl(ga^4\epsilon\!-\!3ka^2\!\Bigr)\!\right),
\nonumber \\
{\frac {dz}{d\sigma_1}}\!=\!z\!\left(\!
{\cal Z}_{\nu_{+}\nu_{-}}^{(0,0)}(g,z)a^4\!+\!
\left(\!{\frac {4p_{{}_\Phi}}3}\right)^{\!2}\!\!g^4 z^4
\!\Bigl(\!ga^4\epsilon\!-\!{\frac 3 2}ka^2\!\Bigr)\!\right),
\nonumber \\
{\frac {da}{d\sigma_1}}\!=\!a^5.\hskip 6.6truecm
\la{Norm3}
\een
If  $6<\nu_{+}$, one has to introduce, 
instead of regularizing parameter $\sigma_1$, 
another one, $\sigma_2$, by 
$d\sigma_1=g^{\nu_{+}-6}d\sigma_2$, and to rewrite
the system (\ref{Norm3}) in a similar form by multiplying
the right hand sides by $g^{\nu_{+}-6}$.
Here ${\cal D}_{\nu_{+}\nu_{-}}^{(0,0)}(g,z)$
and ${\cal Z}_{\nu_{+}\nu_{-}}^{(0,0)}(g,z)$
are polynomials in the variables $g$ and $z$,
described in Appendix~B, and under our assumption that 
$n_{max}=4$, the expression $a^4\epsilon(a)$
is a polynomial in $a$ of degree less than $4$.
The canonical polynomial form of the right-hand sides of
Eqs.~(\ref{Norm3}), and in the analogous system in the second
case, makes transparent the fact that the
point $a=0$, $g=0$, $z=0$ is a complex singular
point which has a rich structure -- 
different for $0<\nu_{+}<6$ and for $6<\nu_{+}$.

For different initial conditions and different
cosmological potentials, one can 
observe numerically different types of solutions: 
bouncing ones, oscillating ones, and solutions that are 
similar to the one we have discussed in 
the case $k=\epsilon=0$. They have inflation 
regime and de Sitter asymptotic for $t\to \infty$ 
(in accordance with Proposition 3).

Hence, in the presence of space-curvature
and matter, a novel approach to the initial singularity
problem is possible in 4D-DG.
The systematic study of properties
of solutions of both systems of type (\ref{Norm3})
which arise for potentials (\ref{Vnu}) is a complicated
mathematical issue and deserves independent investigation.

\vskip .5truecm

\paragraph{Static and Turning Points in the Case $k=+1$\\ }

In the case $k=+1$, there may exist turning points
in the evolution of $a(t)$ with $\dot a=0$ 
and $\ddot a=0$, some of them being unstable
static solutions, $a(t)\equiv {\rm const}$,
similar to the original Einstein static solution
in GR with $\Lambda>0$.
In the $(\Phi,a)$-plane, these points lie on the line
\ben
\epsilon_{pot}=U(\Phi)+\epsilon(a)- {{3 \Phi}\over {a^2}}=0,
\la{Pot0}
\een
which may have a complicated structure 
depending on $U(\Phi)$ and $\epsilon(a)$.
The analytical form of the solutions and the form of
the corresponding level surface of the Lyapunov function
$\eta$ in a vicinity of this line are interesting from
mathematical point of view, but we will not describe
them in the present article.
In the cases $k=-1$ and $k=0$, such phenomena
are not possible in 4D-DG.

\subsection{The Inverse Cosmological Problem in 4D-DG}

Instead of representing the system (\ref{NDE})
in the normal form (\ref{Norm}),
we can exclude the cosmological
potential $U(\Phi)$, and arrive at a {\em linear}
differential equation of second order for the
function $\Phi(x_3)$, 
\ben
\Phi^{\prime\prime}\!+\!
\left({\frac {H^{\prime}} H}\!-\!1\right)\Phi^{\prime}
\!+\!2\left({\frac {H^{\prime}} H}\!-
{\frac k {H^2}}e^{-2x_3}\right)\Phi
\!=\!{\frac {\epsilon^{\prime}} {3H^2}},\,\,\,\, 
\la{ODEPhi}
\een
equivalent to Eq.~(\ref{EqPhi}) because
of the relation $\epsilon^\prime+3(\epsilon+p)=0$
\footnote{This derivation actually proves once again 
that Eq.~(\ref{EqPhi}) follows from Eq.~(\ref{DERWU})
and the energy-momentum conservation law for matter.}.

In terms of the function
$\psi(a) = \sqrt{|H(a)|/a}\,\Phi(a)$ the equation 
(\ref{ODEPhi}) reads 
\ben
\psi^{\prime\prime} + n^2 \psi = \delta,
\la{DEPsi}
\een
where we have introduced the new functions \footnote{For
the coefficient $n$, one can obtain the additional
representation $-n^2 = 3+{\frac{2 k}{a^2 H^2}}+
{\frac S {2H^2}}+ {\frac 5 2}q $, 
where $-q=\ddot a/(aH^2)=1+H^\prime/ H$
is a decelerating parameter, and 
$S\bigl(a(t)\bigr)=\dddot a/\dot a-
{\frac 3 2}\bigl(\ddot a /\dot a\bigr)^2$
is the Schwarzian derivative of the scale factor $a(t)$.
}
\ben
-n^2 = {\frac 1 2}{\frac {H^{\prime\prime}} H}-
{\frac 1 4}\left({\frac {H^{\prime}} H}\right)^2\!-
{\frac 5 2}{\frac {H^{\prime}} H}+{\frac 1 4}
+{\frac {2 k}{H^2}} e^{-2x_3},\nonumber \\
\delta={\frac {\epsilon^\prime\,e^{-x_3/2}}{3\,|H|^{3/2}}}.
\hskip 5.05truecm
\la{n_delta}
\een

The Schr\"odinger-like equation (\ref{DEPsi}) for
$\psi(x_3)$ can be analyzed with some well known 
mathematical tools. 
For example, the condition $n(x_3)<0$ ensures the absence of
a new type of oscillations of the field
$\Phi(x_3)=\sqrt{1/|H(x_3)|}e^{x_3/2}\,\psi(x_3)$
in domains where $H(x_3)$ is a monotonic function.
In the opposite case, $n(x_3)\!>\!0$, 
we do have such oscillations -- only of the dilaton field
$\Phi$ as a function of scale parameter $a=e^{x_3}$.
This type of dilaton oscillations is different
from the one described in Section VI.B.1. 

Now we are ready to consider the inverse cosmological problem:

{\em Find the cosmological potential
$U(\Phi)$ and the dilatonic potential $V(\Phi)$ that yield
a given evolution of the Universe,
determined by the function $a(t)$.}

It is remarkable that in 4D-DG the following result
take place:

\noindent
{\bf \em Proposition 4:}
{\em For any three times differentiable function
$a(t)$ in 4D-DG, there exist a two-parameter family of local
solutions of the inverse cosmological problem.}

Indeed, given $a(t)$, we can construct
the function $H(x_3)$ and find the general solution
$\Phi(x_3;\tilde x_3,\tilde\Phi,\tilde\Phi^\prime)$
of the linear second order differential equation
(\ref{ODEPhi}) in the following Cauchy form:
\ben
\Phi(x_3;\tilde x_3\tilde\Phi,\tilde\Phi^\prime)\!=\!
C_1\Phi_1(x_3)\!+\!C_2\Phi_2(x_3)\!+\!
\Phi_\epsilon(x_3).
\la{sol}
\een
Here
$$
C_1=(\tilde\Phi_2^\prime\tilde\Phi-
\tilde\Phi_2\tilde\Phi^\prime)/\tilde\Delta,\,\,\,\,
C_2=(\tilde\Phi_1\tilde\Phi^\prime-
\tilde\Phi_1^\prime\tilde\Phi)/\tilde\Delta,
$$
the functions $\Phi_1(x_3)$ and $\Phi_2(x_3)$ constitute 
a fundamental system of solutions of the homogeneous 
equation associated with the non-homogeneous
one, (\ref{ODEPhi}), 
(nontrivial examples for such solutions
can be found in \cite{F00}),
$$
\Delta(x_3):=
\Phi_1\Phi_2^\prime - \Phi_2\Phi_1^\prime=
(\tilde\Delta \tilde H)e^{x_3-\tilde x_3}/H(x_3)\neq 0,
$$
and the term
$$
\Phi_\epsilon\!=
\!{\frac {1}{3\tilde H\tilde\Delta}}\!\!\left(\!
\Phi_2\!\int_{\tilde x_3}^{x_3}\limits
\!\Phi_1{\frac {e^{-(x_3-\tilde x_3)}} H}d\epsilon-
\Phi_1\!\int_{\tilde x_3}^{x_3}\limits
\!\Phi_2{\frac {e^{-(x_3-\tilde x_3)}} H}d\epsilon\!\right).
$$
describes the contribution of matter.
The point $\tilde x_3$ lies in the admissible
domain,  and $\tilde \Phi$ and $\tilde \Phi^\prime$
are some initial values at $\tilde x_3$.
The tilde sign shows that the corresponding quantities
are calculated at the initial point $\tilde x_3$.

Then, the cosmological potential and the dilatonic
potential, as functions of $x_3$,
are determined by equations
\ben
U(x_3;\tilde \Phi,\tilde\Phi^\prime)\!=
\hskip 6.3truecm\nonumber \\
3\Phi(x_3;\tilde \Phi,\tilde\Phi^\prime)
\left(H^2\!+\!ke^{-2x_3}\right)\!+\!
3 H^2\Phi^\prime(x_3;\tilde \Phi,\tilde\Phi^\prime)\!-
\!\epsilon,
\hskip .truecm\nonumber \\
V(x_3;\tilde \Phi,\tilde\Phi^\prime)\!=
{\frac 2 3}\Phi(x_3;\tilde \Phi,\tilde\Phi^\prime)
U(x_3;\tilde \Phi,\tilde\Phi^\prime)\!-
\hskip 1.5truecm\nonumber\\
\!2 \int d\Phi(x_3;\tilde \Phi,\tilde\Phi^\prime)
U(x_3;\tilde \Phi,\tilde\Phi^\prime),
\hskip 1.2truecm
\la{UVx3}
\een
which define the functions
$U(\Phi;\tilde \Phi,\tilde\Phi^\prime)$
and $V(\Phi;\tilde \Phi,\tilde\Phi^\prime)$
implicitly, i.e., via the inverse function
$x_3(\Phi;\tilde \Phi,\tilde\Phi^\prime)$
of the function
$\Phi(x_3;\tilde \Phi,\tilde\Phi^\prime)$.

Hence, one can construct a two-parameter family
of cosmological and dilaton potentials
for a given scalar factor $a(t)$.
This way, we see that in the 4D-DG-RW model of Universe
one can find potentials for which there exist
solutions {\em without initial singularities}:
$a(t_{in})=0$ (which are typical for GR), with
{\em any desired kind of inflation}, or with
other needed properties of RW scale factor $a(t)$.
But, in general, one cannot guarantee
all necessary properties of these potentials,
like the existence of only one minimum, or even 
the single-valuedness of the functions (\ref{UVx3}).  
The sufficient conditions on the function $a(t)$ 
that which guarantee such properties of the potentials
$U$ and $V$ are not known at present.

Here we shall make one more step -- to consider 
the choice of the point $\tilde x_3$, and of 
initial values $\tilde \Phi$ and $\tilde\Phi^\prime$.
These have to reflect some known basic properties of
4D-DG. So far, the only established general properties
of this type are the corresponding values for the unique 
de Sitter vacuum (\ref{dSVcommon}) 
(see Section IV.C.3).

\noindent
{\bf \em Proposition 5:}
{\em In the unique dSV-state of Universe:}
\ben
\bar x_3=\infty,\hskip .6truecm\nonumber \\
\bar\Phi=1,\,\,\,\bar\Phi^\prime=0,\nonumber \\
\bar H=1/\sqrt{3},\,\,\,\bar H^\prime = 0, \nonumber \\
\hbox{and}\,\,\,\bar\Delta e^{-\bar x_3}=1.
\la{dSVvalues}
\een
{\em The corresponding quantities  approach these values
in the limit $x_3 \to \infty$.}

The proof is based on the following observations:

1) For dSV-state of Universe we have $\epsilon=0$,
according to definition (\ref{dSV}).
Then, from Eq.~(\ref{epsilon_p}), we see that this
unique state corresponds to $\bar x_3=\infty$.
Hence, other quantities for dSV must be considered in
the limit $x_3 \to \infty$.
The second of Eq.~(\ref{epsilon_p}) shows that $\bar p =0$, 
and, together with the conservation of energy-momentum
of matter, it leads to the relation $\bar\epsilon^\prime=0$.
(Actually, one can weaken the conditions on the behavior
of these quantities simply by requiring that $\bar\epsilon=0$ and
$\bar\epsilon^\prime=0$. This is enough for our
purposes in this section.)

2) Then, from the above equations, we obtain the
values described in Proposition 5:

i) from Eq.~(\ref{Unorm}) -- $\bar \Phi=1$;

ii) from Eq.~(\ref{dSS}) -- $\bar H = 1/\sqrt{3}$;

iii) from Proposition 3 and Eq.~(\ref{dxt}) --
$\bar \Phi^\prime=\bar H^\prime=0$;

iv) from Eq.~(\ref{ODEPhi}), which reduces in this
limit to $\Phi^{\prime\prime}-\Phi^\prime=0$, 
we see that there exist two fundamental solutions, 
$\Phi_{1,\infty}(x_3)\sim 1$ and
$\Phi_{2,\infty}(x_3)\sim e^{x_3}(=a)$. 
Hence, for them
$\Delta(x_3)\sim e^{x_3}$ and this proves the last
relation in Eq.~(\ref{dSVvalues}).

As a result of Proposition 5, we obtain the following
final form of solution (\ref{sol}):
\ben
\Phi_\infty(x_3)=\Phi_{1,\infty}(x_3)+\Phi_{\epsilon,\infty}(x_3),
\la{sol_infty}
\een
where $\Phi_{1,\infty}(x_3)$ 
is the fundamental solution of Eq.~(\ref{ODEPhi})
that has the asymptotic behavior $\Phi_{1,\infty}(x_3)\sim 1$ 
as $x_3 \to \infty$. Then
\ben
\Phi_{\epsilon,\infty}\!=
\!{\frac {1}{\sqrt{3}}}\!\!\left(\!
\Phi_1\!\int_{ x_3}^{\infty}\limits
\!\Phi_2{\frac {e^{-x_3}} {H}}d\epsilon-
\Phi_2\!\int_{ x_3}^{\infty}\limits
\!\Phi_1{\frac {e^{-x_3}} {H}}d\epsilon\!\right).
\la{PhiEinfty}
\een

If we wish to reconstruct the cosmological potential
$U(\Phi)$ and the dilaton potential $V(\Phi)$ according
to relations (\ref{UVx3}), Eq.~(\ref{PhiEinfty})
requires to know the value of the function $H(x_3)$
{\em in the future}! Of course, if we know the exact
dependence of Hubble parameter on $x_3$ in the past, 
we can obtain its values in the future using  standard mathematical 
techniques for analytical continuation.  It is interesting to
reconsider the available observational data from this
point of view, taking into account the theoretical restrictions
(\ref{dSVvalues}), but this problem lies beyond
the scope of the present article.

\section{A Novel  Adjustment Mechanism
for the Cosmological Constant Problem}

\subsection{Value of $\Lambda^{obs}$
           and Number of Degrees of Freedom
           in the Observable Universe}

Concerning the observed small positive value
of the cosmological constant (\ref{Lambda}),
which seems mysterious from point of view
of quantum field theory, we see that the real problem 
one has to solve is to explain the extremely small
value $P\approx 10^{-61}$ of Planck number.
This number appears not only in the formula (\ref{P})
for the cosmological constant,
but also in the formula ({\ref{UnitA}) for the unit of action
and in expression (\ref{TFA_GD}) for the action 
of the gravy-dilaton sector. 
This observation yields a new possible direction for investigations: 
one can try to transform the cosmological constant problem 
to the problem of explaination of the value 
of the total action in the  Universe. 
It seems natural to think that the huge ratio ${\cal A} / \hbar
\sim 10^{122}$ is produced during the long evolution of
the Universe in the time interval
$\Delta t_U \sim 4\times 10^{17} {\rm sec}$.
Then the new question is,
how many degrees of freedom do we have
{\em in the observable
Universe}, and what is the amount of action 
accumulated in them since the Beginning.

To answer this question in 4D-DG, we need estimates
for the amount of matter action and the action
in the gravi-dilaton sector.

1. Amount of action in the matter sector:

We can obtain a crude estimate for the
amount of matter action in the observable Universe
by using the following purely qualitative analysis \cite{PF003}.

First we consider the simplest model of the Universe
built only of Bohr hydrogen atoms in ground state,
i.e., we describe the whole content of the Universe by using
such {\em effective Bohr hydrogen} (EBH) atoms.
Then for the  time of the existence of Universe, 
$\Delta t_U \sim 4\times 10^{17} {\rm sec}$, one
EBH atom with Bohr angular velocity $\omega_B = m_e
e^4\hbar^{-3}\sim 4\times 10^{16} {\rm sec}^{-1}$
accumulates classical action
${\cal A}_{EBH}= 3/2 \,\omega_B T_U\, \hbar \sim 2.4\times
10^{34} \,\hbar$.
Hence, the number of EBH in Universe that is needed to
explain the  value of the present-epoch action 
${\cal A}\sim {\cal A}_c$, must be 
$N_{EBH}\sim 5\times 10^{87}$.
This seems to be quite a reasonable number,
taking into account that the number of barions in
observable Universe is $N_{barions}^{obs}\sim 10^{78}$.

Thus, we see that within our ``action approach'', 
when applied to the observed Universe, the discrepancy
between the above primitive model and observations
is only about $10^9$ times. One has to compare this very crude 
estimate with the corresponding one in quantum field 
``derivation'' of cosmological constant that differs
from observation $10^{54}$ -- $10^{122}$ times.
We have at our disposal some $9$ powers of 10 
to solve the problem by taking into account the contribution
of all other constituents of matter and radiation
(quarks, leptons, gamma quanta, etc.), 
and the contribution of the gravi-dilaton sector
during the evolution of the Universe from
the Beginning to the present epoch.

Neglecting the temperature evolution of the Universe
during the hot Big Bang phase, we obtain
an accumulated action
${\cal A}_\gamma \sim
\omega_\gamma T_U \, \hbar \sim 10^{30}\, \hbar$
for one $\gamma$-quantum of CMB
(which is the most significant part of
present days radiation in Universe).

A simple estimate for the Bohr-like angular
velocity of the constituent quarks in a proton is
$\omega_{B q}= m_e/m_q (r_B/r_p)^2
\omega_B \approx 10^7 \omega_B$
(where the mass of the constituent quark, $m_q \sim 5 {\rm MeV}$,
the standard Bohr radius $r_B$ 
and the known radius of proton $r_p\sim 
8 \times 10^{-13}{\rm cm} $ have been used). 
Then the action accumulated by the constituent
quarks in one proton during the evolution
of the Universe is
${\cal A}_{p}\sim \omega_{B q} T_U\, \hbar \sim 10^{42} \,\hbar$.
This gives an unexpectedly good estimate for
the number of effective protons (ep) in the Universe:
$N_{ep}\sim 10^{80}$.

We may use the two remaining orders of magnitude
to take into account the contribution of the other matter
constituents, of the dark matter (see \cite{Primack} and
the references therein), and of the temperature evolution
of Universe -- during the short-time initial hot phase, 
some additional action must be produced in the matter sector.

Using the same arguments in opposite direction 
(i.e., using them to obtain an estimate 
of the total action in the matter sector instead of
obtaining the number of particles in it), 
we can say that according to our crude analysis,
the amount of action ${\cal A}_{matt}$ 
accumulated in the matter sector
during the evolution of the observable Universe is
between $10^{-2}$ and $1$ 
(in cosmological units).

2) Amount of action in the gravi-dilaton sector:

A similar crude estimate of the amount of action in
the gravi-dilaton sector of 4D-DG, based on completely 
different reasoning, can be derived from formula
(\ref{TFA_GD}).
Unfortunately, at the moment we do not know the form of
the solutions in the 4D-DG-RW Universe during the short
inflation epoch in the presence of matter. But we know them 
during the infinite time interval of de Sitter regime --
see formulae (\ref{dxt}). Ignoring the inflation epoch 
and the small oscillations of the dilaton $\Phi$
during de Sitter regime (which will be averaged
in the integration in expression (\ref{TFA_GD})),
we can substitute the simple de Sitter solution (\ref{dSS}) 
in the integral (\ref{TFA_GD}). 
This gives immediately the value $1$ of the integrand,
independently of the choice of a cosmological potential
$U(\Phi)$ (see (\ref{dSVcommon})).
Hence, for the 4D-DG gravi-dilaton action in
a unit 3-volume, we obtain in cosmological units
${\cal A}_{g,\Phi}\sim \Delta t_U\sim 1$,
independently of the choice of a cosmological potential.

If the inflation epoch gives a small additional
contribution to both ${\cal A}_{matt}$ and
${\cal A}_{g,\Phi}$, then 
${\cal A}_{tot}={\cal A}_{matt}+{\cal A}_{g,\Phi}\sim 1$
(in cosmological units), and we obtain 
a physical explanation of relation (\ref{UnitA})
which defines the value of Planck number in this approach. 
Thus we would have reached an
explanation of small value $\Lambda^{obs}=1$
(in cosmological units) in phenomenological frame
having explanation of the huge value of the present-epoch-action 
in Universe.

We see that, in the framework of 4D-DG, 
the answer to the question,
why the cosmological constant is so small in
the phenomenological frame,
might be:
Because the Universe is old and has a huge,
but limited, number of degrees of freedom in it.
The first part of this answer was proposed
in the quintessence models 
developed in the last decade \cite{Q}. 
As we saw, 4D-DG is physically
an essentially different model, but it leads
to the same conclusion.

In addition, the above consideration gives
us some idea how to explain the cosmological
coincidence problem when reformulated in terms
of action: the actions ${\cal A}_{matt}$
and ${\cal A}_{g,\Phi}$ are of the same order of
magnitude, at least in our crude approximation.
This result will still hold if the inflation epoch
gives comparable contributions to both of actions.

The main conclusion of the above crude arguments 
%based on classical mechanics and simplest application
%of basic semi-classical quantum relations 
is that within the framework
of 4D-DG the observed small positive value 
of the cosmological constant $\Lambda^{obs}$\,
(\ref{Lambda}) actually {\em restricts the number
of degrees of freedom} in the observable Universe.
Similar conclusion was reached recently in
\cite{Fischler} by using completely different arguments.
Thus, it is possible that a small positive $\Lambda^{obs}$ 
forbids the existence of a large number of more fundamental
levels of matter below the quark level.

It is obvious that these considerations need 
deeper investigation based on quantum field theory,
on the Standard Model, on the modern cosmological
results about the evolution of the Universe and the Big Bang.
To make them rigorous, one needs to know
the detailed description of the 4D-DG inflation in the 
presence of matter and space-curvature.
Another possible scenario is an inflation without matter after
dilaton-scale time $t_{{}_\Phi}=\hbar/m_{{}_\Phi}c^2$ 
(described in Section VI.C), accomplished with 
some mechanism of creation of matter by gravi-dilaton 
sector at the end, or after this stage of inflation.

\subsection{A Possible Novel Adjustment Mechanism for $\Lambda$}

Here we describe a possibility to solve
the cosmological constant problem by employing
a novel adjustment mechanism.
The basic idea is very simple:
one can have a huge cosmological constant in the
stringy frame, as a result of vacuum fluctuations
of different quantum fields introduced  by string theory.
In spite of this, the transition to phenomenological
frame  may reduce the value of this huge constant
to the observable one (\ref{Lambda}) 
because of the corresponding Weyl transformation (\ref{Weyl}).

Indeed, taking into account the relations
(\ref{S_FZL}), (\ref{Ephi}), (\ref{Tsigma}),
and (\ref{T_FL}) connecting the SF cosmological
potential and the PhF$\equiv$TF one, we obtain
\ben
\Lambda^{obs} U(\Phi)=
{}_{{}_S\!}\Lambda(\phi)\exp(D{}_{{}_T\!}\sigma_{\!{}_S}(\phi)).
\la{L_SL}
\een
Hence, to solve the cosmological constant problem, 
in dS-vacuum state we must have
\ben
{}_{{}_T\!}\sigma_{\!{}_S}(\bar\phi)=
{\frac 1 D}\ln
\left(\!{\frac {{}_{{}_S\!}\Lambda(\bar\phi)}
{\Lambda^{obs}}}\!\right)\!=\!{\frac 1 D}\ln P^{-2},\,\,\,
\la{CCSol}
\een
where we have used the normalization (\ref{Unorm}).
For $D=4$ and for the value of Planck number given by (\ref{P}), 
we obtain ${}_{{}_T\!}\sigma_{\!{}_S}(\bar\phi)\approx 70$.
Then, in tree-level approximation (\ref{TFU0}),
we see that one needs vacuum values of the SF dilaton
\ben
\bar\phi\approx 45,\,\,\,\hbox{or}\,\,\,
\bar\phi\approx 166
\la{phi_vac}
\een
to rescale the huge value of
${}_{{}_S\!}\Lambda(\bar\phi)$ in SF
to the observed value (\ref{Lambda}) in PhF$\equiv$TF.

At present, the vacuum value of the SF dilaton $\phi$
is not known as physical quantity, and
the above values (\ref{phi_vac})
do not seem to be unacceptable.
These preliminary estimates 
indicate that it is not excluded to find the solution
of cosmological constant problem in this direction.

%%%%%%%%%%%%%%%%%%%%%%%%%%%%%%%%%%%%%%%%%%%%%%%%%%%%%%%%%%%%

\section{Concluding Remarks}

In this section, 
%we do not give a summary of all results
%obtained in the present article since we have already
%discussed them in the basic text.
we would like to stress one general result:
Our analysis of 4D-DG leads us to the conclusion that
the best way to study SUSY breaking
is to look at the sky and to try to reconstruct
the {\em real} time evolution of the Universe.

Below, we discuss some open problems in 4D-DG.
%We discuss here some of the basic open problems in 4D-DG
%which need to be considered
%for the further developments of this model.

The main open physical problem at the moment
seems to be the precise determination of the dilaton mass.
The restriction (\ref{m_Phi}) on it is too weak.
It is convenient to have a dilaton $\Phi$ with mass
$m_{{}_\Phi}$ in the range $10^{-3}$--$10^{-1}\,{\rm eV}$.
In this case, the dilaton will not be able to decay
into other particles of SM,
since they would have greater masses \cite{E-F_P}.
On the other hand, in 4D-DG we do not need such a 
suppressing mechanism since a direct interaction of
the dilaton $\Phi$ with matter of any kind
is forbidden by the WEP.
This gives us the freedom to enlarge
significantly the mass of the dilaton
without contradiction with the known physical experiments.
One of the important conclusions of
the present article is that $m_{{}_\Phi}$ is related
to the time duration of the inflation.
One is tempted to try a new speculation -- 
to investigate a 4D-DG with
dilaton mass, $m_{{}_\Phi}$, in the domain
$100\,{\rm GeV}$--$1\,{\rm TeV}$,
and dilaton Compton wave length, $l_{{}_\Phi}$, 
between $10^{-18}$ and $10^{-16}\,{\rm cm}$.
In this case the time-duration of inflation, $t_{{}_\Phi}$,
will be of order of $10^{-28}\,{\rm sec}$; 
the dimensionless dilaton parameter, $p_{{}_\Phi}$,
will be about $10^{-45}$,
and the ultra-high frequency, $\omega_{{}_\Phi}$,
of dilatonic oscillations during
de Sitter asymptotic regime will be approximately 
$10^{19}\,{\rm GHz}$.
Such new values of the basic dilaton parameters
are very far from the Planck scales.
They seem to be accessible for the particle accelerators
in the near future, and raise new physical problems.

Our 4D-DG is certainly not applicable to time
instants smaller then, or of order of, the Planck time,
$t_{Pl}\sim 10^{-44} \,{\rm sec}$, because 4D-DG is
a low-energy theory and ignores quantum
corrections in (S)ST. One must take into account
these corrections in order to obtain a correct physical
description in this domain.

The most important open problems in the development of
a general theoretical framework of 4D-DG
are the detailed theory of cosmological perturbations,
structure formation, and possible consequences
of our model for the CMB parameters.
The properties of the solutions of the basic equation
for linear perturbations $\delta\Phi$ differ
essentially from the ones in other cosmological models
with one scalar field. This equation yields a strong
``clusterization'' of the dilaton $\Phi$ at very small
distances \cite{E-F_P}.
Actually, the equation for dilaton perturbations
shows the existence of ultra-high frequency
oscillations (described in Section VI.C.1)
and non-stationary gravi-dilaton waves
with length $l_{{}_\Phi}\leq 10^{-2}\,{\rm cm}$.
Such new phenomena cannot be viewed as
a clusterization at astrophysical scales.
Thus, their investigation as unusual cosmological
perturbations is an independent interesting issue.
For example, it is interesting to know whether it is possible
to consider these space-time oscillations as
a kind of dark matter in the Universe.

A more profound description of the inflation in 4D-DG,
both in the absence and in the presence of matter and
space-curvature, is needed.
It requires correct averaging of dilatonic oscillations.

Other open problems are
the development of the theory of binary systems,
relativistic collapse, gravi-dilaton waves,
and other non-static phenomena.
A special attention must be paid
to the search for exact analytical or numerical solutions
like the black holes in 4D-DG,
where we have no asymptotically flat space-time.

For completeness, we would like to mention that in
the present article we have ignored one basic property
of physics in the phenomenological frame,
namely, the well-known barion-anti-barion asymmetry.
It seems to us that this phenomenon can
be naturally connected with an anti-symmetric
(axion-like) field of the universal sector of (S)ST
which we have ignored so far.

We intend to present the corresponding results elsewhere.

\begin{acknowledgments}

I am deeply grateful to Professor Steven Weinberg
for his invitation to join
the Theory Group at the University of Texas at Austin
and for his kind hospitality.
I also wish to thank Prof.\ Steven Weinberg,
Prof.\ Willy Fischler and the members of the Theory Group
for the stimulating and illuminating discussions
and invaluable help during my stay in Austin.

I am indebted to Prof.\ Sergei Odintsov,  to Prof.\ C.~J.~A.~P. Martins,
and to  Prof.\ T. Dent
for bringing important references to my attention
and for some useful suggestions.

I am grateful to Dr.\ Nikola Petrov
who carefully read the manuscript
and made numerous remarks
that enhanced the clarity of the exposition.

My visit to the University of Texas at Austin was
supported by Fulbright Educational Exchange Program,
Grant Number 01-21-01.

This research was supported in part by NSF grant PHY-0071512,
and by the University of Sofia Research Fund, contract 404/2001.

\end{acknowledgments}

\appendix

\section{Proof of Proposition 3}

The proof of Proposition 3 in Section VI.~B is based on
linearization of the system (\ref{Norm})
with respect the small deviations $\delta {\bf x}(\tau)$
from the functions (\ref{dSS}). Note that in the case
$k\neq 0$, $\epsilon \neq 0$, the functions (\ref{dSS})
are not a solution of the system (\ref{Norm}).
Therefore, the corresponding linearized system is
a non-homogeneous one:
\ben
{\frac d {d\tau}}\delta {\bf x}=
\bigl(M + N(\tau)\bigr)\delta {\bf x}+{\bf f}(\tau).
\la{dNorm}
\een
The $3\times 3$ constant matrix $M$ 
is given by the formula:
\ben
M=\pmatrix{-{\frac 4 3} &
{\frac 4 9}\left(1\!+\!{\frac 3 4}p_{{_\Phi}}^{-2}\right) & 0\cr
-1& 1& 0 \cr 1&0&0\cr}.
\la{Mmatr}
\een
It has eigenvalues $\mu_\pm=-{\frac 1 2}\pm i\omega_{{}_\Phi}$
with a huge frequency $\omega_{{}_\Phi}$ 
(given by Eq.~(\ref{omega})), and $\mu_0=0$.

The $3\times 3$ matrix $N(\tau)$ depends on $\tau$ as follows:
\ben
N(\tau)\!=\!
ke^{-{\frac 2 3}\tau}\!\pmatrix{2 & 0 & {\frac 4 3}\cr
0& -1& 2 \cr 0&0&0\cr}\!+\!
{\frac {\epsilon^\prime(\tau)} 3}\!
\pmatrix{0 & 0 & 0\cr
0& 0& 1 \cr 0&0&0\cr}.
\la{Nmatr}
\een
The inhomogeneous term in Eq.~(\ref{dNorm}) is
\ben
{\bf f}(\tau)\!=\!
-ke^{-{\frac 2 3}\tau} \!\pmatrix{2\cr 1 \cr 0\cr}\!+\!
{\frac {\epsilon(\tau)} 3}\!
\pmatrix{0\cr 1 \cr 0\cr}.
\la{fmatr}
\een
In these formulae, in accordance with 
Eq.~(\ref{epsilon_p}) and Eq.~(\ref{dSS}), we have
$\epsilon(\tau) = \sum_n \Phi_n e^{- n\tau/3}$
and
$\epsilon^\prime(\tau)=-\sum_n n\Phi_n e^{-n\tau/3}$.
Hence, for the norms
$||N(\tau)||$ and $||{\bf f(\tau)}||$, 
we obtain an asymptotic behavior 
$\sim e^{-\nu\tau/3}\to 0$
as $\tau\to \infty$, where $\nu=\min\{2,n\}>0$.
We have $\nu=2$, and the asymptotics of the above two norms
are dominated by space-curvature term if $k\neq 0$, and the
condition $n\in [3,4]$ for the matter is fulfilled.
(Note that it is possible to prove Proposition 3
under weaker requirements, then the condition
$n\in [3,4]$ for matter.)

The solution of the system (\ref{dNorm}) can be represented in the form
$\delta {\bf x}(\tau)=e^{M\tau}\delta {\bf y}(\tau)$, where
$$\delta {\bf y}(\tau)=\tilde U_{\tau\tau_0}\delta {\bf y}_0
+\int_{\tau_0}^\tau \tilde U_{\tau\tau^\prime}
\tilde {\bf g}(\tau^\prime) d\tau^\prime.$$
Here
$\tilde U_{\tau\tau_0}=
T\!\!-\!\exp\left(\int_{\tau_0}^\tau
\tilde N(\tau^\prime)d\tau^\prime\right)$, 
and $\tilde N(\tau)= e^{-M\tau}N(\tau)e^{M\tau}$.
Hence, we have the estimate
$||\tilde U_{\tau\tau_0}||\leq
\exp\left(\int_{\tau_0}^\tau
\tilde ||N(\tau^\prime)||d\tau^\prime\right)\leq
\exp\left({\rm const}\int_{\tau_0}^\tau
e^{\left(1/2-\nu/3\right)\tau}d\tau \right)
\to {\rm const} < \infty$ as $\tau\to \infty$ if $\nu> 3/2$.
Under the same condition, we obtain
$||\int_{\tau_0}^\tau \tilde U_{\tau\tau^\prime}
\tilde {\bf g}(\tau^\prime) d\tau^\prime||\leq
\int_{\tau_0}^\tau ||\tilde U_{\tau\tau^\prime}||\,
||\tilde {\bf g}(\tau^\prime)|| \,d\tau^\prime\leq
{\rm const} \int_{\tau_0}^\tau e^{\left(1/2-\nu/3\right)\tau}d\tau
\to 0$.

As a result, under the condition $\nu> 3/2$, we obtain
the asymptotic behavior
$$\delta {\bf y}(\tau)\to \delta {\bf z}_0={\rm const} \,\,\,
\hbox{with}\,\,\,\,\,\, ||\delta {\bf z}_0||<\infty,$$
which in turn yields
$$\delta {\bf x}(\tau)\sim e^{M\tau}\delta{\bf z}_0.$$
This is a solution of the homogeneous modification of
the system (\ref{dNorm}) with $N=0$ and ${\bf f}=0$ under
the initial condition $\delta {\bf z}_0$. This solution
is described in Proposition 3.

Let us mention that in the mathematically 
simpler case of one-dimensional equation (\ref{dNorm})
(when $M$ and $N$ are numbers instead of non-commutative matrices), 
the condition $\nu > 3/2$ is not needed for the proof. 
However, since this condition is more than enough 
to cover all physically interesting cases of standard matter,
we will not look for stronger mathematical results.
\\
%\vskip .5truecm

\section{Some Properties of the Solutions of Eq.~(\ref{zg})}

Here we give some properties of the solutions of Eq.~(\ref{zg})
for potentials (\ref{Vnu}) with $\nu_{\pm}>0$.
In this case we obtain
\ben
v_{\nu_{+}\nu{-}}(g)\!=\!{\frac {16}{3(\nu_{+}\!+\!\nu_{-})}}
\left({\frac {g^{-\nu_{+}}\!-\!1} {\nu_{+}} }\!+\!
{\frac {g^{\nu_{-}}\!-\!1} {\nu_{-}} }\right),
\nonumber\\
w_{\nu_{+}\nu{-}}(g)\!=\!{\frac {8}{(\nu_{+}\!+\!\nu_{-})}}
\Bigl({\frac {g^{1\!-\!\nu_{+}}} {\nu_{+}\!-\!3}}\!+\!
{\frac {g^{1\!+\!\nu_{-}}} {\nu_{-}\!+\!3} }\!-\!
\nonumber \\
{\frac {(\nu_{+}\!+\!\nu_{-})g^{-2}}
{(\nu_{+}\!-\!3)(\nu_{-}\!+\!3)}}\Bigr):
\hbox{for}\,\nu_{+}\neq 3,
\nonumber\\
w_{3\nu{-}}(g)\!=\!{\frac {8 g^{-2}}{(3\!+\!\nu_{-})}}
\Big(
{\frac {g^{3\!+\!\nu_{-}}\!-\!1} {3\!+\!\nu_{-}}}\!-\!
\ln g\Bigr):\hbox{for}\,\nu_{+}\!=\!3.\,
\la{vw}
\een
Then, 
\ben
{\cal Z}_{\nu_{+}\nu_{-}}^{(0,0)}(g,z)=
-{\frac{4z^4}{3(\nu_{+}\!+\!\nu_{-})}}
\left(g^{6-\nu_{+}} - g^{6+\nu_{-}}\right)
\la{Znu}
\een
and
\ben
{\cal D}_{\nu_{+}\nu_{-}}^{(0,0)}(g,z)\!=\!
1\!-\!{\frac {8z^4}{3(\nu_{+}\!+\!\nu_{-})}}
\Bigl({\frac {g^{6\!-\!\nu_{+}}} {\nu_{+}\!-\!3}}\!+\!
{\frac {g^{6\!+\!\nu_{-}}} {\nu_{-}\!+\!3} }\!-\!
\nonumber \\
{\frac {(\nu_{+}\!+\!\nu_{-})g^{3}}
{(\nu_{+}\!-\!3)(\nu_{-}\!+\!3)}}\Bigr)
\!-\!\left({\frac{4p_{{}_\Phi}}{3}}\right)^2 z^4 g^3:
\hbox{for}\,\nu_{+}\neq 3,
\nonumber\\
{\cal D}_{3\nu_{-}}^{(0,0)}(g,z)\!=\!
1\!-\!{\frac {8z^4g^{3}}{3(3\!+\!\nu_{-})}}
\Big({\frac {g^{3\!+\!\nu_{-}}\!-\!1} {\nu_{-}\!+\!3}}\!-\!
\ln g\Bigr)\!-
\nonumber \\
\!\left({\frac{4p_{{}_\Phi}}{3}}\right)^2 z^4 g^3:\,\,\,\,\,\,\,
\hbox{for}\,\nu_{+}\!=\!3.\hskip .7truecm
\la{Dnu}
\een
Here the upper index $(0,0)$ indicates
the case $k\!=\!0$, $\epsilon\!=\!0$.

a) As we see, for all values $0<\nu_{+}<6$
$$\lim_{g\to 0}{\cal D}_{\nu_{+}\nu_{-}}^{(0,0)}(g,z)=1,$$
and, taking into account only the leading
terms, we obtain from Eq.~(\ref{Norm2}), (\ref{Ddef}), (\ref{TN}),
(\ref{zg}), and (\ref{D00}) the following results in the limit $g\to 0$:
\ben
a(g)\sim g,\hskip 5.2truecm
\nonumber\\
z(g)\!=\!z_0\!-
{\frac{4 z_0^5 g^{6-\nu_{+}}}{3(\nu_{+}\!+\!\nu_{-})(6-\nu_{+})}}
+{\cal O}_{7-\nu_{+}}(g),\nonumber\\
t(g)\!=\!{\frac{4p_{{}_\Phi}}{\sqrt{3}}}z_0^2 g^2
\!+\!{\cal O}_3(g).\hskip 3.05truecm
\la{azt_g<6}
\een

b) For $6<\nu_{+}$, we have 
$$\lim_{g\to 0}g^{\nu_{+}-6}
{\cal D}_{\nu_{+}\nu_{-}}^{(0,0)}(g,z)=
{\frac {8z^4}{3(\nu_{+}\!+\!\nu_{-})(\nu_{+}\!-\!3)}},$$
and the leading terms are different. Therefore, 
in this case we obtain in limit $g\to 0$
\ben
a(g)\sim g,\,\,\,
z(g)=z_1g^{\frac{\nu_{+}-3}{2}}
+{\cal O}_{\frac{\nu_{+}-1}{2}}(g),\nonumber\\
t(g)\!=\!{\frac { 4p_{{}_\Phi} } {\sqrt{3}} }
{\frac {z_1^2} {\nu_{+}-1} } g^{\nu_{+}-1}\!+
\!{\cal O}_{\nu_{+}}(g).\hskip 0.8truecm
\la{azt_g>6}
\een

These formulae show:

1) the existence of the Beginning,
i.e., the existence of a time instant $t=0$ at which $a(0)=0$;

2) the zero value of gravity at the Beginning: $g(0)=0$;

3) the finiteness of the time interval needed for
reaching nonzero values of $a$ and $g$,
starting from the Beginning;

4) the constancy of $z(t)$ at the Beginning;

5) when solved with respect to time $t$, they give
Eqs. (\ref{nu<6}) and (\ref{nu>6}).

In addition, we can derive another important result:
the time $t$ and the number of e-folds ${\cal N}$
increase by a finite amount when the solution of
Eq.~(\ref{zg}) crosses the zero line $Z[2]$ -- 
${\cal D}=0$, although the denominator
in the integrals (\ref{N}) and (\ref{TN}) vanishes.
Indeed, taking into account that the points where
the solution crosses the curve $Z[2]$ are extreme
points of function $g(z)$, and using the expansion
$\Delta z=
\left(-{\frac 3 8}z{\frac {gv{,_g}}u}\Delta g\right)^{1/2}
+{\cal O}_1(\Delta g)$ in a vicinity of
such a point, we easily obtain 
\ben
\Delta {\cal N}= z^{-4}g^{-6}u^{-1}
\sqrt{-{\frac{6u}{gv{,_g}}} \Delta g}+{\cal O}_1(\Delta g),
\nonumber \\
\Delta t={\frac{4p_{{}_\Phi}}{\sqrt{3}}}z^{-2}g^{-4}u^{-1}
\sqrt{-{\frac{6u}{gv{,_g}}} \Delta g}+{\cal O}_1(\Delta g)
\la{DeltaNt}
\een
for potentials $u(g)$ and $v(g)$ of the most general type.
Here the values of the coefficients are taken on the curve
${\cal D}=0$.

Combined with the previous results, this proves that the time
intervals of inflation $t_{infl}^{(i)}$ are finite for all values
of $i=0,1,\ldots$.

\end{document}